\documentclass[twocolumn]{aastex631}

\usepackage{footmisc} 

\newcommand{\nova}{PGIR22akgylf}

\shorttitle{{\em TESS} photometry of \nova{}}
\shortauthors{Sokolovsky et al.}

\begin{document}
\title{{\em TESS} detection of periodic brightness variations during the rise of classical nova \nova{}}

\correspondingauthor{Kirill~Sokolovsky}
\email{kirx@kirx.net}

\author[0000-0001-5991-6863]{Kirill~V.~Sokolovsky}
\affiliation{Department of Astronomy, University of Illinois at Urbana-Champaign, 1002 W. Green Street, Urbana, IL 61801, USA}
\affiliation{Department of Physics \& Astronomy, Texas Tech University, Box 41051, Lubbock, TX 79409-1051, USA}

\author[0000-0001-8525-3442]{Elias~Aydi}
\affiliation{Department of Physics \& Astronomy, Texas Tech University, Box 41051, Lubbock, TX 79409-1051, USA}

\author[0000-0001-7179-7406]{Konstantin Malanchev}
\affiliation{Department of Astronomy, University of Illinois at Urbana-Champaign, 1002 W. Green Street, Urbana, IL 61801, USA}
\affiliation{McWilliams Center for Cosmology \& Astrophysics, Department of Physics, Carnegie Mellon University, Pittsburgh, PA 15213, USA}

\author[0000-0002-8286-8094]{Koji~Mukai}
\affiliation{CRESST and X-ray Astrophysics Laboratory, NASA/GSFC, Greenbelt, MD 20771, USA}

\author{Jennifer~L.~Sokoloski}
\affiliation{Columbia Astrophysics Laboratory, Columbia University, New York, NY 10027, USA}

\author[0000-0002-8400-3705]{Laura~Chomiuk}
\affiliation{Center for Data Intensive and Time Domain Astronomy, Department of Physics and Astronomy, Michigan State University, 567 Wilson Rd, East Lansing, MI 48824, USA}

\author[0000-0002-3673-0668]{Peter~Allen~Craig}
\affiliation{Center for Data Intensive and Time Domain Astronomy, Department of Physics and Astronomy, Michigan State University, 567 Wilson Rd, East Lansing, MI 48824, USA}

\author[0000-0002-0476-4206]{Rebekah~A.~Hounsell}
\affiliation{University of Maryland, Baltimore County, Baltimore, MD 21250, USA}
\affiliation{NASA Goddard Space Flight Center, Greenbelt, MD 20771, USA}

\author[0000-0002-3873-5497]{Justin~D.~Linford}
\affiliation{National Radio Astronomy Observatory, Domenici Science Operations Center, 1003 Lopezville Road, Socorro, NM 87801, USA}

\author[0009-0004-4418-0645]{Isabella~Molina}
\affiliation{Center for Data Intensive and Time Domain Astronomy, Department of Physics and Astronomy, Michigan State University, 567 Wilson Rd, East Lansing, MI 48824, USA}

\author{Montana~N.~Williams}
\affiliation{National Radio Astronomy Observatory, Domenici Science Operations Center, 1003 Lopezville Road, Socorro, NM 87801, USA}

\nocollaboration{32}

\author[0000-0002-8989-0542]{Kishalay De}
\affiliation{Department of Astronomy and Columbia Astrophysics Laboratory, Columbia University, New York, NY, USA}
\affiliation{Center for Computational Astrophysics, Flatiron Institute, New York, NY, USA}

\author[0000-0002-5619-4938]{Mansi M. Kasliwal}
\affiliation{Cahill Center for Astrophysics, California Institute of Technology, Pasadena, CA 91125, USA}

\author[0000-0001-6627-9903]{Nicholas Earley}
h\affiliation{Cahill Center for Astrophysics, California Institute of Technology, Pasadena, CA 91125, USA}

\collaboration{3}{(Gattini IR)}

\author[0000-0002-6097-8719]{David~J.~Lane}
\affiliation{Abbey Ridge Observatory, 45 Abbey Rd, Stillwater Lake, NS, B3Z1R1 Canada}
\affiliation{Burke-Gaffney Observatory, Saint Mary's University, 923 Robie Street, Halifax, NS B3H 3C3 Canada}
 
\author[0000-0002-5268-7735]{Filipp~D.~Romanov}
\affiliation{American Association of Variable Star Observers (AAVSO), 185 Alewife Brook Parkway, Suite 410, Cambridge, MA 02138, USA}
\affiliation{Abbey Ridge Observatory, 45 Abbey Rd, Stillwater Lake, NS, B3Z1R1 Canada}

\author{Richard Schmidt}
\affiliation{AAVSO Observer}
 
\collaboration{3}{(AAVSO)}




\begin{abstract}
Classical novae are transient events powered by thermonuclear burning in
a layer of hydrogen-rich material accreted by a white dwarf from its binary
companion. Most classical novae reach optical maximum within $\sim1$\,d, but 
a rare few rise far more slowly.
We probe the envelope structure and ejection mechanism of the
slowly-rising nova \nova{} with {\em TESS} photometry spanning 
3 to 16\,d after the nova discovery, supplemented by ground-based 
observations that cover its full $\sim133$\,d ascent to maximum.
We detect a $0.1802 \pm 0.0012$\,d periodic brightness modulation with
a peak-to-peak amplitude of $\sim0.02$\,mag, identified with \nova{}
via temporal and spatial coincidence.
The period is stable over the two weeks of {\em TESS} coverage,
suggesting an orbital origin.
Whether this period corresponds to the full or half orbital period, it
implies a dwarf donor companion.
At the time of the {\em TESS} observations the nova was $\gtrsim6$\,mag
above quiescence (but still 4\,mag below peak), 
so its light should be dominated by the expanding photosphere.
We interpret the periodic signal as arising from the binary orbital motion
distorting the nova envelope while its size remains comparable to 
the binary separation. 
This interpretation points to common-envelope interaction as a contributor to shell
ejection in \nova{} and demonstrates that the slow-rise phenomenon is not
exclusive to thermonuclear eruptions in symbiotic binaries, where the
large orbital separation of the giant companion inhibits such interaction.
\end{abstract}



\section{Introduction}
\label{sec:intro}

Classical novae are luminous transient phenomena arising from thermonuclear runaway 
on the surfaces of white dwarfs accreting hydrogen-rich material from companion stars in close binary systems
\citep{2016PASP..128e1001S}.
As the accreted layer accumulates on the white dwarf surface, 
increasing pressure and temperature at its base eventually trigger a thermonuclear runaway 
(explosive fusion of hydrogen into helium), causing rapid energy release and dramatic expansion 
and eventual ejection of the envelope with typical velocities in the range
500-5000\,km\,s$^{-1}$. 
The optical brightness of the binary rises by 8--18 magnitudes, 
reaching peak absolute magnitudes of $-4$ to $-10$\,mag \citep{1975Natur.258..501L,1990ApJ...356..609V,2021ApJ...910..120K}
and then declines on a timescale of days to months \citep{2010AJ....140...34S}. 
The nuclear burning on the white dwarf, manifested by ``super-soft source'' (SSS; \citealt{1997ARA&A..35...69K}) X-ray emission, 
continues on similar or even longer timescales until the hydrogen fuel is exhausted 
\citep{2011ApJS..197...31S,2013A&A...559A..50N,2024A&A...689A.335T}.
On average $11 \pm 1$ novae were discovered per year over the last
decade\footnote{\url{https://asd.gsfc.nasa.gov/Koji.Mukai/novae/novae.html}} 
with the total Galactic nova rate estimated to be around 30  
\citep{2017ApJ...834..196S,2022ApJ...937...64K,2022ApJ...936..117R}
to 45 \citep{2021ApJ...912...19D,2023MNRAS.523.3555Z} events per year with
the majority of them being hidden by interstellar dust.

While the thermonuclear reactions restarted on the white dwarf are clearly
the primary energy source of a nova, the actual physical processes acting to
eject the nova envelope and their relative role are actively debated
\citep[e.g, section~2.2 of][]{2021ARA&A..59..391C}.
An initial, {\it impulsive ejection} is thought to occur almost contemporaneously with the thermonuclear runaway.
Convection transports short-lived $\beta$-unstable nuclei ($^{13}$N, $^{14}$O, $^{15}$O, $^{17}$F; \citealt{2016PASP..128e1001S})
from the reaction zone to the surface. 
As these nuclei decay on timescales of minutes to hours, they release additional heat throughout the
envelope. The decays drive a continuing expansion even after the initial runaway, often pushing the luminosity above the Eddington limit 
and initiating the envelope's escape from the white dwarf's gravity. 
As the system rises to maximum, the photosphere expands along with the expanding envelope
- this is referred to as ``fireball stage'' \citep[e.g.,][]{aydi2018,2020PhDT........48P}.

\cite{2022ApJ...938...31S} argue that only a small fraction of the envelope
can be ejected this way, while most of the envelope is expanding 
slowly remaining bound to the white dwarf until it reaches the companion
star that may use a fraction of its orbital energy to unbind the envelope
via a {\it common envelope} interaction \citep{1983ApJ...273..280K,1990ApJ...356..250L,2021ApJ...914....5S}.
Such interaction may produce an inhomogeneous low-velocity outflow concentrated in the orbital plane
of the binary \citep{2016MNRAS.461.2527P}. The outflow velocity is expected
to be comparable to the orbital velocity of the binary.
Direct observational support for such delayed, binary-shaped ejection has 
recently come from near-infrared interferometric imaging of the very slow nova
V1405\,Cas, in which \citet{2026NatAs..10..271A} resolved a compact photosphere
engulfing the binary for more than 50\,days during the rise to visible peak,
before the bulk of the envelope was expelled.

Finally, the continuing nuclear burning on the white dwarf makes it a source
of a fast, {\it radiation-driven wind}
\citep{1990LNP...369..244F,1994ApJ...437..802K,2004BaltA..13..116F,2001MNRAS.326..126S,2002ASPC..261..585S}.
A nova reaches its maximum optical brightness when the photosphere reaches
its maximum radius. Spread beyond this maximum radius the ejecta becomes
too disperse and transparent, so while the nova shell continues its
expansion, the photosphere detaches from it and starts to shrink as the rate
of mass loss (and thus the density of the wind at a given radius) decreases.

An interface between the fast wind and the slow, orbital-plane-focused envelope 
is a natural site to expect formation of shocks \citep{2014Natur.514..339C,2021ARA&A..59..391C}.
Shocks produce non-thermal emission in novae observed at TeV \citep{2022NatAs...6..689A,2022Sci...376...77H,2025A&A...695A.152A} and 
GeV $\gamma$-rays \citep{2010Sci...329..817A,2014Sci...345..554A,2018A&A...609A.120F}, 
as well as in radio alongside thermal emission \citep{2021ApJS..257...49C}. 
Reprocessed thermal X-rays from shocks may be a major contribution to 
nova optical light \citep{2017NatAs...1..697L,2017MNRAS.469.4341M,2020NatAs...4..776A}.

The onset of the thermonuclear runaway is manifested by a short-lived
(hours-long) X-ray flare \citep{2022Natur.605..248K}. As the white dwarf
atmosphere expands and cools, the peak of the nova's spectral energy
distribution moves into optical band producing the initial rise in the nova
lightcurve.
The rising portion of a nova lightcurve may potentially serve as diagnostics of 
dominating mass-loss mechanism. The transition from convection-limited
energy transport in a still-bound envelope to radiation-driven mass
loss may manifest itself as the pre-maximum halt
\citep{1995PASP..107.1201P,2014MNRAS.437.1962H,2017MNRAS.467.2684E}.
The transition from a slow, equatorially focused
common-envelope outflow to a faster optically thick wind
\citep{2021ARA&A..59..391C,2022ApJ...938...31S} may produce
additional inflections, multiple peaks, or shock-powered flares as
the fast and slow components collide. Each of these features is
short-lived and easily missed by ground-based, diurnally interrupted
monitoring. 

In most classical novae the initial optical brightening occurs very rapidly.
They climb most of the way to maximum in less than 24 hours (the recent
well-documented examples of such fast rise include V1405\,Cas,
\citealt{2023ApJ...958..156T}, RS\,Oph, \citealt{2022ApJ...935...44C}). This fast rise
may be interrupted by a pre-maximum halt \citep{2004ApJ...612L..57H,2018arXiv180707947P} pronounced in lightcurves of
some novae and completely indistinguishable in others, 
that is followed by a slower final 1-2\,mag climb to maximum that may take anywhere from hours 
to weeks and longer 
($\sim300$\,d for HR\,Del \citealt{1967ATsir.447....7S,1970PASP...82..889B,1974A&AS...15..107T}; 
$\sim100$\,d for V723\,Cas, \citealt{1997CoSka..27...53C,2005Ap&SS.296..431S}; 
$\sim14$\,d for V5856\,Sgr, \citealt{2017NatAs...1..697L,2022ApJ...941..138W}; 
$\sim 50$\,d for V1405\,Cas; \citealt{2023arXiv230204656V,2023ApJ...958..156T}).
The rapid rise is often missed and has been observed only for a few novae
that erupted within the field of view of the instruments continuously
imaging a large portion of the sky from space 
\citep{2014MNRAS.438.3483H,2016ApJ...820..104H,2017MNRAS.467.2684E,2017MNRAS.470.4061T,2020NatAs...4..776A}
or from the ground \citep{2024ApJ...977...17Q}.
\cite{2009ApJ...690.1148S} refer to a nova that took more than five days to
reach its peak as ``slowly rising''. 
\cite{2024ApJ...977...17Q} report an exceptionally detailed rise lightcurve
of a very fast nova V1674\,Her that shows no pre-maximum halt, but instead
reveals a presence of yet another slow brightening phase prior to the onset
of the fast rise.

A few novae have been documented having no obvious rapid rise stage 
and brighten very gradually over timescales of up to months,
while spectroscopically presenting themselves as slow classical novae.
Recent examples of such very slow-rising novae: 
V2891\,Cyg (PGIR\,19brv) climbed for about 50\,d to its first peak \citep{2020JAVSO..48...13S}, 
Gaia22alz that took 180\,d to climb to
its peak \citep{2023MNRAS.524.1946A}. 

If the accreted hydrogen envelope is non-degenerate 
(possibly because of the low mass of the white dwarf) the onset of
thermonuclear reactions might be gradual, non-eruptive \citep{2025CoSka..55c..47M}. 
This is indeed observed as years- to decades-long thermonuclear-powered eruptions in symbiotic binaries
containing a white dwarf accreting from a giant (rather than dwarf) companion
\citep{1983ApJ...273..280K,2008ASPC..401...42M,2012BaltA..21....5M}. 
In the absence of impulsive ejection and common envelope interaction, a low-mass white dwarf
may fail to eject the envelope via the wind action alone, retaining nuclear fuel and 
prolonging the nuclear burning phase leading to much longer decline times observed 
in thermonuclear-powered symbiotic eruptions compared to classical novae \citep{2009ApJ...699.1293K}.

Some symbiotic binaries host classical (fast, explosive) nova eruptions \citep{2025CoSka..55c..47M}, 
sometimes referred to as ``embedded novae'' \citep{2021ApJ...910..134G,2021ARA&A..59..391C} 
with V407\,Cyg \citep{2010Sci...329..817A} and RS\,Oph \citep{2025A&A...695A.152A} being prime examples. 
The existence of these embedded novae suggests, that the envelope ejection
in what looks like a fairly typical fast nova eruption may be achieved
without the contribution of the common envelope interaction.
According to \cite{2016ApJ...817..143W}, 30\% of novae might actually be
hosted by giant-donor systems, based on their search for quiescent 
counterparts of M31 novae in archival {\em HST} images. 
This suggest the possibilities that the very slow rising novae may either be 
hosted by symbiotic binaries (not yet recognized as such). Alternatively they 
may be dwarf-donor analogs to non-explosive thermonuclear eruptions in symbiotic binaries
by resembling them in some way: non-degenerate hydrogen shell resulting in no impulsive ejection, 
ineffective common envelope interaction with wind alone being unable to
produce a rapidly expanding photosphere.

In this paper we try to tap these questions by investigating an exceptionally detailed lightcurve of nova
\nova{} obtained during its very slow rise to maximum by {\em TESS} space photometer. 
We report the detection of the periodic modulation in 
the pre-peak lightcurve of the nova and discuss its implications for the
structure of the nova envelope following the thermonuclear runaway.
%
{\em TESS} data have previously been used to study novae during and after
eruption: \cite{2023arXiv231002220L} analyzed the {\em TESS} lightcurve of
V1674\,Her, an exceptionally fast nova hosted in an intermediate polar system
\citep{2022ApJ...940L..56P,2024CoSka..54b.128D,2026arXiv260506917O}; \cite{2026A&A...708A.352L}
reported post-eruption spin-down of the white dwarf in V1405\,Cas and orbital
modulation in V1716\,Sco; \cite{2026arXiv260522802L} measured the white dwarf
rotation period in YZ\,Ret, confirming it as an intermediate polar; and
\cite{2023arXiv231104903S} identified periodic modulation in the slow nova
V606\,Vul, which, unlike \nova{}, did undergo a rapid rise phase.
\cite{2022MNRAS.517.3640S,2023MNRAS.525..785S,2023MNRAS.519..352B,2023MNRAS.525.1953B}
and \cite{2026NewA..12602540Q} used {\em TESS} photometry to
characterize orbital periods of multiple nova-hosting systems outside of an eruption.
In Section~\ref{sec:obs} we present the discovery history of \nova{} and our
analysis of the {\em TESS} observations. In Section~\ref{sec:discussion} we
discuss the implications of the new observations for our understanding of
nova eruptions. We summarize our conclusions in Section~\ref{sec:conclusions}.

\section{Observations and analysis}
\label{sec:obs}

\subsection{\nova{} - a highly reddened classical nova}

\nova{} also known as MASTER OT\,J200029.27$+$345309.1, ZTF22abazrjk, and AT\,2022sfe 
was discovered in Cygnus on 2022-08-16.1900\,UTC ($t_0 = {\rm JD(UTC)} 2459807.6900$) by Palomar Gattini-IR survey
\citep{2020PASP..132b5001D} having the near-infrared magnitude of $J = 14.28$. 
The same night the transient was independently discovered by MASTER survey 
\citep{2010AdAst2010E..30L,2012ExA....33..173K} at the unfiltered optical
magnitude 15. A few nights later it appeared in the public transient
candidates stream of Zwicky Transient Facility \citep{2019PASP..131a8002B,2019PASP..131g8001G}
having $g = 19.2 \pm 0.1$ and $r = 17.93 \pm 0.04$ on 2022-08-22.26 ($t_0 + 6$\,d).
\cite{2022ATel15587....1D} obtained a high-resolution spectrum of \nova{} on 2022-08-29
using the Magellan Clay telescope classifying the transient as a classical nova.
%
%

As the transient position we adopt the median of 227 individual ZTF detections:
\begin{verbatim}
20:00:29.254 +34:53:09.17 J2000
\end{verbatim}
The scatter of individual measurements is $0^{\prime\prime}.06$ (median absolute
deviation scaled to standard deviation), consistent with the expectations for
a moderately bright source \citep{2019PASP..131a8003M}. As ZTF astrometry is
calibrated against Gaia, the systematic offset between the two coordinate
systems is expected to be less than $0^{\prime\prime}.014$ \citep{2019PASP..131e4504O}.
Visual inspection of Pan-STARRS \citep{2016arXiv161205560C} images reveals a blended pair of faint
sources with the northern source coinciding with the ZTF position and 
the nearest cataloged source PS1~149863001219723424 ($r = 21.7$, $i = 20.1$)
located $0^{\prime\prime}.6$ south of the ZTF position representing the combined light of the blended pair.
The galactic coordinates of \nova{} are $l = 71.29366$, $b = 2.53075$. 
The total line of sigh extinction in this direction is $A(V)= 9.22$ 
\citep{1998ApJ...500..525S}.

\subsection{Ground-based photometry}
\label{sec:groundphot}

\begin{figure*}
        \includegraphics[width=1.0\linewidth,clip=true,trim=0.0cm 0cm 0cm 0cm,angle=0]{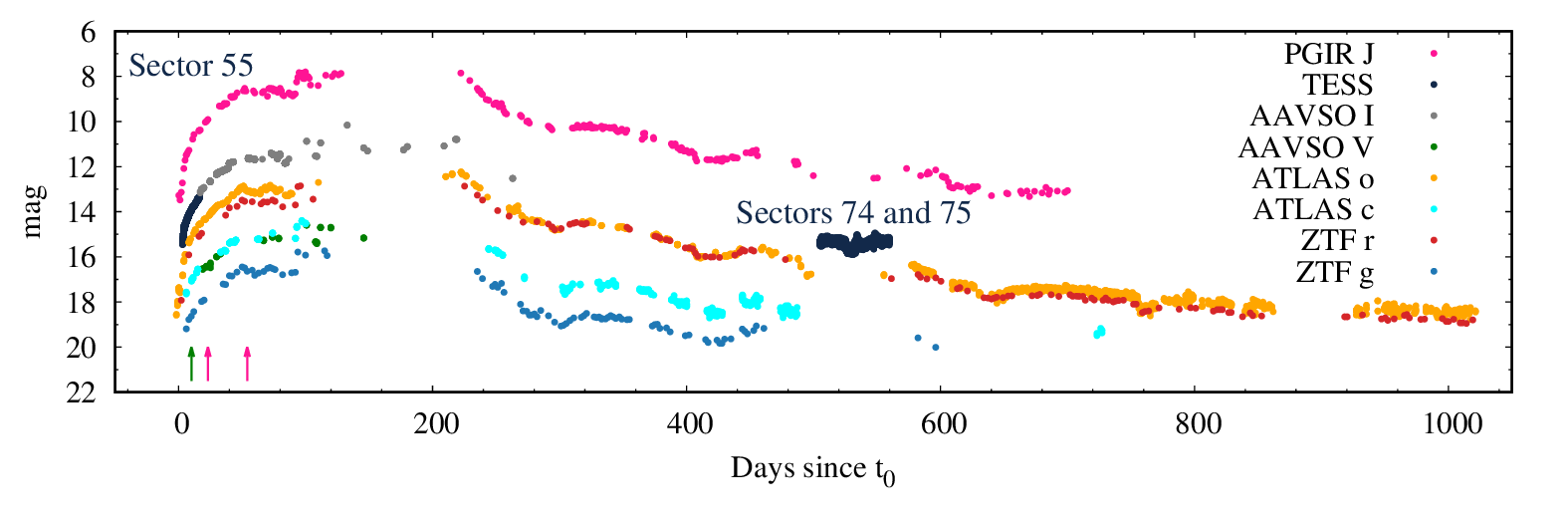}
\caption{The combined lightcurve of \nova{} presenting {\em TESS} photometry
of the nova (\S~\ref{sec:tessphot}) in the context of ground-based measurements
(\S~\ref{sec:groundphot}). 
The arrows mark times of optical (green; Figure~\ref{fig:optical_spectrum}) and infrared 
(magenta; Figure~\ref{fig:IR_spectra}) spectroscopic observations described in \S~\ref{sec:spec}.}
    \label{fig:lcall}
\end{figure*}

To construct the overall lightcurve of \nova{} (Figure~\ref{fig:lcall}) we
combine the Palomar Gattini-IR $J$ band data with 
publicly available ZTF $gr$ photometry, ATLAS cyan and orange 
filter photometry \citep{2018PASP..130f4505T,2020PASP..132h5002S} with 
$VI$ measurements collected by AAVSO observers \citep{AAVSODATA}.
ZTF PSF-fit difference image photometry and astrometry of \nova{} were obtained through the \texttt{FINK} Science Portal \citep{2021MNRAS.501.3272M}.
ATLAS PSF-fit difference image photometry was extracted from the forced photometry
server\footnote{\url{https://fallingstar-data.com/forcedphot/}}.

\nova{} was observed at $\approx 1$\,d cadence in $J$-band as part of routine 
operations 
of the 
Palomar Gattini-IR (PGIR; \citealt{2020PASP..132b5001D}) -- a wide-field near-infrared survey at Palomar Observatory
using a 30-cm telescope equipped with a Teledyne~H2RG HgCdTe CMOS detector
\citep{2011ASPC..437..383B}. 
We extracted the lightcurve of \nova{} by performing forced aperture photometry at 
the nova position with a radius of 2\,pixels ($\approx 8.7$\arcsec, corresponding to 
the characteristic size of the PGIR point spread function). 
The magnitudes are calibrated to the 2MASS system \citep{2006AJ....131.1163S}. 

The $VI$ photometric measurement shared via the AAVSO International Database were
collected by two observers, RS (1602 $I$ band measurements collected over 22
nights) and FDR (75 $I$ band points over 27 nights, 42 $V$ band points in 14 nights).
RS used a 0.32-m PlaneWave CDK astrograph equipped with a SBIG STL-1001E CCD camera and Astrodon $I_C$ filter.
The telescope is housed in a rooftop observatory in Washington~DC urban area \citep{2016MPBu...43..129S}. 
An analysis of a large fraction of RS observations was reported earlier by \cite{2022JAVSO..50..260S}.
FDR conducted observations remotely using a 0.36-m Celestron~C14 Schmidt-Cassegrain telescope
equipped with a SBIG ST-8XME CCD camera and Astrodon filters. The telescope was located in Abbey Ridge Observatory (owned by DJL) in Stillwater Lake, NS, Canada.


The combined lightcurve presented in Figure~\ref{fig:lcall} shows the nova
peaking on 2022-12-26.9296 ($t_0 + 133$\,d) at $I = 10.16$. No data in other
photometric bands are available around the peak.
The time of decline by two magnitudes in $I$ is $t_2 \approx 120$\,d.
%
Table~\ref{tab:tesscolor} presents optical colors of \nova{} computed from ZTF $gr$ and {\em TESS} $T$ magnitudes (see below) 
measured within 0.1\,d of each other. There is no significant change in color during the initial rise 
of \nova{} covered by {\em TESS}.

\begin{deluxetable}{ccccc}
\tablecaption{Optical color evolution of \nova{} during {\em TESS} Sector~55}
\label{tab:tesscolor}
\tablewidth{0pt}
\tablehead{
\colhead{Days since $t_0$} & \colhead{$T$} & \colhead{$g-T$} & \colhead{$r-T$} & \colhead{$g-r$} }
\startdata
6.1                        & 14.43         & $4.76 \pm 0.09$ &                 &                 \\
8.1                        & 14.16         & $4.61 \pm 0.10$ & $1.75 \pm 0.03$ & $2.86 \pm 0.11$ \\
10.0                       & 13.93         & $4.70 \pm 0.06$ &                 &                 \\
12.0                       & 13.75         & $4.67 \pm 0.08$ &                 &                 \\
16.1                       & 13.38         &                 & $1.72 \pm 0.03$ &                 \\
18.1                       &               &                 &                 & $3.02 \pm 0.07$ \\
\enddata
\end{deluxetable}

\subsection{Optical/IR spectroscopy}
\label{sec:spec}

\begin{figure*}
\begin{center}
  \includegraphics[width=1.0\textwidth]{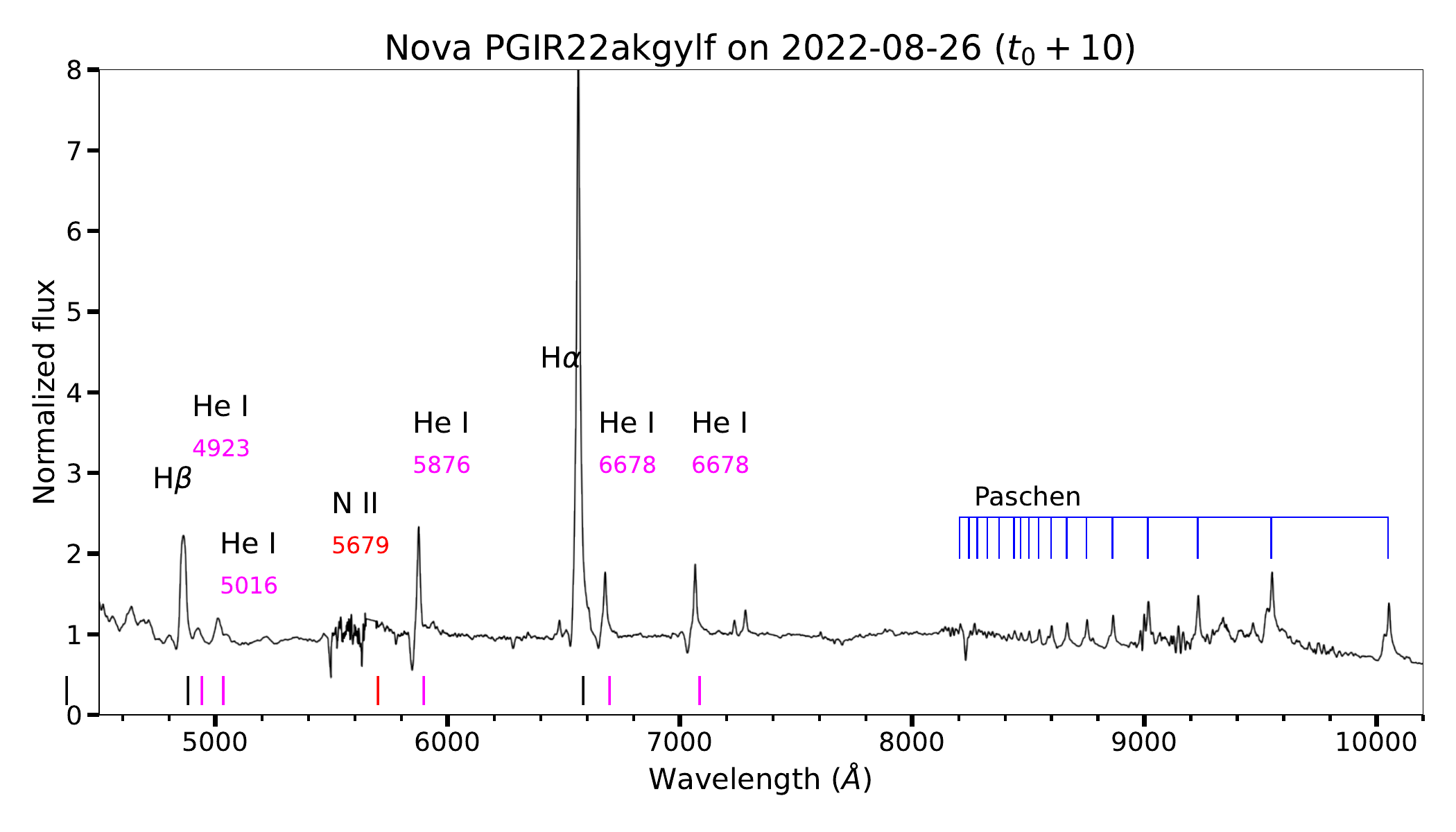}
\caption{The optical spectrum of Nova \nova{} obtained on 2022-08-26 ($t_0 + 10$\,d)
with the Double Beam spectrograph at the Palomar 200-inch telescope (\S~\ref{sec:spec}).
Spectral features are marked with colored line identifications for clarity.} 
\label{fig:optical_spectrum}
\end{center}
\end{figure*}

We obtained an optical spectrum of PGIR\,22akgylf using the Double Beam spectrograph 
(DBSP; \citealt{1982PASP...94..586O}) at the Palomar 200-inch telescope on 2022-08-26 UT ($t_0 + 10$\,d), 
for a total exposure time of 600\,s at $R\approx 1000$. 
The data were reduced using standard methods involving flat-fielding, 
wavelength calibration using lamps followed by flux calibration using a standard star. 
In Figure~\ref{fig:optical_spectrum}, we present the optical spectrum normalized to unity. 
The raw spectrum shows a red continuum with little signal below $4500\,\AA$, due to the substantial reddening toward the nova. 
The spectrum is dominated by P~Cygni lines from the Balmer series, He I, and N II. 
In the near-infrared, the Paschen series is clearly detected. 
The absorption trough of the H$\alpha$ P~Cygni profile corresponds to 
a blueshifted velocity of approximately $1500\,\mathrm{km\,s^{-1}}$. 
The spectrum is typical of a nova before peak, 
during the He/N spectral \textit{phase} \citep{2020ApJ...905...62A,2023arXiv230907097A}. 
This is consistent with the very slow light-curve evolution, with the nova reaching peak brightness more than 130 days after $t_0$
(Figure~\ref{fig:lcall}).

\begin{figure*}
\begin{center}
  \includegraphics[width=1.0\textwidth]{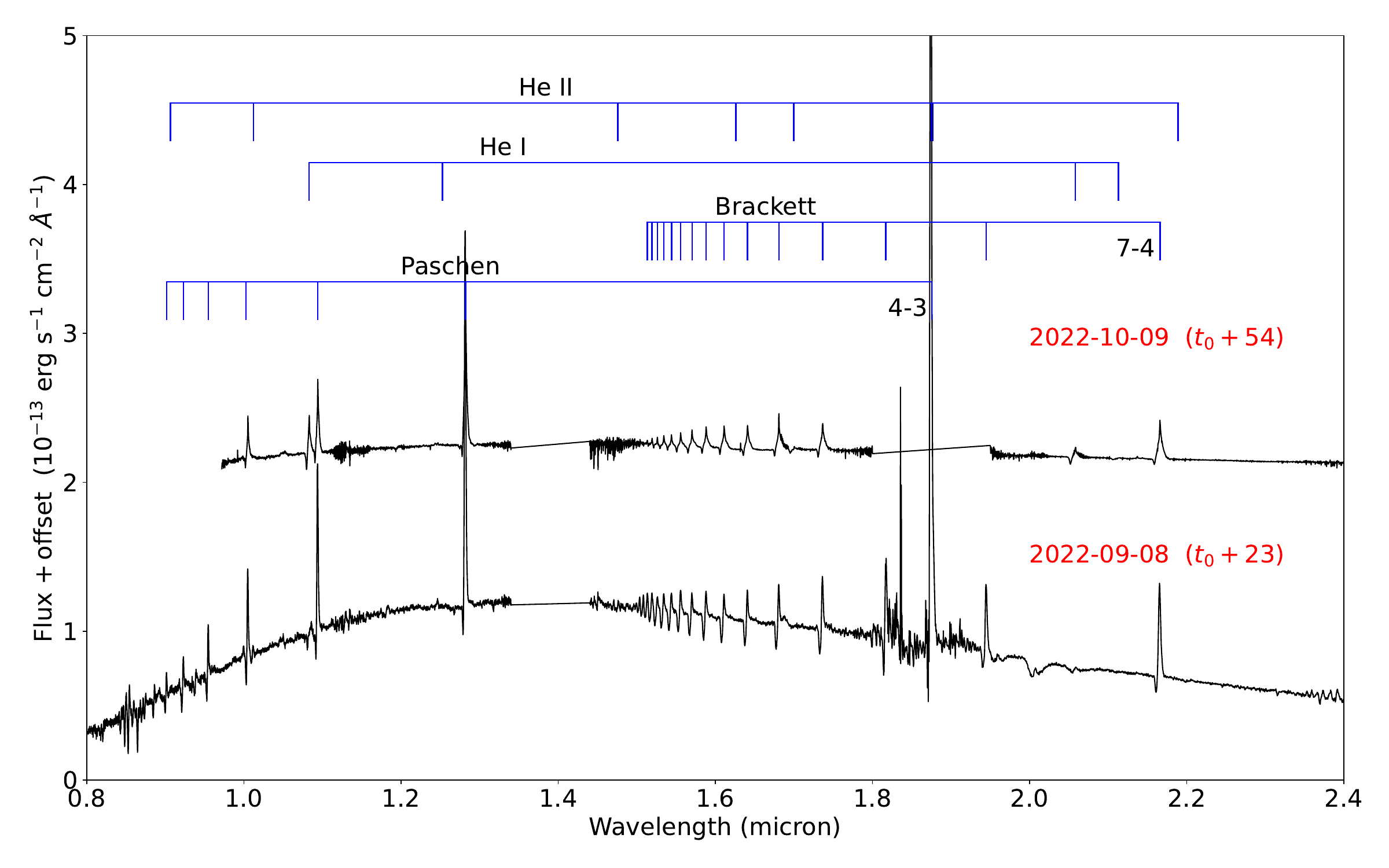}
\caption{Near-infrared spectra of Nova \nova{} obtained on 2022-09-08 ($t_0 + 23$\,d) 
with the TripleSpec spectrograph on the Palomar 200-inch telescope 
and 2022-10-09 ($t_0 + 54$\,d) using the SpeX spectrograph on the NASA Infrared Telescope Facility. 
A vertical offset has been applied to separate the spectra visually. Line identifications are marked to guide the reader.} 
\label{fig:IR_spectra}
\end{center}
\end{figure*}

On 2022-09-08 ($t_0 + 23$\,d) and 2022-10-09 ($t_0 + 54$\,d), we obtained medium resolution ($R\approx 3000$) near-IR spectra 
using the TripleSpec spectrograph on the Palomar 200-inch telescope \citep{2008SPIE.7014E..0XH} 
and the SpeX spectrograph (SXD mode; Program ID: 2022B052, PI: De) on 
the NASA Infrared Telescope Facility \citep{1998SPIE.3354..468R} respectively. 
The observations consisted of dithered exposures in the ABBA pattern amounting to a total exposure time 
of $\approx 600$\,s each. The data were reduced using the \textsc{spextool} package \citep{2004PASP..116..362C} 
followed by flux calibration using the \textsc{xtellcor} tool \citep{2003PASP..115..389V}. The near-infrared spectra, presented in
Figure~\ref{fig:IR_spectra} also shows P~Cygni lines of H~I and He~I, consistent with the optical spectra 
and indicating that the nova was still in the early He/N phase \citep{2023arXiv230907097A}.
We expect that, as the nova approaches peak, Fe II lines from multiplets (42), (38), and (39) will emerge, 
while the He and N lines will weaken relative to them. 
This behavior is commonly observed in slow novae \citep{Shore_etal_2014,2023arXiv230907097A}.

\citet{2020ApJ...905...62A} showed that most novae exhibit P~Cygni lines in their optical spectra during the rise to peak. 
Typically, the absorption components of these P~Cygni profiles have velocities ranging from a few hundred $\mathrm{km\,s^{-1}}$ 
to $\sim3000\,\mathrm{km\,s^{-1}}$ in some extreme cases \citep{2026NatAs..10..271A}. 
During the rise to peak, these absorption components decelerate as the photosphere recedes inward in velocity space. 
After peak, the spectra show faster emission components coexisting with the pre-maximum P~Cygni lines. 
\citet{2020ApJ...905...62A} associated these two distinct spectral features with separate outflows/ejecta characterized by different velocities. 
The interaction between these outflows can produce powerful shocks, which are thought to be responsible for the $\gamma$-ray emission detected 
from a growing sample of novae \citep{2020NatAs...4..776A,2026MNRAS.546f2270C}. Due to limited spectroscopic monitoring, we do not have post-peak spectra 
of this nova and therefore cannot measure the velocity of the faster component/outflow. Nevertheless, the presence of pre-maximum
P~Cygni absorption indicates that this nova followed the early spectroscopic evolution commonly seen in novae.

\subsection{{\em TESS} photometry of \nova{}}
\label{sec:tessphot}

The {\em Transiting Exoplanet Survey Satellite} \citep[{\em TESS};][]{2015JATIS...1a4003R} 
is equipped with four 105\,mm aperture f/1.4 focal ratio lenses each projecting 
a $24^\circ \times 24^\circ$ field of view on a $2\times2$ 
mosaic of $2048 \times 4096$ frame transfer CCDs 
\citep{2019AcAau.160...46K}. The cameras are red-sensitive covering 
the wavelength range of 6000--10000\,\AA.
Due to the large image scale of $20^{\prime\prime}$/pix, 
the depth of {\em TESS} images at low galactic latitudes is limited by
source confusion rather than sky and detector background - a situation
typically found in observations with single-dish radio telescopes \citep{1974ApJ...188..279C} 
but also encountered in deep optical or near-infrared imaging of crowded
fields \citep{2001AJ....121.1207H}.

A {\em TESS} sector is a continuous observation of the same
$24\degr \times 96\degr$ strip of sky, tiled by the four onboard
cameras (each imaging its own 
portion of the strip) and interrupted only by brief pauses for data downlink. 
(The high-gain antenna is fixed to the body of the spacecraft, so the
spacecraft has to be rotated so the antenna points to a ground station.)
The full-frame image cadence has been progressively shortened
over the course of the mission: 1800\,s in Sectors~1--26
(2018--2020), 600\,s in Sectors~27--55 (2020--2022), and 200\,s in
Sectors~56 onward (2022--present). Each full-frame image is constructed onboard
by stacking $N=10$ shorter sub-exposures, with the two extreme
samples in each pixel dropped for cosmic-ray rejection, so the
effective on-source integration time is reduced by a factor of
$(N-2)/N = 0.8$ relative to the nominal cadence
\citep{tesshandbook}.

\subsubsection{Periodic signal in Sector~55}

\begin{figure*}
\centering
        \includegraphics[width=0.48\textwidth,clip=true,trim=5.0cm 0.5cm 2.0cm 0.8cm,angle=0]{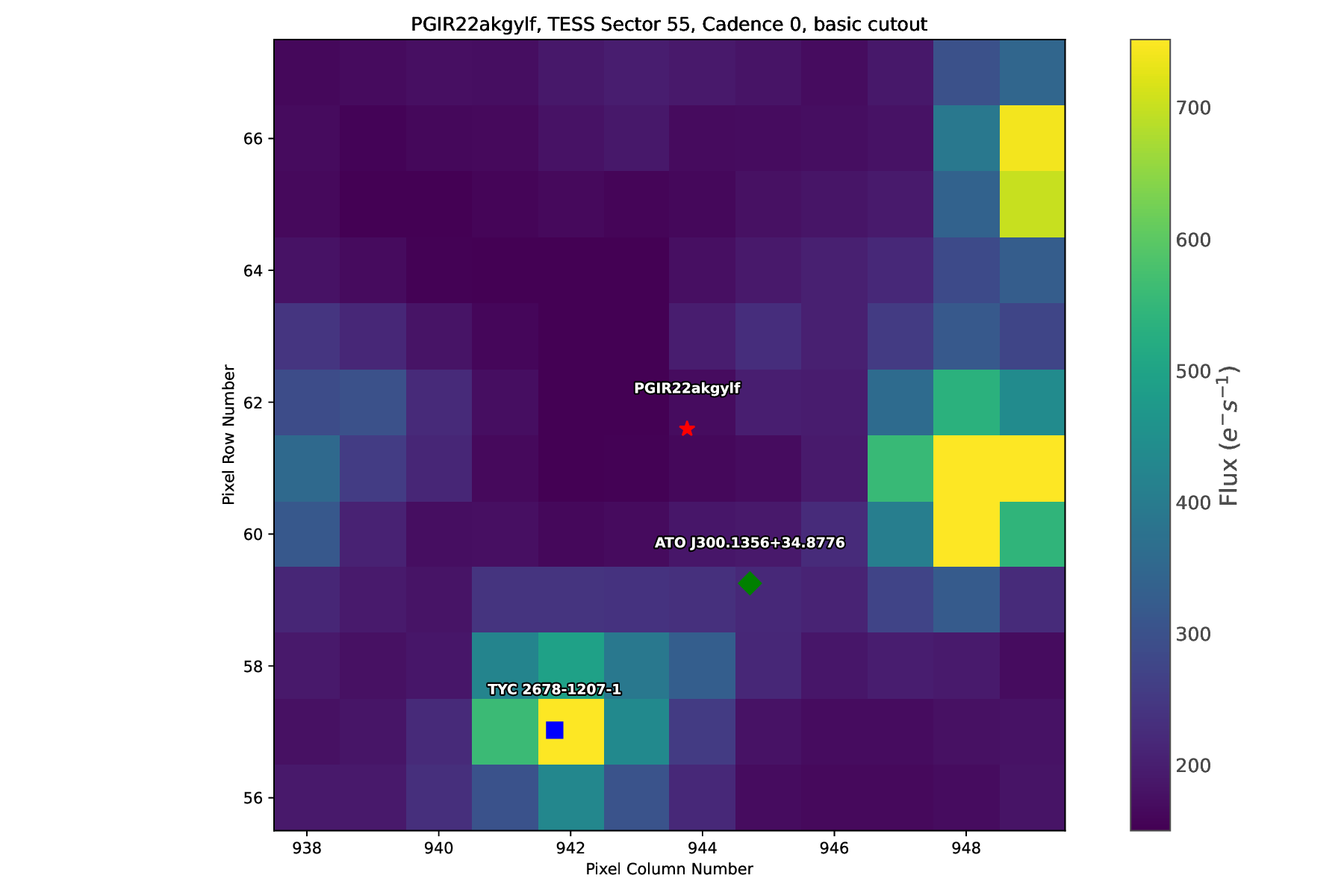}~~~~
        \includegraphics[width=0.48\textwidth,clip=true,trim=5.0cm 0.5cm 2.0cm 0.8cm,angle=0]{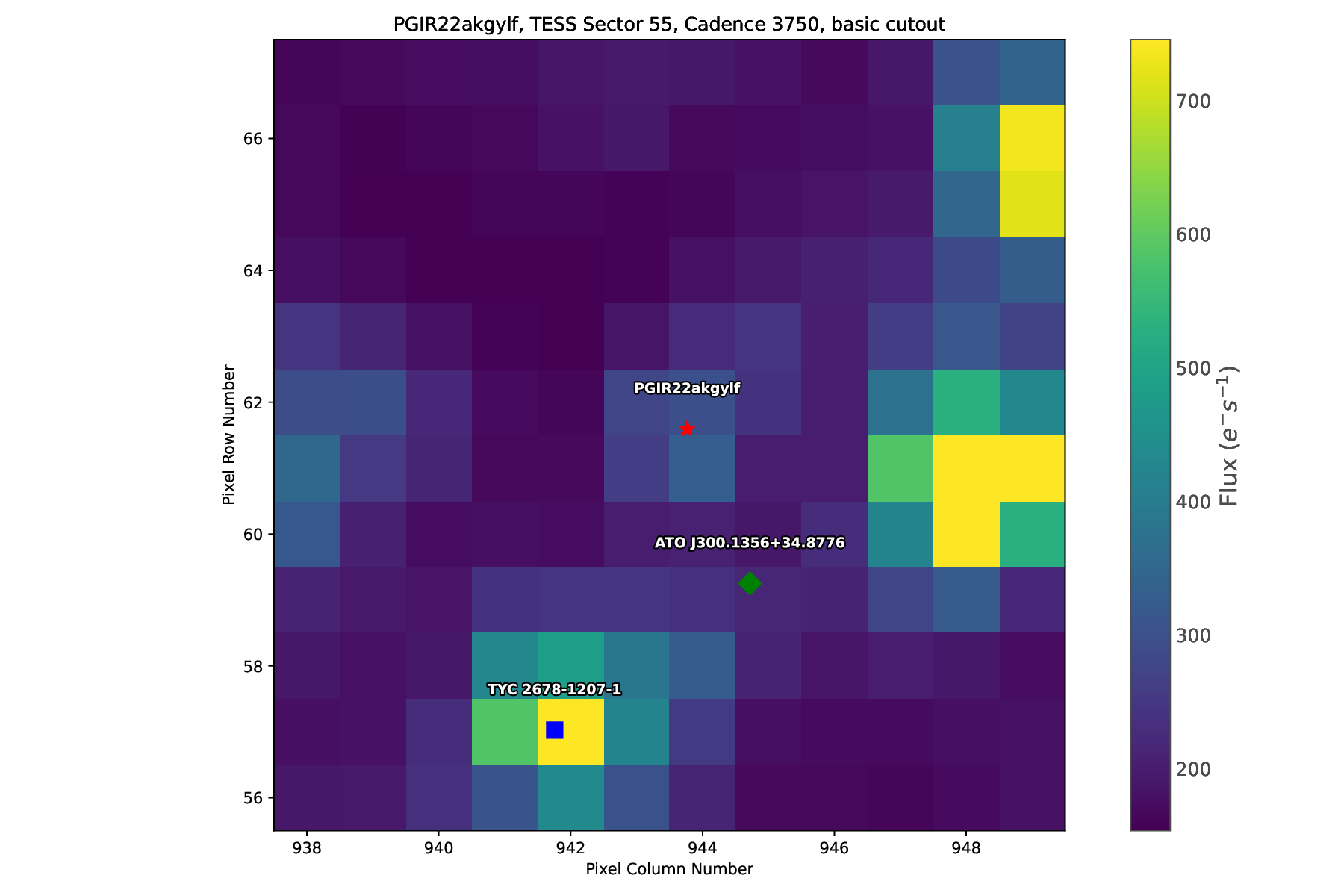}
\caption{{\em TESS} images of \nova{} obtained at the beginning on Sector~55
observations on 2022-08-05.6106~TDB when \nova{} was faint (left) and around the
end of Sector~55 on 2022-09-01.6241 when the nova was bright (right). Positions
of \nova{}, the nearby eclipsing binary ATO\,J300.1356$+$34.8776 ($T_{\rm mag} = 15.20$) 
and TYC\,2678-1207-1 ($T_{\rm mag} = 11.76$) serving as the check star are marked.}
    \label{fig:img}
\end{figure*}

\begin{figure}
\centering
        \includegraphics[width=0.48\textwidth,clip=true,trim=5.0cm 0.5cm 2.0cm 0.8cm,angle=0]{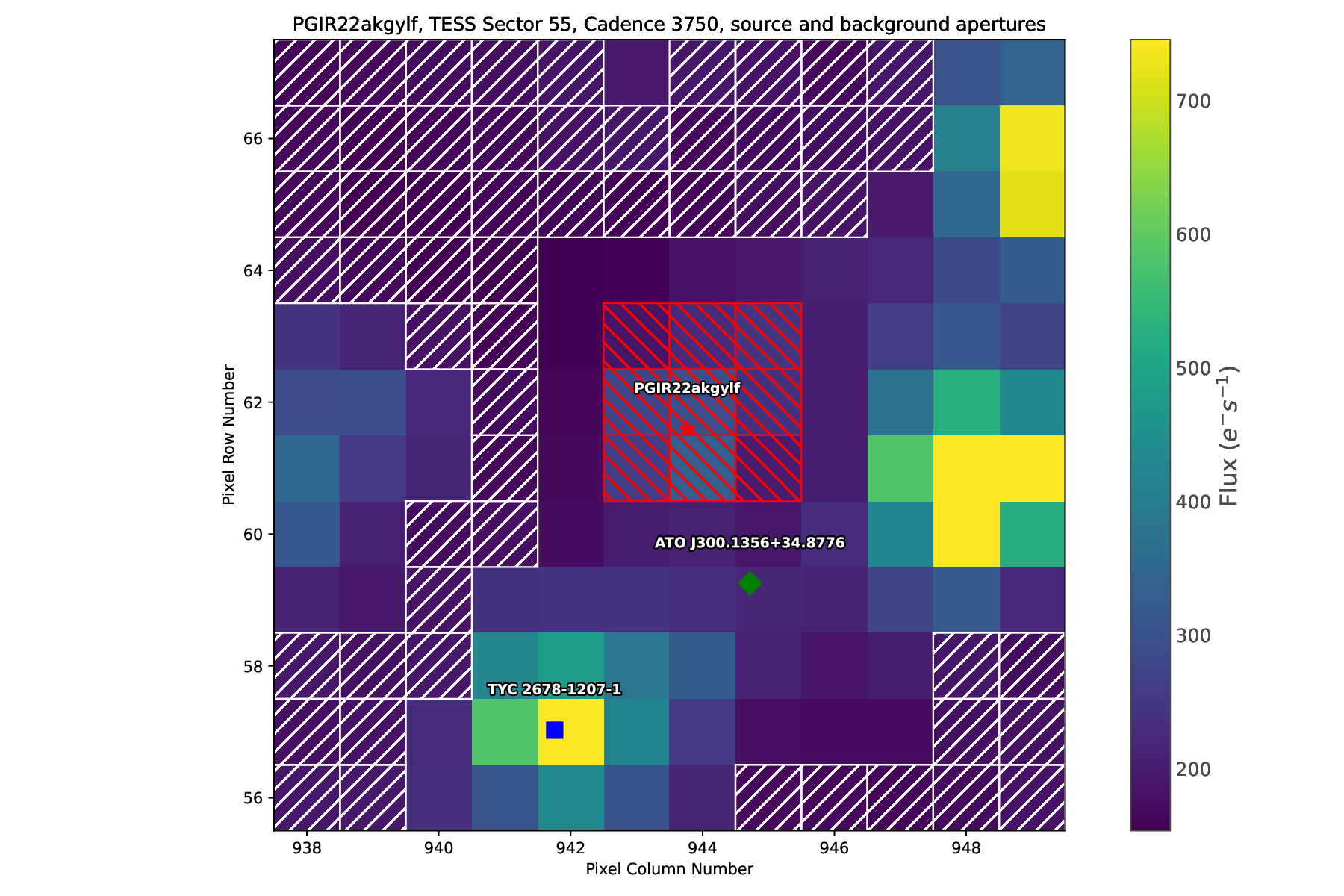}
\caption{Same as the right panel of Figure~\ref{fig:img} with the source and background
extraction apertures indicated as red and white hatched areas, respectively.}
    \label{fig:imgap}
\end{figure}

\begin{figure*}
\centering
        \includegraphics[width=0.48\textwidth,clip=true,trim=1.0cm 0.0cm 2.0cm 1.15cm,angle=0]{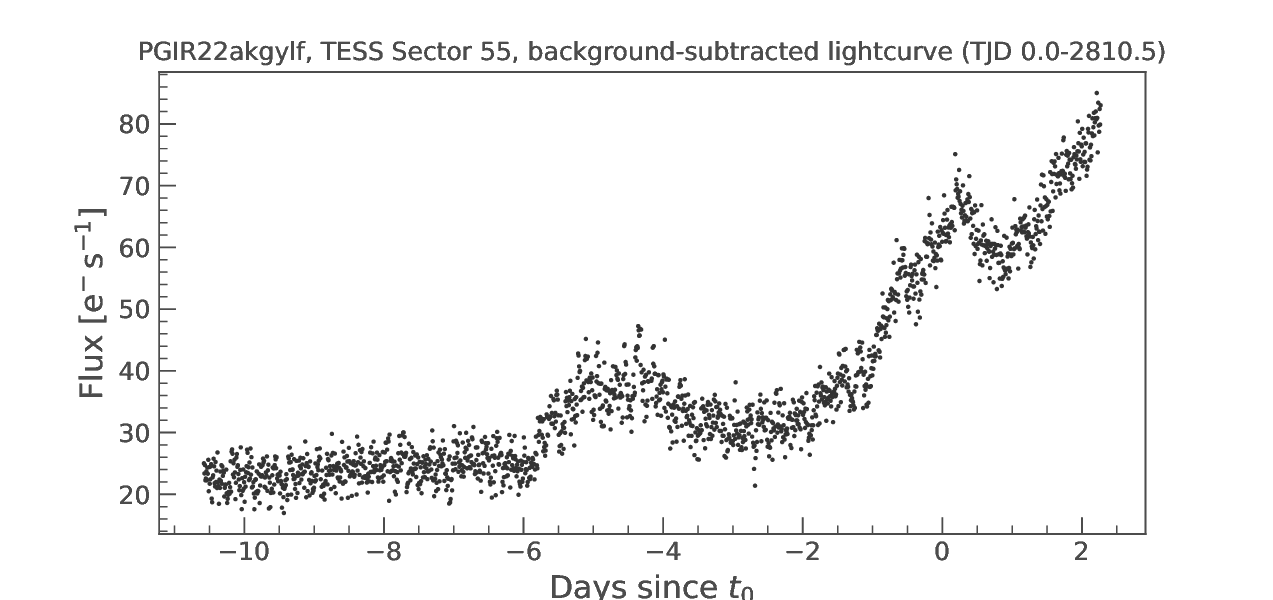}~~~~
        \includegraphics[width=0.48\textwidth,clip=true,trim=1.0cm 0.0cm 2.0cm 1.15cm,angle=0]{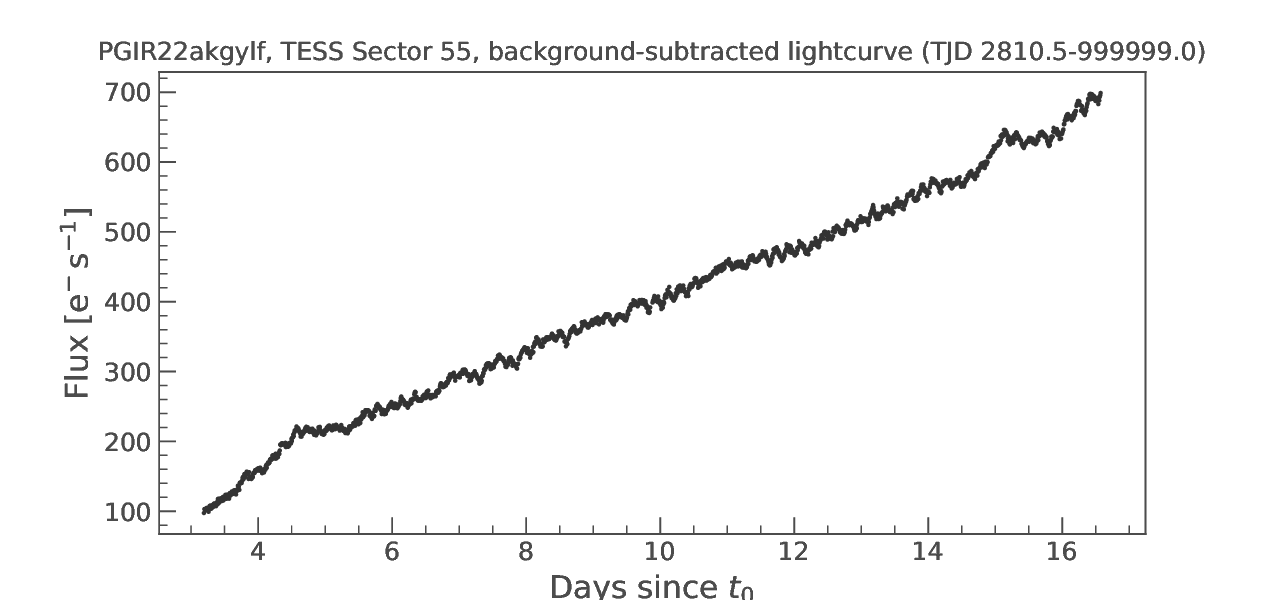}
        \includegraphics[width=0.48\textwidth,clip=true,trim=0.0cm 0.0cm 0.0cm 0.82cm,angle=0]{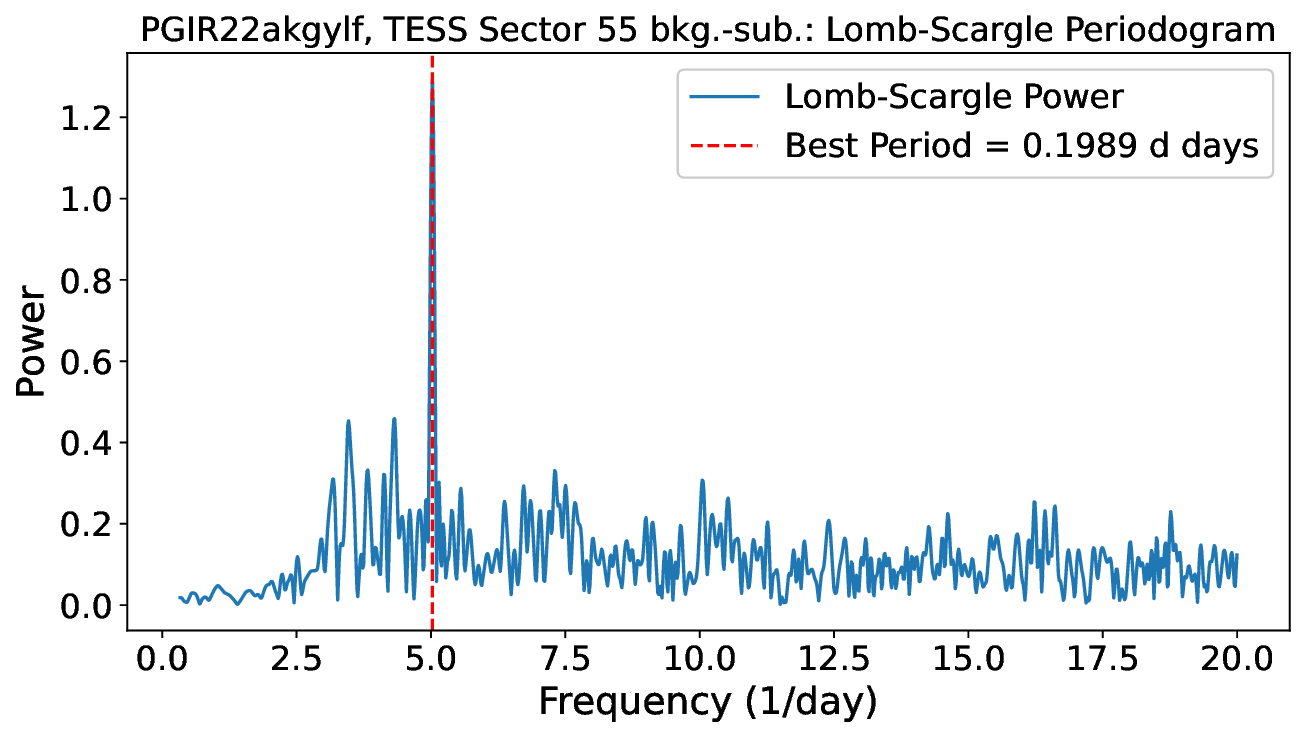}~~~~
        \includegraphics[width=0.48\textwidth,clip=true,trim=0.0cm 0.0cm 0.0cm 0.82cm,angle=0]{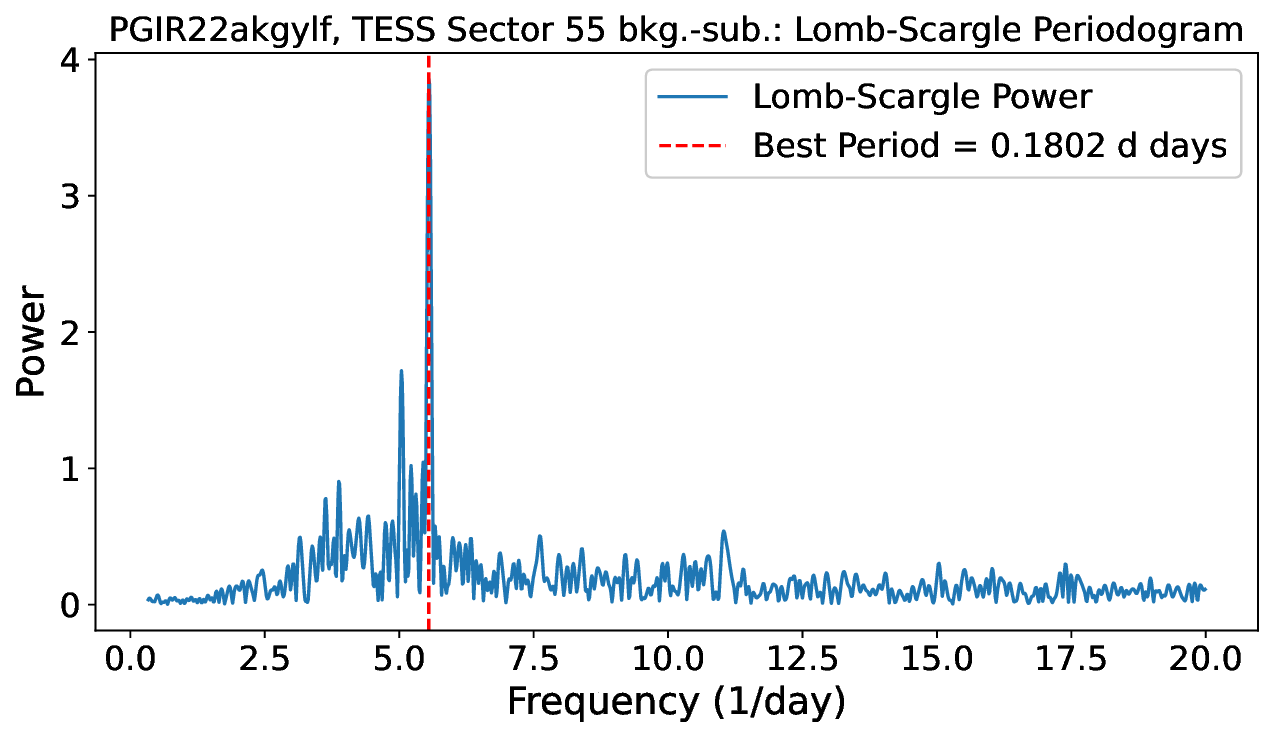}
\caption{{\em TESS} lightcurve of \nova{} obtained during the first (top left)
and second (top right) orbits of Sector~55. The corresponding Lomb-Scargle periodogram
plots are presented at the bottom panels. The drop in power at frequencies below $\sim 2.5$\,d$^{-1}$
is the result of detrending applied to the lightcurve before computing the
periodogram.}
    \label{fig:lcandperiodogram1}
\end{figure*}

\begin{figure}
\centering
        \includegraphics[width=0.48\textwidth,clip=true,trim=0.0cm 0.0cm 0.0cm 0.0cm,angle=0]{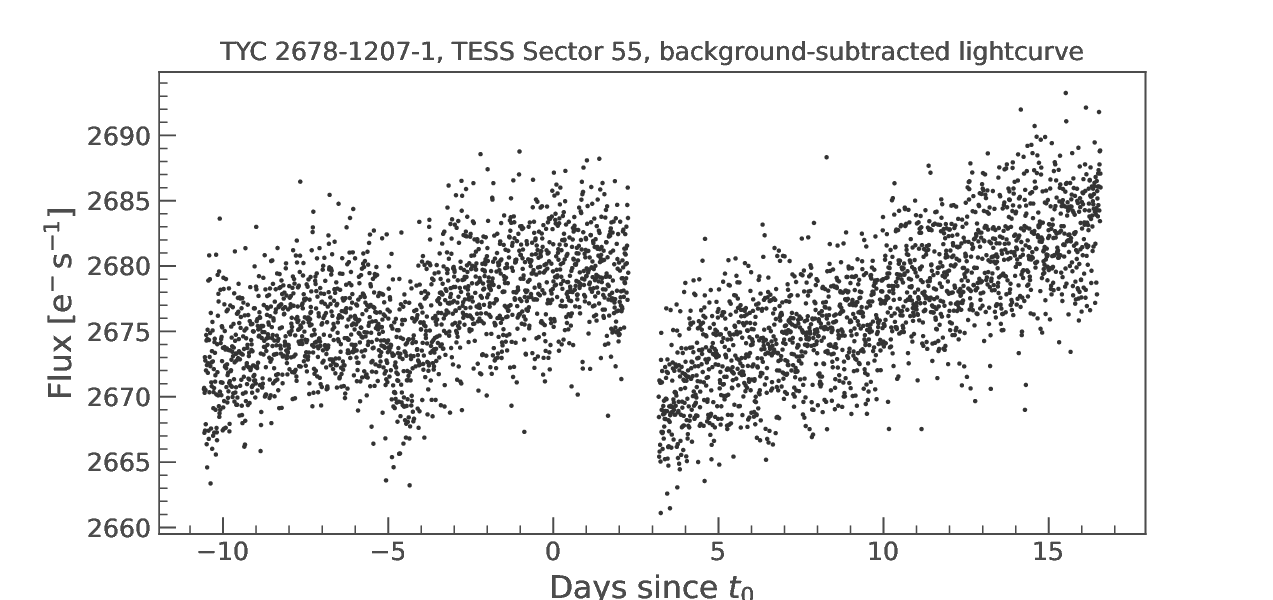}
        \includegraphics[width=0.48\textwidth,clip=true,trim=0.0cm 0.0cm 0.0cm 0.0cm,angle=0]{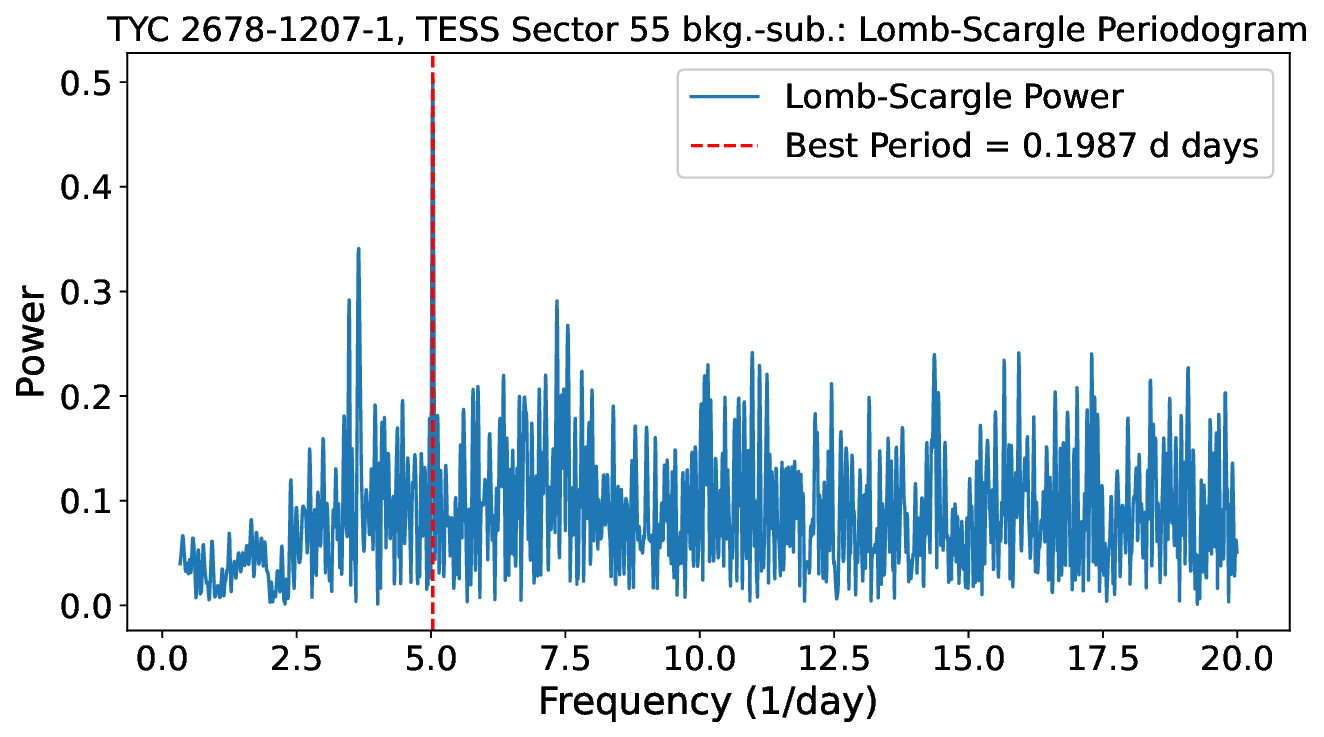}
\caption{{\em TESS} Sector~55 lightcurve (top) and Lomb-Scargle periodogram
(bottom) of the check star TYC\,2678-1207-1 (Figure~\ref{fig:img}).
Detrending with the same parameters as were used for the nova lightcurve was applied
to the check star lightcurve before computing the periodogram. The
periodogram peak is produced by periodic variations of light leaking in the
aperture from the nearby eclipsing binary ATO\,J300.1356$+$34.8776, see
Figure~\ref{fig:img}.}
    \label{fig:checkstar}
\end{figure}

\nova{} erupted within the field of view of Camera~3 CCD~1 during
observations of {\em TESS} Sector~55 (2022-08-05 to 2022-09-01; 
the last sector observed with 600\,s cadence).
Figure~\ref{fig:img} presents cutouts of the {\em TESS} images centered on
the nova position that were obtained at the beginning and the end of
Sector~55. We use the \textsc{Lightkurve} code \citep{2018ascl.soft12013L}
to perform aperture photometry with Figure~\ref{fig:imgap} illustrating the
placement of the source and background extraction apertures.
We applied \texttt{hard} \textsc{Lightkurve} quality cuts to reject images affected by stray light. 
There is one data downlink gap in the middle of Sector~55 observations, between
approximately $t_0 + 2.5$ and $t_0 + 3$\,days.
The top panels of Figure~\ref{fig:lcandperiodogram1} present the lightcurve
extracted from \nova{} position before and after the gap.
The same procedure is applied to extract the lightcurve of a check star (Figure~\ref{fig:checkstar})
TYC\,2678-1207-1 (TIC\,103617911; \citealt{2021arXiv210804778P}) located $1^\prime35^{\prime\prime}$
from \nova{} and $1^\prime11^{\prime\prime}$ from the nearby variable star ATO\,J300.1356$+$34.8776 (TIC\,103617800) 
discussed below.


The {\em TESS} lightcurve of \nova{} reveals that the nova may have started brightening 
at an uneven rate around $t_0 - 6$\,days (Figure~\ref{fig:lcandperiodogram1}, top left panel).
The brightening of the nova may also have started closer to $t_0$ as the
lightcurve of the presumably non-variable check star (Figure~\ref{fig:checkstar}) has a feature around $t_0 - 6$\,days
pointing to its possible instrumental origin (possibly a pointing glitch).
The brightening of \nova{} continues until the end of Sector~55 observations. 
A short-term modulation is clearly visible in the nova lightcurve (Figure~\ref{fig:lcandperiodogram1}, top right panel).
To investigate it further we detrend the nova lightcurve by applying the 
\cite{1964AnaCh..36.1627S} 
low-pass filter with the window width of 101 points and the fifth degree
polynomial fitting the points within the window. The smoothed version of the
nova lightcurve produced by the Savitzky-Golay filter is then subtracted from
the original lightcurve to retain only high-frequency variability. 
Iterative $3 \sigma$ clipping implemented in \textsc{Lightkurve}'s
\texttt{.remove\_outliers()} function is applied to the light curve before and after the Savitzky-Golay detrending.
The calculation is performed in flux units (electrons per second).
Finally, we construct the Lomb-Scargle periodogram 
\citep{1976Ap&SS..39..447L,1982ApJ...263..835S,2018ApJS..236...16V} of the
detrended and $\sigma$-clipped lightcurve. The \textsc{Jupyter} notebooks implementing the
analysis described here are available
online\footnote{\label{fn:ghjupynotebooks}\url{https://github.com/kirxkirx/PGIR22akgylf_lightkurve}}.
We have also repeated the period analysis detrending the
\textsc{Lightkurve}-extracted nova lightcurve using a piecewise linear function 
\citep[as implemented in \textsc{VaST} lightcurve viewer;][]{2018A&C....22...28S}
and the \citet{1975Ap&SS..36..137D} power spectrum for period search.
The results were consistent with the \textsc{Jupyter} notebooks analysis.

The periodograms for the first and the second halves of Sector~55 reveal something unexpected. 
During the first half of Sector~55, the signal in the source aperture was
mostly dominated by the local background while the nova was still faint. 
However its periodogram (Figure~\ref{fig:lcandperiodogram1}, bottom left panel) shows a clear peak corresponding to the period of 0.1991\,d.
We attribute it to the source aperture being contaminated by the light from a nearby
($50^{\prime\prime}$ from \nova{}) variable star 
ATO\,J300.1356$+$34.8776 also known as
ZTF\,J200032.54$+$345239.5, KISO\,J200032.55$+$345239.4 and Gaia~DR3~2059317182140623872 
-- an eclipsing binary of W~UMa type with a period of 
 0.39754\,d \citep{2018AJ....156..241H,2020ApJS..249...18C,2021AJ....161..176R}
and 
$I$ band variability amplitude of 14.921 to 15.203 \citep{2021AJ....161..176R}.
The periodogram peak corresponds to the half of the eclipsing binary's orbital period.

\begin{figure}
\centering
        \includegraphics[width=0.48\textwidth,clip=true,trim=0.0cm 0.0cm 0.0cm 0.0cm,angle=0]{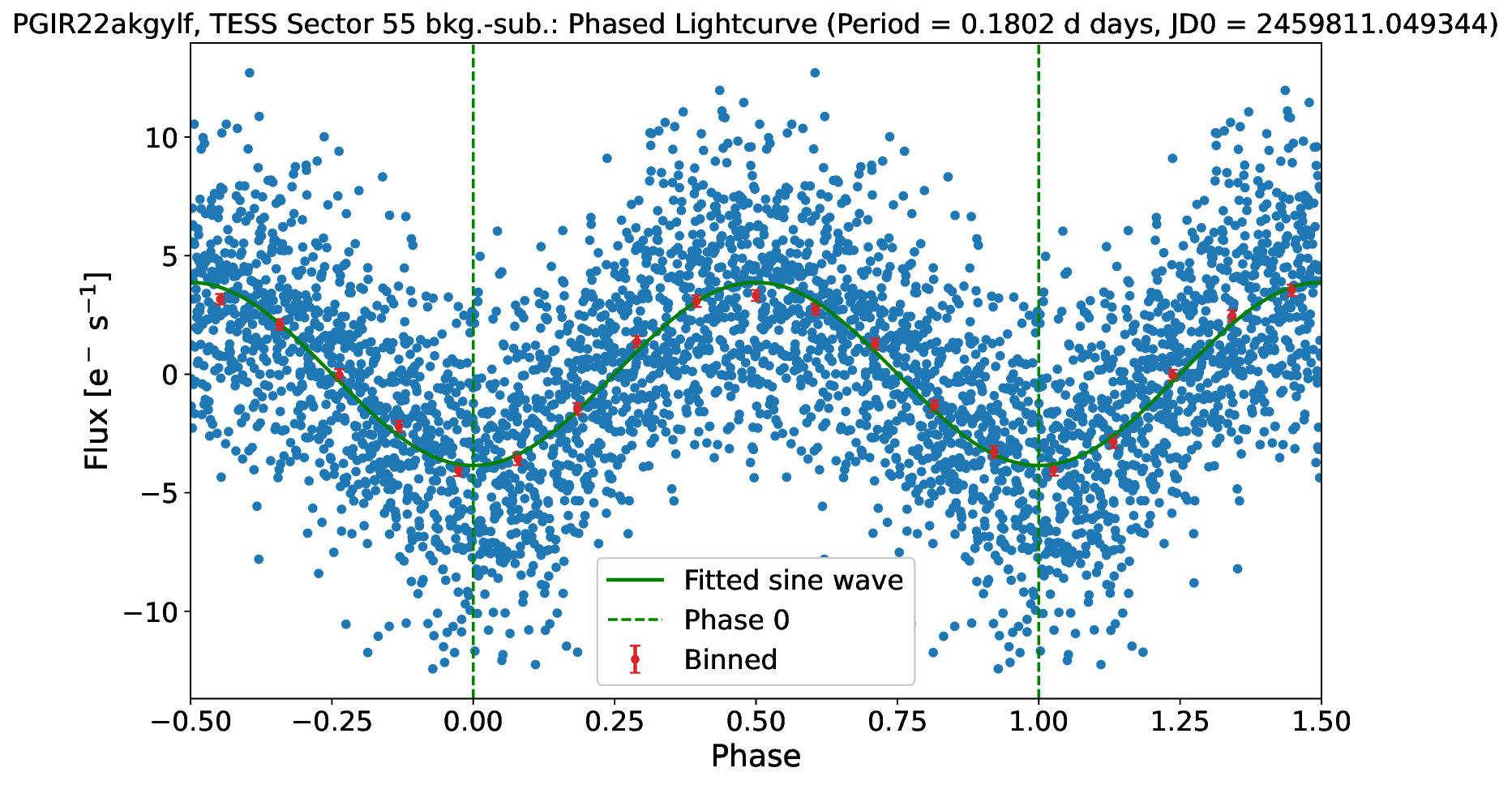}
\caption{The detrended {\em TESS} Sector~55 lightcurve of \nova{} phased with light elements from equation~(\ref{eq:lightelements}).}
    \label{fig:phaselc}
\end{figure}

During the second half of Sector~55, when the signal in the source aperture is
dominated by the brightening nova, the periodogram peak corresponding to the
nearby eclipsing binary is still present, but the new stronger peak appears at 
the period of 0.1802\,d (Figure~\ref{fig:lcandperiodogram1}, bottom right panel), 
attributed to the modulation seen in the nova lightcurve (Figure~\ref{fig:lcandperiodogram1}, top right panel). 
The phased lightcurve is presented in Figure~\ref{fig:phaselc} and the
corresponding light elements are
\begin{equation}
{\rm HJD(TDB)}_{\rm min} = 2459811.0493 + (0.1802 \pm 0.0012) \times E
\label{eq:lightelements}
\end{equation}
Here 2459811.0493 is the reference minimum epoch derived from fitting a sine
wave to the phased lightcurve, $E$ is the integer epoch number. The period
uncertainty is conservatively estimated as 
\begin{equation}
P_{\rm err} = 0.5 \frac{P^2}{\Delta T},
\label{eq:perr}
\end{equation}
where $P$ is the period and $\Delta T$ is the lightcurve duration (for
\nova{} - the second half of Sector~55) and 0.5 is the phase
shift between the first and the last point of the lightcurve if the true
period is off by $P_{\rm err}$, see the discussion by \cite{2022ApJ...934..142S}.

\begin{figure*}
\centering
        \includegraphics[width=1.0\textwidth,clip=true,trim=0.0cm 0.0cm 0.0cm 2.0cm,angle=0]{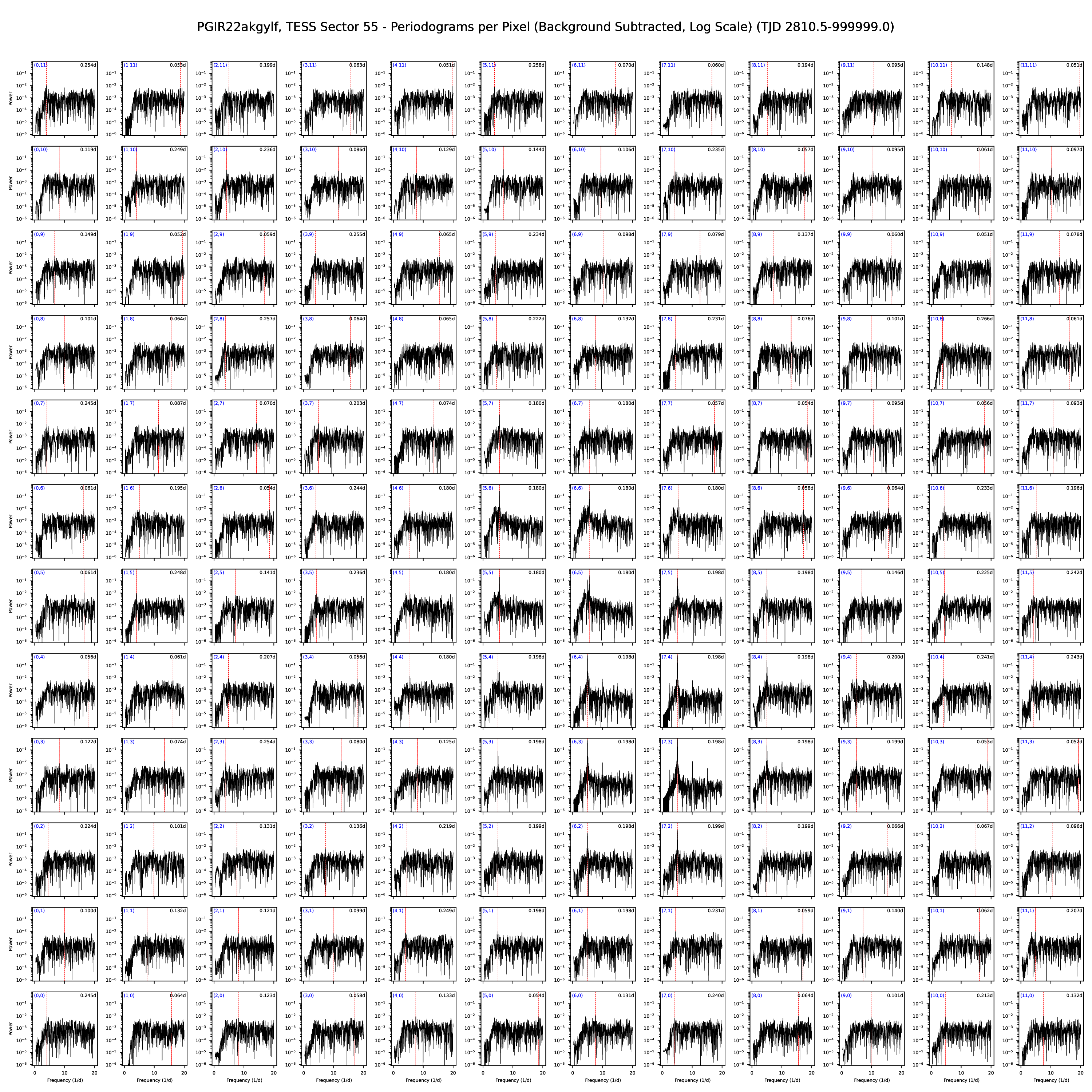}
\caption{The grid of plots presenting periodograms of background-subtracted
detrended lightcurves extracted from each individual pixel of the cutout shown in Figures~\ref{fig:img} and \ref{fig:imgap}.
The lightcurves are restricted to the second half of Sector~55, same as 
the aperture lightcurve presented in the right panel of Figure~\ref{fig:lcandperiodogram1}. 
The periodograms are computed in the range of trial period from 0.05 to 3.0\,d.
The red dashed line in each plot marks the position of the highest
periodogram peak and the corresponding period value is presented in the top
right corner of each plot. The pair of numbers in the top left corner of
each plot represents coordinates of the pixel within the cutout corresponding
to axes labels in Figure~\ref{fig:perimap}. The cutout pixel (0,0)
corresponds to detector coordinates (938,56) in Figures~\ref{fig:img} and \ref{fig:imgap}.
While periodograms of most pixels have low peak power (as expected for noise),
a group of pixels in the center shows peaks at 0.180\,d (period of the nova
modulation) while a group of pixels to the lower right of the center shows
high peaks at 0.198\,d (half-period of the W~UMa type eclipsing binary).}
    \label{fig:perigrid}
\end{figure*}

\begin{figure*}
\centering
        \includegraphics[width=0.48\textwidth,clip=true,trim=2.0cm 0.0cm 2.0cm 2.4cm,angle=0]{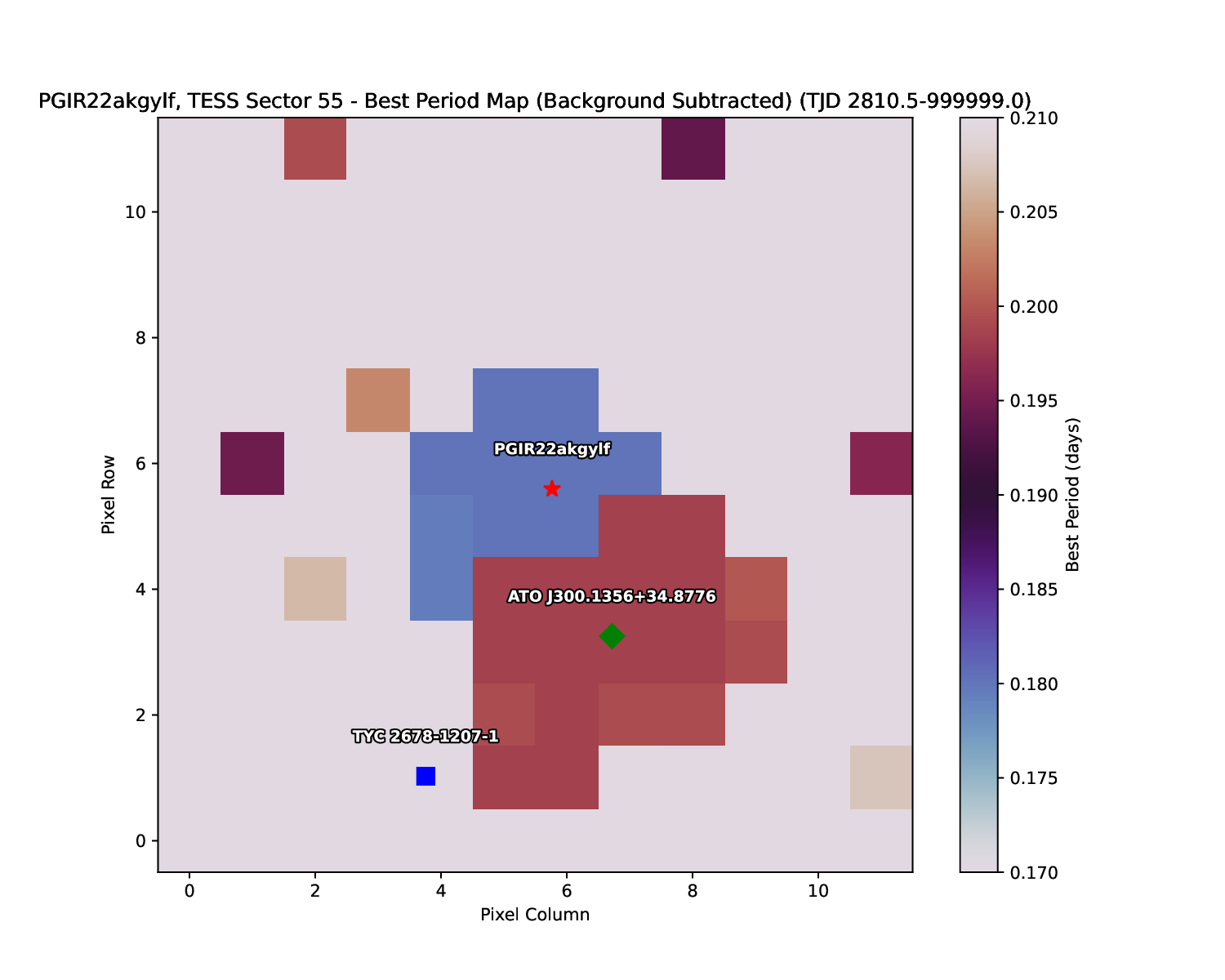}
        \includegraphics[width=0.48\textwidth,clip=true,trim=2.0cm 0.0cm 2.0cm 2.4cm,angle=0]{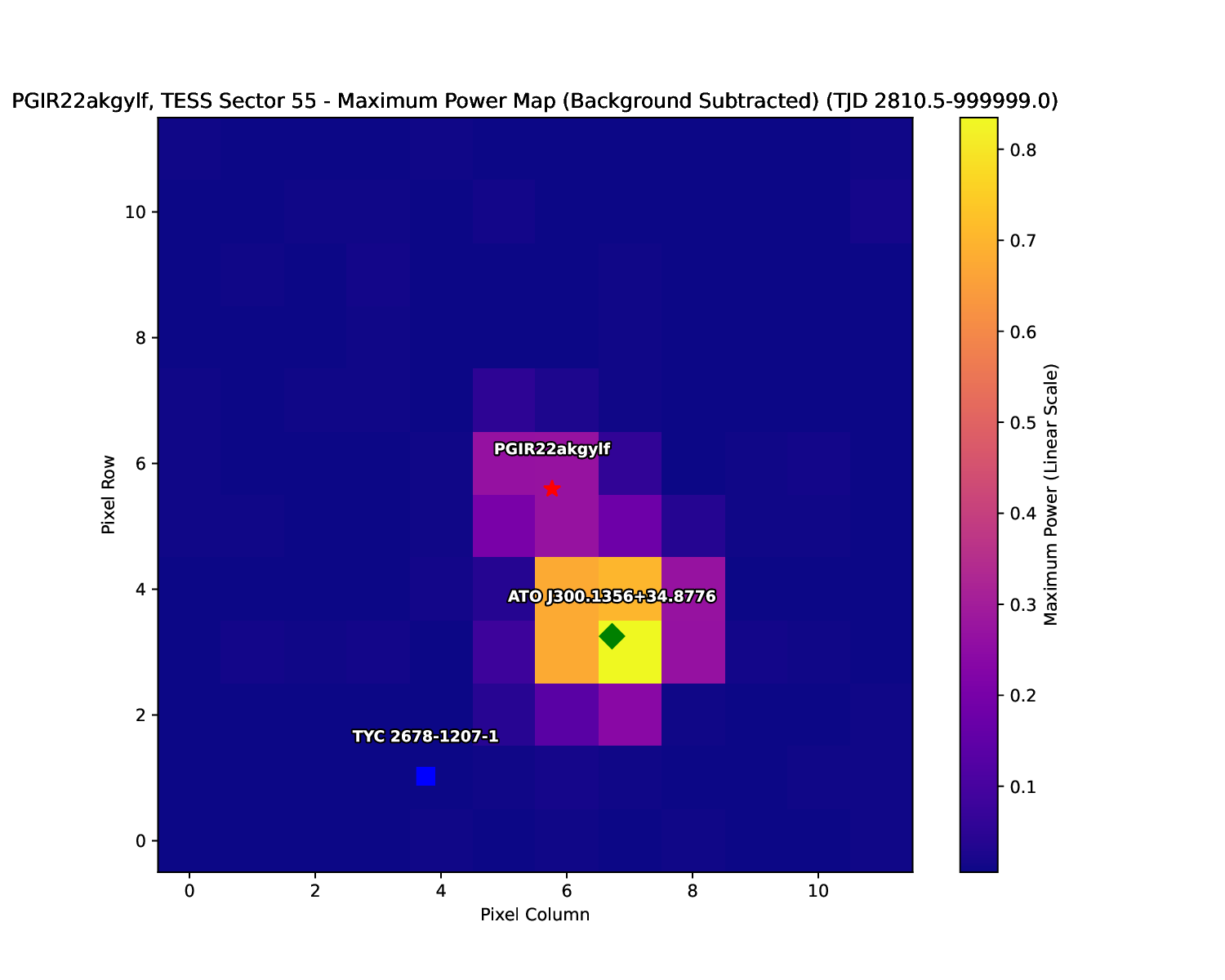}
\caption{The best period map (left) and the highest periodogram power map (right)
corresponding to the periodogram plots presented in Figure~\ref{fig:perigrid}.
The pixel grid corresponds to {\em TESS} image cutouts shown in Figures~\ref{fig:img} and \ref{fig:imgap}.}
    \label{fig:perimap}
\end{figure*}

To spatially localize the origin of the two periodic signals we construct
background-subtracted lightcurves of individual pixels.
The grid of periodogram plots presented in Figure~\ref{fig:perigrid} corresponds to
the cutout images presented in Figures~\ref{fig:img} and \ref{fig:imgap}.
Figure~\ref{fig:perimap} is an alternative way to summarize the results of period search in single-pixel lightcurves.

The left panel of Figure~\ref{fig:perimap} presents the map of periods corresponding to the highest periodogram peak in each pixel.
There are two clusters of pixels with two different periods corresponding to \nova{} and ATO\,J300.1356$+$34.8776, respectively, 
while the majority of background pixels have best periods (corresponding to periodogram noise peaks) much higher 
or lower than the two variables. The right panel of Figure~\ref{fig:perimap} presents the map of the highest periodogram peak power. 
Here again, two connected islands of high power are visible corresponding to the eclipsing binary and the nova.

\begin{figure}
\centering
        \includegraphics[width=0.48\textwidth,clip=true,trim=0.0cm 0.0cm 0.0cm 0.0cm,angle=0]{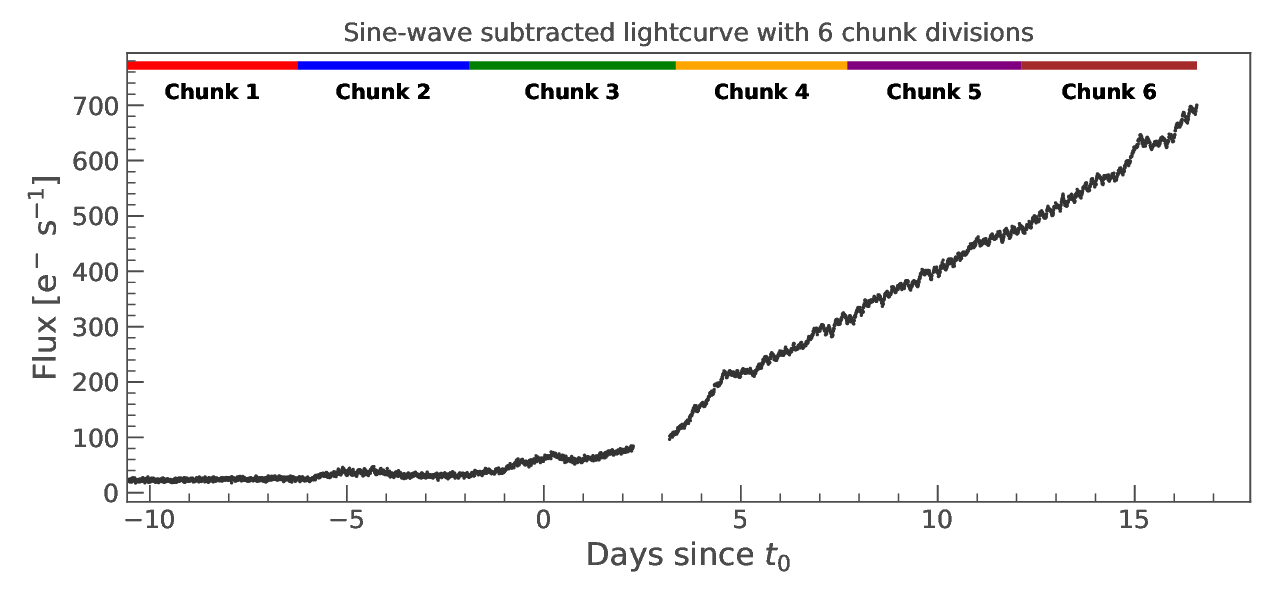}
        \includegraphics[width=0.48\textwidth,clip=true,trim=0.0cm 0.0cm 0.0cm 0.0cm,angle=0]{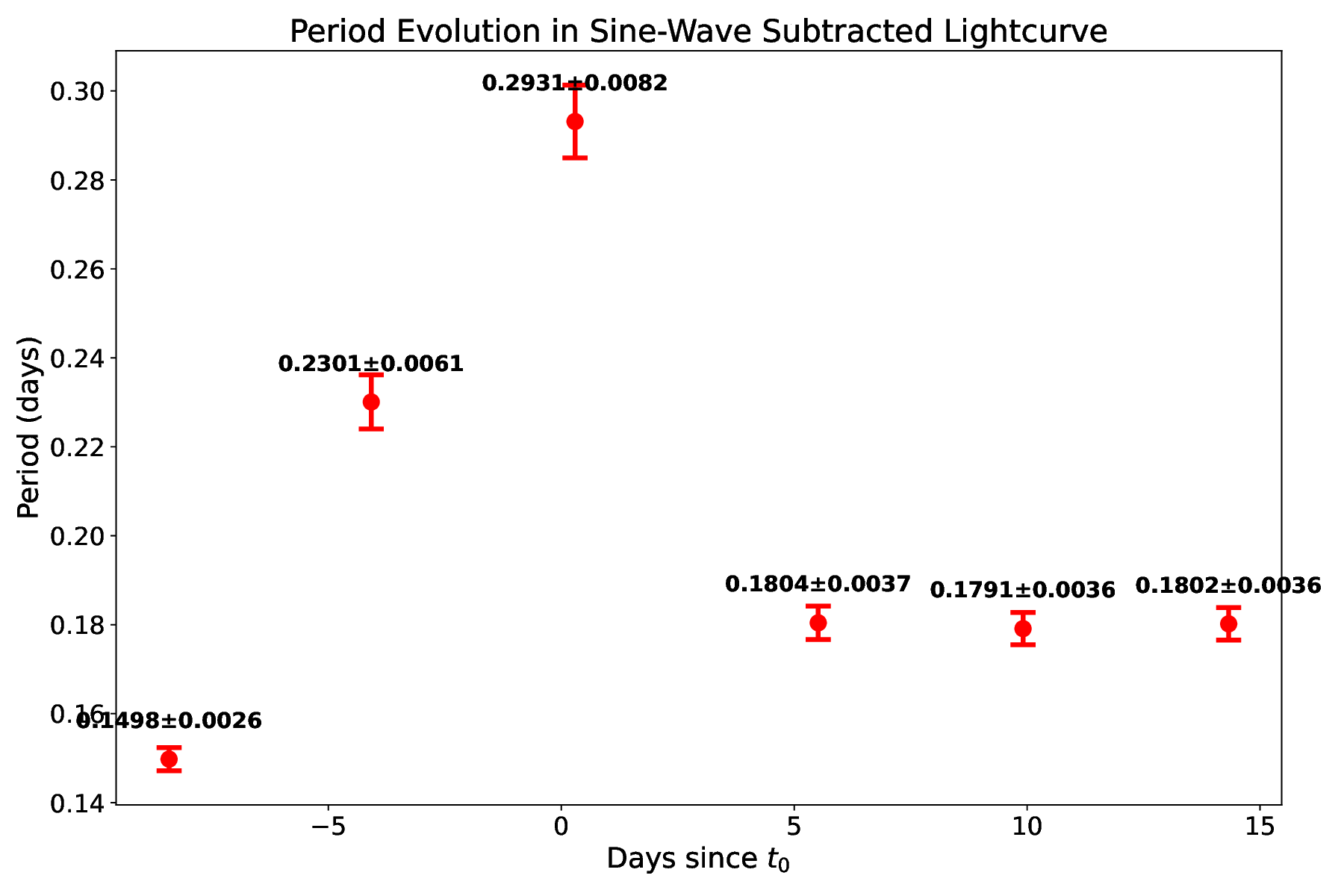}
        \includegraphics[width=0.48\textwidth,clip=true,trim=0.0cm 0.0cm 0.0cm 0.0cm,angle=0]{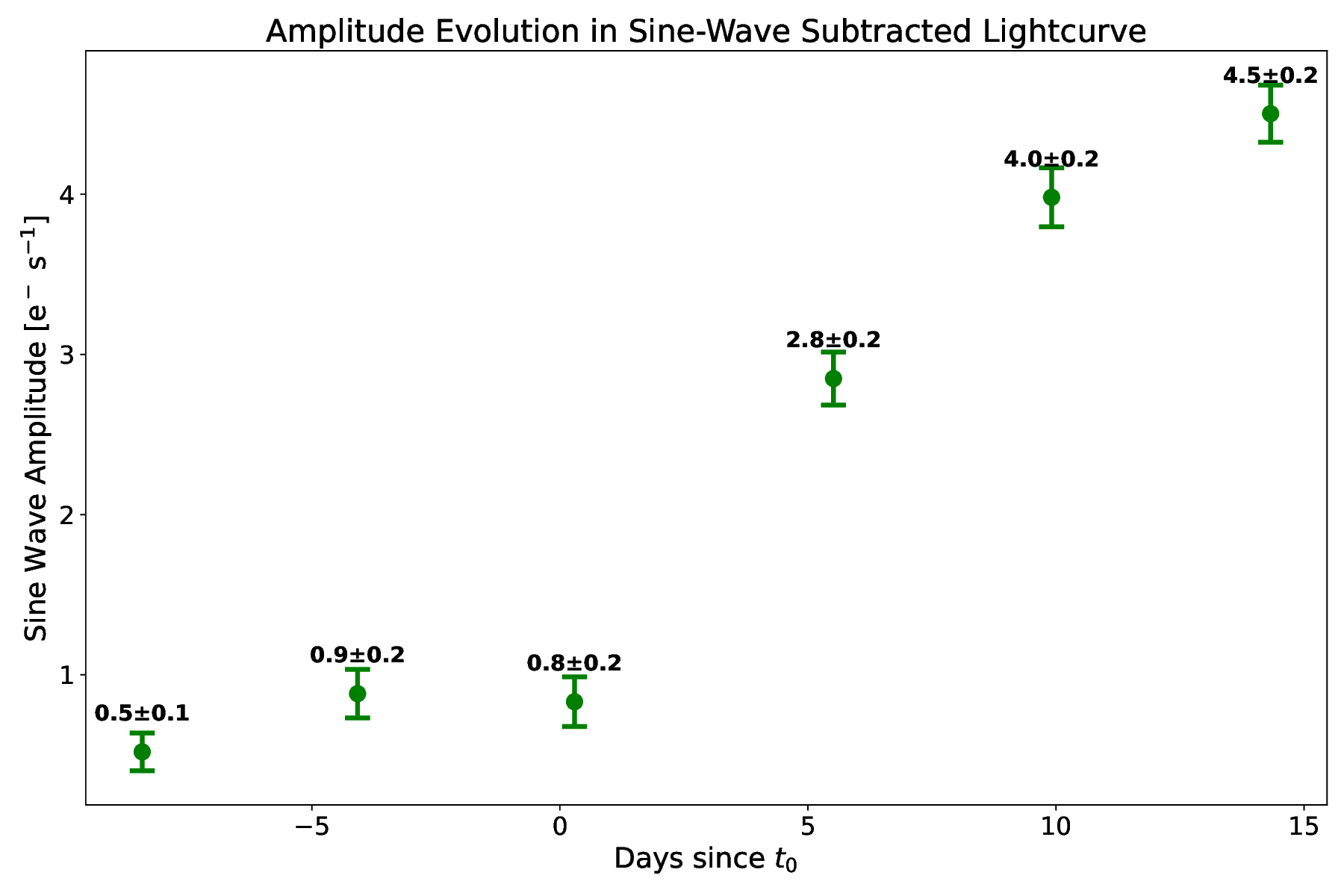}
\caption{{\em TESS} Sector~55 lightcurve split in six chunks for independent
period analysis (top). Period (middle) and peak-to-mean amplitude (bottom) as a function of time.}
    \label{fig:peramptime}
\end{figure}

\begin{deluxetable*}{ccccccc}
\tablecaption{Summary of 6-chunk analysis for PGIR22akgylf Sector 55 lightcurve}
\tablewidth{0pt}
\tablehead{
\colhead{Chunk} & 
\colhead{Center Time} & 
\colhead{Duration} & 
\colhead{Period} & 
\colhead{Amplitude} & 
\colhead{Mean Flux} & 
\colhead{Fractional variation} \\
\colhead{} & 
\colhead{(days since $t_0$)} & 
\colhead{(days)} & 
\colhead{(days)} & 
\colhead{(e$^-$ s$^{-1}$)} & 
\colhead{(e$^-$ s$^{-1}$)} & 
\colhead{(peak-to-peak)}
}
\startdata
1$^*$ & $-8.420$ & $4.319$ & $0.1498 \pm 0.0026$ & $0.5 \pm 0.1$ & $23.9$ & $0.043$ \\
2$^*$ & $-4.079$ & $4.347$ & $0.2301 \pm 0.0061$ & $0.9 \pm 0.2$ & $32.7$ & $0.054$ \\
3$^*$ & $0.295$ & $5.243$ & $0.2931 \pm 0.0082$ & $0.8 \pm 0.2$ & $58.9$ & $0.028$ \\
4 & $5.516$ & $4.347$ & $0.1804 \pm 0.0037$ & $2.8 \pm 0.2$ & $228.7$ & $0.025$ \\
5 & $9.912$ & $4.410$ & $0.1791 \pm 0.0036$ & $4.0 \pm 0.2$ & $401.9$ & $0.020$ \\
6 & $14.326$ & $4.451$ & $0.1802 \pm 0.0036$ & $4.5 \pm 0.2$ & $578.4$ & $0.016$ \\
\enddata
\tablecomments{Periods are derived from Lomb-Scargle periodograms of detrended, 
eclipsing binary light subtracted lightcurve chunks. 
$^*$ symbol marks the chunks with no significant periodic signal.
Period errors are calculated as $P_{\rm err} = 0.5 P^2 / T$, where $T$ is the chunk duration. 
Peak-to-middle amplitudes and their errors are derived from sine wave fitting uncertainties
(see also Figure~\ref{fig:peramptime}). 
The fractional variation represents peak-to-peak variation (2 $\times$ peak-to-middle amplitude) divided by mean flux.}
\label{tab:chunk_analysis}
\end{deluxetable*}

We also construct the full {\em TESS} Sector~55 lightcurve of \nova{}
(Figure~\ref{fig:lcandperiodogram1} left and right panel combined) and use
the first four days (while the nova was still faint) to quantify 
contribution of the eclipsing binary light leaking into the nova aperture.
We perform a period search in this narrow time window and model 
the eclipsing binary as a sine wave with a period of 0.1986\,d and a
peak-to-middle amplitude of 1.3\,e$^-$\,s$^{-1}$. We then subtract this sine
wave model from the full Sector~55 lightcurve. The period analysis of the
eclipsing-binary-light-subtracted lightcurve is consistent with equation
(\ref{eq:lightelements}).

To see when the periodic modulation appears and how its amplitude changes
with time we split the eclipsing-binary-light-subtracted lightcurve in six
chunks and perform period search in each of them independently. The results
are summarized in Table~\ref{tab:chunk_analysis} and Figure~\ref{fig:peramptime}. 
The first three chunks (covering the first half of Sector~55) show random periods and low amplitudes
arising from noise (the eclipsing binary light has already been subtracted).
The consistent periodic signal appears in the three chunks covering the second half of Sector~55, right after the data downlink gap. 

\subsubsection{Sectors~14, 15, 54, 74, 75, 81, and 82}

\nova{} was within the {\em TESS} field of view in Sectors~14, 15, 54 prior
to eruption and Sectors~74, 75, 81, and 82 after the eruption recorded in
Sector~55 (Table~\ref{tab:tesslog}).
We extracted the lightcurve at the nova position using the same
technique as in Sector~55. 
When detrending the lightcurves before computing the Lomb-Scargle periodogram, 
we scale the Savitzky-Golay window size (which is set as the number of data points) 
to account for the difference in observing cadence between sectors (Table~\ref{tab:tesslog}) 
and maintain approximately constant window size in time units.
In Sectors~74 and 75 the signal in the nova aperture is still well above the
local background (set by confusion noise) so we add these measurements to
the combined lightcurve presented in Figure~\ref{fig:lcall}. The
measurements are more noisy than those in Sector~55 due to higher observing
cadence (shorter exposures). The other sectors have signal within the nova
aperture consistent with or below the local confusion-limited background. 
Applying detrending and period search to all these sectors we consistently
see the periodic signal leaking from the eclipsing binary to the nova
aperture. Surprisingly, the nova modulation signal is completely absent in
sectors other than Sector~55. Specifically, it is not present in Sectors~74 and 75
where some slow irregular variations attributed to the nova are seen.
The full analysis results for all sectors listed in Table~\ref{tab:tesslog} are available online\footref{fn:ghjupynotebooks}.

\begin{deluxetable}{cccc}
\tablecaption{{\em TESS} Observations}
\label{tab:tesslog}
\tablewidth{0pt}
\tablehead{
\colhead{Sector} & \colhead{Observation Dates} & \colhead{$t_0 +$ (d)} & \colhead{Exposure (s)} }
\startdata
14 & 2019 Jul 18 -- Aug 14 & $-1125$ -- $-1098$ & 1426 \\
15 & 2019 Aug 15 -- Sep 10 & $-1097$ -- $-1071$ & 1426 \\
54 & 2022 Jul 09 -- Aug 05 & $-38$ -- $-11$ & 475 \\
55 & 2022 Aug 05 -- Sep 01 & $-11$ -- 16 & 475 \\
74 & 2024 Jan 03 -- Jan 30 & 505 -- 532 & 158 \\
75 & 2024 Jan 30 -- Feb 26 & 532 -- 559 & 158 \\
81 & 2024 Jul 15 -- Aug 10 & 699 -- 725 & 158 \\
82 & 2024 Aug 10 -- Sep 05 & 725 -- 751 & 158 \\
\enddata
\end{deluxetable}

\section{Discussion}
\label{sec:discussion}

\subsection{Observed properties of the pre-peak periodic modulation}

The periodic signal is real and associated with \nova{} as its appearance
coincides in time (Figures~\ref{fig:lcandperiodogram1} and \ref{fig:peramptime}) 
and space (Figures~\ref{fig:perigrid} and \ref{fig:perimap}) with the nova eruption. 
We perform a lightcurve simulation to test whether the complex temporal evolution of
the nova eruption can somehow interfere with pre-existing periodic signals from the eclipsing binary. 
For the simulation we use real {\em TESS} Sector~55 observations of \nova{} to extract
the intrinsic trend using Savitzky-Golay filter. 
A synthetic 0.200-day periodic signal with amplitude matching the real variability 
is then added to this smooth trend and tested for period recovery using the
same detrending and clipping technique used for real data analysis. 
The resulting periodogram has a single prominent peak corresponding to the injected period.
We successfully recover the injected period, demonstrating that the complex
lightcurve of the rising nova does not significantly alter the frequency of
a pre-existing periodic signal.

Finding a periodic modulation in a nova on the way to its peak is highly unusual. 
It apparently contradicts the picture of an expanding fireball outlined in \S~\ref{sec:intro}.
We are aware of two previous reports of pre-peak periodic modulation in
novae V723\,Cas \citep{2007AstBu..62..125G} and V606\,Vul \citep{2023arXiv231104903S}.
Both were slow novae, but initially experienced a rapid rise stage, so the
periodic modulation in them was observed between the rapid rise and the peak
as well as after the peak. In both cases the modulation was interpreted as orbital.

As \nova{} brightens the amplitude of modulation in flux units increases,
but somewhat slower than the overall flux rise, so the fractional modulation
decreases from 0.025 to 0.016 as the flux rises by one magnitude in nine days 
(between the middle of the 4th and 6th chunk in
Table~\ref{tab:chunk_analysis}, the full amplitude between the first point
of chunk 4 and the last point of chunk 6 is about 2 magnitudes).
No change in period is found between chunks 4, 5 and 6 that contain the
periodic signal.

The absence of the periodic modulation during {\em TESS} Sector~74 and 75 observations (200\,s cadence),
when the nova was about as bright as shortly after $t_0$ (Figure~\ref{fig:lcall}) is notable. 
It could be understood if the source of the periodic signal during Sector~55 was 
the yet-to-be fully ejected nova envelope. The nova was well in the decline
phase by the time of Sector~74 and 75 observations, by which time 
the envelope was ejected, transparent and unaffected by the host binary orbital motion.

\subsection{Possible origin of the pre-peak modulation}

We attempt to explain the few-percent-level periodic modulation observed during the nova rise to optical maximum. 
This modulation may undergo period changes before disappearing entirely around or shortly after the nova peak. 
Following \cite{1977MNRAS.180..749F}, \cite{1978SvAL....4..141I} and \cite{2023arXiv231104903S}, 
we propose that it may originate in the photosphere of a gradually expanding common envelope, 
stirred by the orbital motion of the underlying binary.

Models of common envelope interaction suggest that 
the ejected matter is concentrated towards the orbital plane of the binary 
\citep{2012ApJ...746...74R} 
and features two tails or plumes which extend from near the outer
$L_2$ (near the secondary; \citealt{1941ApJ....93..133K,1979ApJ...229..223S,2016MNRAS.455.4351P}) and $L_3$ (near the primary) Lagrange points
\citep{2018ApJ...863....5M}. 
A photosphere is formed somewhere in this non-axisymmetric structure. 
If viewed not directly pole on, the rotation of the envelope may produce
the photometric modulation.
The clear detection of the modulation together with the absence of eclipses
in the {\em TESS} lightcurve points to an intermediate inclination of
the binary orbit, neither pole on nor edge on.

Models of the common envelope as the binary evolution phase highlight the
importance of recombination energy that may produce a more spherical ejecta
compared to models that ignore this \citep{2013Sci...339..433I,2022MNRAS.517.3181G}. 
However, this may be irrelevant for novae that presumably keep
their envelopes ionized thanks to continuing irradiation by the nuclear
burning white dwarf, so a substantial orbital plane concentration and
asymmetry are expected in novae.
%

For an average nova peak absolute magnitude $M = -7.5$
\citep{2022MNRAS.517.6150S}, the corresponding $10^4$\,K blackbody radius is 
$100 R_\Sun$, while the component separation in a $1.5 M_\Sun$ total mass
binary in a 5\,h (8\,h) orbit is $1.5 R_\Sun$ ($2.5 R_\Sun$). 
The periodic modulation observed in {\em TESS} Sector~55 was seen while the
nova was about 5 to 3 magnitudes below its peak (Figure~\ref{fig:lcall}).
Assuming the same temperature, the blackbody radius during the {\em TESS}
observations was 10 to $25 R_\Sun$ -- between 4 and 16 times larger than the
binary separation (depending on photometric modulation corresponding to the
full or half the orbital period and contingent on the assumed total mass of
the binary).
%
%
A few percent departure from spherical symmetry may be expected for such a ratio 
of photosphere radius to binary separation. 
%
%
\citet{1976MNRAS.175..305B} pointed out 
that classical-nova binaries ``must be interacting strongly
during outburst,'' and that the oscillations and irregular behavior of the
transition phase set in once the photosphere within the evaporating envelope shrinks to dimensions
comparable to the binary separation. We find \nova{} in a similar regime, with
the photosphere within an order of magnitude of the orbital separation, except
that we detect its photometric signature during the rise rather than during the
post-maximum transition phase. 

It is unclear if the observed 0.1802\,d modulation corresponds to the full
orbital period or half the orbital period.
The modulated wind model proposed to explain periodic variations in
V606\,Vul predicts two lightcurve humps per orbital period \citep{2023arXiv231104903S}.
V1309\,Sco, the one well-traced common envelope event (luminous red 
nova) had one hump per orbital period shortly before the merger
\citep{2011A&A...528A.114T,2014ApJ...788...22P} marking a transition from 
a two hump per period lightcurve (typical for contact
binaries) displayed by the binary in the previous
years to the absence of periodic modulation during its final rise to the peak
\citep{2017ApJ...850...59P}. 
A systematic expansion or contraction of the
photosphere may produce observable period changes without changing the
period of the underlying binary \citep{1977MNRAS.180..749F,1978SvAL....4..141I}.


According to \cite{2022ApJ...938...31S}, the location of the iron-opacity bump that drives
mass loss from the white dwarf migrates inward as the envelope mass decreases.
Early in the eruption the acceleration zone lies outside the white dwarf's
Roche lobe, so the outflow is necessarily shaped by the companion - this is
the regime in which common-envelope interaction dominates and an asymmetric,
orbital-plane-focused photosphere is naturally expected.
Only later, once the iron bump has retreated inside the Roche lobe, does
mass loss transition to a quasi-spherical, optically thick wind 
which may be less affected by the binary motion. 

The common-envelope outflow is expected to leave the binary at a velocity
comparable to the orbital velocity \citep{2016MNRAS.461.2527P}. 
For $P_{\rm orb} = 0.1802$\,d ($2\times0.1802$\,d) and a total mass of
$1.5\,M_\Sun$, the orbital velocity
is $\sim440$ ($\sim340$)\,km\,s$^{-1}$. This is about four times slower than 
the observed 1500\,km\,s$^{-1}$ H$\alpha$ P~Cygni absorption (\S~\ref{sec:spec}). 
The absorption may be tracing a separate, faster outflow coexisting with the
slower expanding envelope.
That the bulk of the envelope is much slower follows from the long rise alone: 
free expansion at 1500\,km\,s$^{-1}$ for 133\,d would reach $\sim115$\,AU
($2.5\times10^4\,R_\Sun$),
while the maximum photospehere radius (expected to correspond to the peak
brightness) is $\sim100\,R_\Sun$ ($0.5\,AU$).
The slow envelope thus sets the continuum photosphere,
drives the rise, and carries the orbital asymmetry responsible for the
photometric modulation, while the fast outflow produces the high-velocity
line absorption.

This also explains the absence of modulation in Sectors~74 and 75. By then 
the slow envelope has dispersed and the photosphere is formed in the faster, more spherical
wind. The nova's {\em TESS}-band brightness is comparable to that early in
Sector~55 (Figure~\ref{fig:lcall}), so the photospheric size may not differ
much, but the fast wind may be less affected by the orbit-locked asymmetry 
and may not produce a detectable photometric modulation considered 
by \cite{1977MNRAS.180..749F}, \cite{1978SvAL....4..141I} and \cite{2023arXiv231104903S}. 
The comparison is only approximate, since much of the late-time flux emerges in lines rather than the continuum.

Our interpretation must confront a long-standing argument against
companion-driven ejection. \citep{2011ApJ...743..157K} noted that if friction
with the companion were an important mass-ejection channel, nova light-curve
shapes should correlate with orbital period, yet no such dependence is seen:
the recurrent nova RS\,Oph develops as fast as U\,Sco despite RS\,Oph's 456\,d
orbit, in which the photosphere never reaches the giant companion, while
U\,Sco's 1.23\,d companion is deeply engulfed. This argument, however, is drawn
from recurrent novae, which host near-Chandrasekhar white dwarfs
\citep{2001MNRAS.327.1323T,2017ApJ...847...99M,2018ApJ...860..110S,2026ApJS..283...24S}.
\citet{2022ApJ...938...31S} showed that such massive white dwarfs readily launch
unbound optically thick winds and eject their envelopes with little help from
the companion, whereas lower-mass white dwarfs ($\lesssim 0.8\,M_\Sun$) do not
develop unbound winds at any stage and must rely on the binary companion to expel 
the envelope \citep[see also the discussion of wind launching conditions by][]{2009ApJ...699.1293K}. 
The role of the companion is therefore mass-dependent: negligible for
the fast, massive-white-dwarf novae that dominate the recurrent-nova sample,
but decisive for slow, low-mass-white-dwarf novae. 
As the relation between nova speed class and white dwarf mass is well established 
\citep{1995ApJ...445..789P,2005ApJ...623..398Y,2015gacv.workE..52K,2025ApJ...981..198C,2025ApJ...993..232S}, 
a photometric orbital imprint during the eruption is thus expected 
in slow novae originating at low-mass white dwarfs.
If a similar eruption occurs on  a low-mass white dwarf in 
a symbiotic binary where the companion is far away and cannot assist 
with envelope ejection \citep{2011ApJ...743..157K}, 
a slow symbiotic nova (Z\,And type event lasting decades;
\citealt{1983ApJ...273..280K,2003ASPC..303..177I}) will occur instead
of a classical slow nova eruption lasting months. 
No photometric modulations with hours-long period is expected in symbiotic novae 
characterized by much longer orbital periods.

The common envelope that we invoke to explain the modulation in \nova{} 
has recently been imaged directly in another very slow nova V1405\,Cas.
V1405\,Cas behaved differently from \nova{}: it rose quickly within days,
settled 
about two magnitudes below its eventual peak, then varied irregularly and reached its
visible maximum during a flare seven weeks after discovery. Using
near-infrared interferometry with the CHARA Array, \citet{2026NatAs..10..271A}
resolved its photosphere near maximum light (53 to 67\,days after eruption
start) as a source of radius $\approx 0.85$\,AU.
Had the envelope been impulsively ejected at the start of the eruption and expanded
freely at the observed P~Cygni velocities (700 to 1500\,km\,s$^{-1}$), the
emitting region would have reached 23--46\,AU by that epoch. 
This 
led \citet{2026NatAs..10..271A} to conclude
that the bulk of the envelope had not yet been expelled but continued to engulf
the binary in a common-envelope phase, and was ejected only after more than
55\,days, triggering the delayed GeV emission. 
These images independently establish that, 
before and around the time a slow nova reaches its peak brightness, 
the emission arises from a photosphere still surrounding the binary rather 
than from freely expanding ejecta. 
The bound envelope is the configuration favorable for the orbital motion 
to imprint the photometric modulation we detect in \nova{}.

\subsection{The ground-based period and its difference from {\em TESS}}

\cite{2022JAVSO..50..260S} reported a $0.1725 \pm 0.0001$ day period
(0.008\,mag amplitude) from ground-based $I$-band photometry covering the time range from 
2022-09-10.02 ($t_0 + 25$\,d, 8\,d after the end of {\em TESS} Sector~55 observations) 
to 2022-10-12.05 ($t_0 + 57$\,d). 
This period is 11\,min shorter (a $6 \sigma$ difference) 
than the one we derived from {\em TESS} Sector~55 data, equation (\ref{eq:lightelements}).

A difference of this size cannot be a real change in the binary orbital period.
The orbital period change scales with the fraction of the binary mass carried beyond 
the orbit. For an isotropic outflow a relative period {\it increase}
$\Delta P / P \sim 2 (1+q)^{-1}\,\Delta M / (M_{\rm WD} + M_{\rm comp})$, 
where $\Delta M$ is the lost mass, $M_{\rm WD}$ ($M_{\rm comp}$) is 
the masses of the white dwarf (comanion star) and $q = M_{\rm comp} / M_{\rm WD}$
is the binary mass ratio \citep{2020MNRAS.492.3323S,2023MNRAS.525..785S}.
The reservoir available is the lifted envelope, $\sim 10^{-5}$ to $10^{-3}\,M_\Sun$ 
including the dredged-up core material that the observed CNO enrichment limits to 
at most a few times the accreted mass. This is one to three orders of magnitude below the
$\sim 10^{-2}\,M_\Sun$ that a 4\% change would require, whether or not the mass
is ultimately unbound. 

The 4\% difference can instead be a real change in the {\it observed} period,
without any change in the orbital period. During the rise we see not the binary
but the surrounding envelope, whose photosphere is an order of magnitude larger
than the orbit. \citet{1977MNRAS.180..749F} showed that the binary imprints a
spiral density pattern on the outflowing wind, observable where it crosses the
scattering surface, and that the period of the resulting brightness variations
can shift as this extended pattern evolves while the underlying orbital period
stays fixed. A drift of a few percent between the {\em TESS} ($t_0 + 3$ to
$t_0 + 16$\,d) and ground-based ($t_0 + 25$ to $t_0 + 57$\,d) epochs is then a
property of the evolving wind pattern rather than of the binary.

It is worth noting that the ground-based analysis includes 
uncertainties 
not reflected in the formal period error. These
include the choice of detrending algorithm (nightly mean magnitude
subtraction) and bad data rejection (removal of the second and the very
last observing nights of those reported to the AAVSO). These analysis
choices have a major impact on the periodogram constructed from ground-based data.

\subsection{Alternative scenarios: nova envelope oscillations and magnetic white dwarf spin}

While above we consider rotation as the driver of the periodic variations,
an analogy with other types of variable stars suggest a potentially
similarly-looking alternative: pulsations of the nova envelope.
\cite{2014MNRAS.437.1962H} present nova eruption models that display
pre-peak oscillations on hours to days timescales 
``caused by the restructuring and rebalancing that the envelope 
undergoes as it begins to expand''. However, \cite{2014MNRAS.437.1962H}
models predict no more than 5-10 oscillation cycles, while the {\em TESS}
Sector~55 lightcurve traces more than 70 oscillation cycles (Figure~\ref{fig:lcandperiodogram1} top right). 
These oscillations may be more relevant for explaining large-scale irregular
variations near the nova peak in \nova{} (Figure~\ref{fig:lcall}) and many
other slow novae, rather than the periodic signal seen in \nova{} during its rise \citep{2022MNRAS.515.1404H}.

\cite{1998ASPC..137..483S,1998ASPC..135..116S,2002ASPC..259..580S} point out
the similarities between the structure of a nova during eruption (a nuclear
burning shell surrounding the core with an extended low-mass envelope on top
of them) that resembles giant stars prone to pulsation instabilities. 
However, this same analogy suggests longer timescales of the pulsations and
the authors predict the pulsations to be visible long after the nova peak.

A periodic signal in a nova could also arise from the spin of a magnetic white
dwarf. While intermediate polars like V1674\,Her, V1405\,Cas \citep{2026A&A...708A.352L} 
and YZ\,Ret \citep{2026arXiv260522802L} have spin periods of a few minutes, 
polars (characterized by even stronger magnetic fields) have the white dwarf
spin synchronized with the orbital period due to magnetic coupling with the
comapnion \citep{1983ApJ...274L..71L,1996Ap&SS.241..263W,2004ApJ...614..349N}.
The binary-modulated wind models proposed by \citet{1977MNRAS.180..749F} and
\cite{1978SvAL....4..141I} to explain the variable $\sim 3$\.h periodicity observed since shortly after eruption of 
nova V1500\,Cyg \citep{1975IBVS.1052....1T,1976IBVS.1157....1S} was
disfavored once the host system was recognized as an asynchroneous polar \citep{1988ApJ...332..282S}. 
While it is consvable that an extreme magnetic field of a white dwarf,
rather than binary companion could shape an expanding envelope of a nova, we
disfavor this scenario for \nova{} because of the transient nature of the
modulation. In V1500\,Cyg the modulation was clearly visible long after 
the eruption when the dominant system light and photometric variability were 
due to the heated companion star \citep{1988ApJ...332..282S,2018MNRAS.479..341P}, 
even if the first days after eruption it was magnetically-shaped photosphere that
produced modulation, later the white dwarf accretion columns and finally the
irradiated secondary - the modulation may have changed its emisison mechanism 
and period (spin to orbital that diverged but may have later synchronised; \citealt{2016MNRAS.459.4161H}) 
-- once detected a few days past maximum, the periodic signal in the lightcurve never disappeared \citep{1979ApJ...231..789P}. 
This behavior of V1500\,Cyg is in contrast 
with that of \nova{} where no periodic modulations where seen by {\em TESS}
$\sim500$ days after eruption.

\section{Conclusions}
\label{sec:conclusions}

We used {\em TESS} full-frame images to trace the initial rise of a slow
nova \nova{}. Unexpectedly, we found a clear periodic modulation in the nova
lightcurve as it rises to its peak. We speculate that the modulation may be
produced by a common envelope engulfing the binary and distorted by the
binary motion while its size is still comparable to the binary orbital
separation. 
The elongated common envelope gets dispersed by the time \nova{} enters {\em TESS} field
of view again during Sector~74 ($t_0 + 505$\,d) observations, as
suggested by the absence of the periodic signal that appeared together with
the nova in Sector~55 ($t_0 + 3$ to $t_0 + 16$\,d).

Instead of the expanding and cooling fireball used to explain rapid rise of most novae, 
\nova{} presents a gradually and isothermally expanding photosphere. 
The periodic modulation in the lightcurve suggests that interaction with 
the binary companion is important in shaping the envelope and take this as an
indication that the envelope expansion is likely to be powered in part by
common-envelope interaction.
Possibly, the low mass of the white dwarf resulted in the accreted hydrogen envelope being non-degenerate 
and the onset of thermonuclear reactions being non eruptive with no impulsive ejection of matter.
We expect that the 0.1802\,d photometric modulation is close to the orbital period
(or half of the orbital period), so the nova host system should have 
a dwarf donor (a giant donor would not fit into such a tight orbit). The gradually expanding white dwarf envelope encounters 
the donor and gets expelled via the common-envelope interaction on a timescale of hundreds of days, 
appropriate for a slow classical nova eruption. Had this nova ignited in a binary system with a distant giant donor 
providing no companion interaction to remove the envelope and starve the nova of its fuel, the thermonuclear eruption would 
have lasted much longer (years or decades), as is often observed in symbiotic binaries.
%

\begin{acknowledgments}
\begin{small}
We acknowledge with thanks the variable star observations from the AAVSO International Database contributed by observers worldwide and used in this research.

This paper includes data collected with the {\em TESS} mission, obtained from the MAST
archive at 
STScI: \dataset[10.17909/0cp4-2j79]{http://dx.doi.org/10.17909/0cp4-2j79}.
Funding for the {\em TESS} mission is provided by the NASA Explorer Program. 
STScI is operated by the Association of Universities for Research in Astronomy, Inc., under NASA contract NAS~5-26555.
Based on observations 
\dataset[10.26131/IRSA539]{http://dx.doi.org/10.26131/IRSA539} 
obtained with the Samuel Oschin Telescope $48^{\prime}$ and the $60^{\prime}$ Telescope at the Palomar
Obs. as part of the ZTF project 
supported by the 
NSF Grants No. AST-1440341 and AST-2034437 and a collaboration including current partners Caltech, IPAC, 
the Weizmann Inst. for Science, the Oskar Klein Center at Stockholm U., the U. of Maryland, 
Deutsches Elektronen-Synchrotron and Humboldt U., the TANGO Consortium of Taiwan, 
the U. of Wisconsin at Milwaukee, Trinity College Dublin,
Lawrence Livermore National Lab., IN2P3, U. of Warwick, Ruhr U. Bochum, Northwestern U. and
former partners the U. of Washington, Los Alamos National Lab., and Lawrence Berkeley National Lab.
Operations are conducted by COO, IPAC, and UW.
This work has made use of data from the Asteroid Terrestrial-impact Last Alert System (ATLAS) project. The Asteroid Terrestrial-impact Last Alert System (ATLAS) project is primarily funded to search for near earth asteroids through NASA grants NN12AR55G, 80NSSC18K0284, and 80NSSC18K1575; byproducts of the NEO search include images and catalogs from the survey area. This work was partially funded by Kepler/K2 grant J1944/80NSSC19K0112 and HST GO-15889, and STFC grants ST/T000198/1 and ST/S006109/1. The ATLAS science products have been made possible through the contributions of the University of Hawaii Institute for Astronomy, 
the Queen's University Belfast, the Space Telescope Science Institute, the South African Astronomical Observatory, and The Millennium Institute of Astrophysics (MAS), Chile.


This work is supported by NASA grant 80NSSC24K0363. 
\end{small}
\end{acknowledgments}

%

\vspace{5mm}
\facilities{AAVSO, ADS, Hale, IRTF, TESS.}


\software{
 \textsc{adstex},
 \textsc{Astropy} \citep{2013A&A...558A..33A,2018AJ....156..123A,2022ApJ...935..167A},
 \textsc{Gnuplot},
 \textsc{Lightkurve} \citep{2018ascl.soft12013L},
 \textsc{Photutils} \citep{2016ascl.soft09011B}, 
 \textsc{SNAD Viewer} \citep{2023PASP..135b4503M},
 \textsc{spextool} \citep{2004PASP..116..362C}, 
 \textsc{TESSCut} \citep{2019ascl.soft05007B},
 \textsc{VaST} \citep{2018A&C....22...28S},
 \textsc{xtellcor} \citep{2003PASP..115..389V}.
}



%

\bibliography{PGIR22akgylf}{}

\begin{thebibliography}{}
\expandafter\ifx\csname natexlab\endcsname\relax\def\natexlab#1{#1}\fi
\providecommand{\url}[1]{\href{#1}{#1}}
\providecommand{\dodoi}[1]{doi:~\href{http://doi.org/#1}{\nolinkurl{#1}}}
\providecommand{\doeprint}[1]{\href{http://ascl.net/#1}{\nolinkurl{http://ascl.net/#1}}}
\providecommand{\doarXiv}[1]{\href{https://arxiv.org/abs/#1}{\nolinkurl{https://arxiv.org/abs/#1}}}

\bibitem[{{Abdo} {et~al.}(2010){Abdo}, {Ackermann}, {Ajello}, {Atwood},
  {Baldini}, {Ballet}, {Barbiellini}, {Bastieri}, {Bechtol}, {Bellazzini}, \&
  et~al.}]{2010Sci...329..817A}
{Abdo}, A.~A., {Ackermann}, M., {Ajello}, M., {et~al.} 2010, Science, 329, 817,
  \dodoi{10.1126/science.1192537}

\bibitem[{{Abe} {et~al.}(2025){Abe}, {Abe}, {Abhishek}, {Acero},
  {Aguasca-Cabot}, {Agudo}, {Alispach}, {Alvarez Crespo}, {Ambrosino},
  {Antonelli}, {Aramo}, {Arbet-Engels}, {Arcaro}, {Asano}, {Aubert}, {Baktash},
  {Balbo}, {Bamba}, {Baquero Larriva}, {Barres de Almeida}, {Barrio}, {Barrios
  Jim{\'e}nez}, {Batkovic}, {Baxter}, {Becerra Gonz{\'a}lez}, {Bernardini},
  {Bernete}, {Berti}, {Bezshyiko}, {Bhattacharjee}, {Bigongiari}, {Bissaldi},
  {Blanch}, {Bonnoli}, {Bordas}, {Borkowski}, {Brunelli}, {Bulgarelli},
  {Bunse}, {Burelli}, {Burmistrov}, {Buscemi}, {Cardillo}, {Caroff}, {Carosi},
  {Carrasco}, {Cassol}, {Castrej{\'o}n}, {Cerasole}, {Ceribella}, {Chai},
  {Cheng}, {Chiavassa}, {Chikawa}, {Chon}, {Chytka}, {Cicciari}, {Cifuentes},
  {Contreras}, {Cortina}, {Costantini}, {Da Vela}, {Dalchenko}, {Dazzi}, {De
  Angelis}, {de Bony de Lavergne}, {De Lotto}, {de Menezes}, {Del Burgo}, {Del
  Peral}, {Delgado}, {Delgado Mengual}, {della Volpe}, {Dellaiera}, {Di Piano},
  {Di Pierro}, {Di Tria}, {Di Venere}, {D{\'\i}az}, {Dominik}, {Dominis
  Prester}, {Donini}, {Dore}, {Dorner}, {Doro}, {Eisenberger}, {Els{\"a}sser},
  {Emery}, {Escudero}, {Fallah Ramazani}, {Ferrarotto}, {Fiasson}, {Foffano},
  {Freixas Coromina}, {Fr{\"o}se}, {Fukazawa}, {Garcia L{\'o}pez}, {Gasbarra},
  {Gasparrini}, {Geyer}, {Giesbrecht Paiva}, {Giglietto}, {Giordano}, {Gliwny},
  {Godinovic}, {Grau}, {Green}, {Green}, {Gunji}, {G{\"u}nther}, {Hackfeld},
  {Hadasch}, {Hahn}, {Hassan}, {Hayashi}, {Heckmann}, {Heller}, {Herrera
  Llorente}, {Hirotani}, {Hoffmann}, {Horns}, {Houles}, {Hrabovsky}, {Hrupec},
  {Hui}, {Iarlori}, {Imazawa}, {Inada}, {Inome}, {Inoue}, {Ioka}, {Iori},
  {Iuliano}, {Jahanvi}, {Jimenez Martinez}, {Jimenez Quiles}, {Jurysek},
  {Kagaya}, {Kalashev}, {Karas}, {Katagiri}, {Kataoka}, {Kerszberg},
  {Kobayashi}, {Kohri}, {Kong}, {Kubo}, {Kushida}, {Lacave}, {Lainez},
  {Lamanna}, {Lamastra}, {Lemoigne}, {Linhoff}, {Longo}, {L{\'o}pez-Coto},
  {L{\'o}pez-Moya}, {L{\'o}pez-Oramas}, {Loporchio}, {Lorini}, {Lozano Bahilo},
  {Luciani}, {Luque-Escamilla}, {Majumdar}, {Makariev}, {Mallamaci}, {Mandat},
  {Manganaro}, {Manic{\`o}}, {Mannheim}, {Marchesi}, {Mariotti}, {Marquez},
  {Marsella}, {Mart{\'\i}}, {Martinez}, {Mart{\'\i}nez}, {Mart{\'\i}nez},
  {Mas-Aguilar}, {Maurin}, {Mazin}, {M{\'e}ndez-Gallego}, {Menon}, {Mestre
  Guillen}, {Micanovic}, {Miceli}, {Miener}, {Miranda}, {Mirzoyan}, {Mizuno},
  {Molero Gonzalez}, {Molina}, {Montaruli}, {Moralejo}, {Morcuende},
  {Morselli}, {Moya}, {Muraishi}, {Nagataki}, \&
  {Nakamori}}]{2025A&A...695A.152A}
{Abe}, K., {Abe}, S., {Abhishek}, A., {et~al.} 2025, \aap, 695, A152,
  \dodoi{10.1051/0004-6361/202452447}

\bibitem[{{Acciari} {et~al.}(2022){Acciari}, {Ansoldi}, {Antonelli}, {Arbet
  Engels}, {Artero}, {Asano}, {Baack}, {Babi{\'c}}, {Baquero}, {Barres de
  Almeida}, \& et~al.}]{2022NatAs...6..689A}
{Acciari}, V.~A., {Ansoldi}, S., {Antonelli}, L.~A., {et~al.} 2022, Nature
  Astronomy, 6, 689, \dodoi{10.1038/s41550-022-01640-z}

\bibitem[{{Ackermann} {et~al.}(2014){Ackermann}, {Ajello}, {Albert}, {Baldini},
  {Ballet}, {Barbiellini}, {Bastieri}, {Bellazzini}, {Bissaldi}, {Blandford},
  {Bloom}, {Bottacini}, {Brandt}, {Bregeon}, {Bruel}, {Buehler}, {Buson},
  {Caliandro}, {Cameron}, {Caragiulo}, {Caraveo}, {Cavazzuti}, {Charles},
  {Chekhtman}, {Cheung}, {Chiang}, {Chiaro}, {Ciprini}, {Claus},
  {Cohen-Tanugi}, {Conrad}, {Corbel}, {D'Ammando}, {de Angelis}, {den Hartog},
  {de Palma}, {Dermer}, {Desiante}, {Digel}, {Di Venere}, {do Couto e Silva},
  {Donato}, {Drell}, {Drlica-Wagner}, {Favuzzi}, {Ferrara}, {Focke},
  {Franckowiak}, {Fuhrmann}, {Fukazawa}, {Fusco}, {Gargano}, {Gasparrini},
  {Germani}, {Giglietto}, {Giordano}, {Giroletti}, {Glanzman}, {Godfrey},
  {Grenier}, {Grove}, {Guiriec}, {Hadasch}, {Harding}, {Hayashida}, {Hays},
  {Hewitt}, {Hill}, {Hou}, {Jean}, {Jogler}, {J{\'o}hannesson}, {Johnson},
  {Johnson}, {Kerr}, {Kn{\"o}dlseder}, {Kuss}, {Larsson}, {Latronico},
  {Lemoine-Goumard}, {Longo}, {Loparco}, {Lott}, {Lovellette}, {Lubrano},
  {Manfreda}, {Martin}, {Massaro}, {Mayer}, {Mazziotta}, {McEnery},
  {Michelson}, {Mitthumsiri}, {Mizuno}, {Monzani}, {Morselli}, {Moskalenko},
  {Murgia}, {Nemmen}, {Nuss}, {Ohsugi}, {Omodei}, {Orienti}, {Orlando},
  {Ormes}, {Paneque}, {Panetta}, {Perkins}, {Pesce-Rollins}, {Piron}, {Pivato},
  {Porter}, {Rain{\`o}}, {Rando}, {Razzano}, {Razzaque}, {Reimer}, {Reimer},
  {Reposeur}, {Saz Parkinson}, {Schaal}, {Schulz}, {Sgr{\`o}}, {Siskind},
  {Spandre}, {Spinelli}, {Stawarz}, {Suson}, {Takahashi}, {Tanaka}, {Thayer},
  {Thayer}, {Thompson}, {Tibaldo}, {Tinivella}, {Torres}, {Tosti}, {Troja},
  {Uchiyama}, {Vianello}, {Winer}, {Wolff}, {Wood}, {Wood}, {Wood},
  {Charbonnel}, {Corbet}, {De Gennaro Aquino}, {Edlin}, {Mason}, {Schwarz},
  {Shore}, {Starrfield}, {Teyssier}, \& {Fermi-LAT
  Collaboration}}]{2014Sci...345..554A}
{Ackermann}, M., {Ajello}, M., {Albert}, A., {et~al.} 2014, Science, 345, 554,
  \dodoi{10.1126/science.1253947}

\bibitem[{{Astropy Collaboration} {et~al.}(2013){Astropy Collaboration},
  {Robitaille}, {Tollerud}, {Greenfield}, {Droettboom}, {Bray}, {Aldcroft},
  {Davis}, {Ginsburg}, {Price-Whelan}, {Kerzendorf}, {Conley}, {Crighton},
  {Barbary}, {Muna}, {Ferguson}, {Grollier}, {Parikh}, {Nair}, {Unther},
  {Deil}, {Woillez}, {Conseil}, {Kramer}, {Turner}, {Singer}, {Fox}, {Weaver},
  {Zabalza}, {Edwards}, {Azalee Bostroem}, {Burke}, {Casey}, {Crawford},
  {Dencheva}, {Ely}, {Jenness}, {Labrie}, {Lim}, {Pierfederici}, {Pontzen},
  {Ptak}, {Refsdal}, {Servillat}, \& {Streicher}}]{2013A&A...558A..33A}
{Astropy Collaboration}, {Robitaille}, T.~P., {Tollerud}, E.~J., {et~al.} 2013,
  \aap, 558, A33, \dodoi{10.1051/0004-6361/201322068}

\bibitem[{{Astropy Collaboration} {et~al.}(2018){Astropy Collaboration},
  {Price-Whelan}, {Sip{\H{o}}cz}, {G{\"u}nther}, {Lim}, {Crawford}, {Conseil},
  {Shupe}, {Craig}, {Dencheva}, {Ginsburg}, {VanderPlas}, {Bradley},
  {P{\'e}rez-Su{\'a}rez}, {de Val-Borro}, {Aldcroft}, {Cruz}, {Robitaille},
  {Tollerud}, {Ardelean}, {Babej}, {Bach}, {Bachetti}, {Bakanov}, {Bamford},
  {Barentsen}, {Barmby}, {Baumbach}, {Berry}, {Biscani}, {Boquien}, {Bostroem},
  {Bouma}, {Brammer}, {Bray}, {Breytenbach}, {Buddelmeijer}, {Burke},
  {Calderone}, {Cano Rodr{\'\i}guez}, {Cara}, {Cardoso}, {Cheedella}, {Copin},
  {Corrales}, {Crichton}, {D'Avella}, {Deil}, {Depagne}, {Dietrich}, {Donath},
  {Droettboom}, {Earl}, {Erben}, {Fabbro}, {Ferreira}, {Finethy}, {Fox},
  {Garrison}, {Gibbons}, {Goldstein}, {Gommers}, {Greco}, {Greenfield},
  {Groener}, {Grollier}, {Hagen}, {Hirst}, {Homeier}, {Horton}, {Hosseinzadeh},
  {Hu}, {Hunkeler}, {Ivezi{\'c}}, {Jain}, {Jenness}, {Kanarek}, {Kendrew},
  {Kern}, {Kerzendorf}, {Khvalko}, {King}, {Kirkby}, {Kulkarni}, {Kumar},
  {Lee}, {Lenz}, {Littlefair}, {Ma}, {Macleod}, {Mastropietro}, {McCully},
  {Montagnac}, {Morris}, {Mueller}, {Mumford}, {Muna}, {Murphy}, {Nelson},
  {Nguyen}, {Ninan}, {N{\"o}the}, {Ogaz}, {Oh}, {Parejko}, {Parley}, {Pascual},
  {Patil}, {Patil}, {Plunkett}, {Prochaska}, {Rastogi}, {Reddy Janga},
  {Sabater}, {Sakurikar}, {Seifert}, {Sherbert}, {Sherwood-Taylor}, {Shih},
  {Sick}, {Silbiger}, {Singanamalla}, {Singer}, {Sladen}, {Sooley},
  {Sornarajah}, {Streicher}, {Teuben}, {Thomas}, {Tremblay}, {Turner},
  {Terr{\'o}n}, {van Kerkwijk}, {de la Vega}, {Watkins}, {Weaver}, {Whitmore},
  {Woillez}, {Zabalza}, \& {Astropy Contributors}}]{2018AJ....156..123A}
{Astropy Collaboration}, {Price-Whelan}, A.~M., {Sip{\H{o}}cz}, B.~M., {et~al.}
  2018, \aj, 156, 123, \dodoi{10.3847/1538-3881/aabc4f}

\bibitem[{{Astropy Collaboration} {et~al.}(2022){Astropy Collaboration},
  {Price-Whelan}, {Lim}, {Earl}, {Starkman}, {Bradley}, {Shupe}, {Patil},
  {Corrales}, {Brasseur}, {N{\"o}the}, {Donath}, {Tollerud}, {Morris},
  {Ginsburg}, {Vaher}, {Weaver}, {Tocknell}, {Jamieson}, {van Kerkwijk},
  {Robitaille}, {Merry}, {Bachetti}, {G{\"u}nther}, {Aldcroft},
  {Alvarado-Montes}, {Archibald}, {B{\'o}di}, {Bapat}, {Barentsen},
  {Baz{\'a}n}, {Biswas}, {Boquien}, {Burke}, {Cara}, {Cara}, {Conroy},
  {Conseil}, {Craig}, {Cross}, {Cruz}, {D'Eugenio}, {Dencheva}, {Devillepoix},
  {Dietrich}, {Eigenbrot}, {Erben}, {Ferreira}, {Foreman-Mackey}, {Fox},
  {Freij}, {Garg}, {Geda}, {Glattly}, {Gondhalekar}, {Gordon}, {Grant},
  {Greenfield}, {Groener}, {Guest}, {Gurovich}, {Handberg}, {Hart},
  {Hatfield-Dodds}, {Homeier}, {Hosseinzadeh}, {Jenness}, {Jones}, {Joseph},
  {Kalmbach}, {Karamehmetoglu}, {Ka{\l}uszy{\'n}ski}, {Kelley}, {Kern},
  {Kerzendorf}, {Koch}, {Kulumani}, {Lee}, {Ly}, {Ma}, {MacBride}, {Maljaars},
  {Muna}, {Murphy}, {Norman}, {O'Steen}, {Oman}, {Pacifici}, {Pascual},
  {Pascual-Granado}, {Patil}, {Perren}, {Pickering}, {Rastogi}, {Roulston},
  {Ryan}, {Rykoff}, {Sabater}, {Sakurikar}, {Salgado}, {Sanghi}, {Saunders},
  {Savchenko}, {Schwardt}, {Seifert-Eckert}, {Shih}, {Jain}, {Shukla}, {Sick},
  {Simpson}, {Singanamalla}, {Singer}, {Singhal}, {Sinha}, {Sip{\H{o}}cz},
  {Spitler}, {Stansby}, {Streicher}, {{\v{S}}umak}, {Swinbank}, {Taranu},
  {Tewary}, {Tremblay}, {Val-Borro}, {Van Kooten}, {Vasovi{\'c}}, {Verma}, {de
  Miranda Cardoso}, {Williams}, {Wilson}, {Winkel}, {Wood-Vasey}, {Xue},
  {Yoachim}, {Zhang}, {Zonca}, \& {Astropy Project
  Contributors}}]{2022ApJ...935..167A}
{Astropy Collaboration}, {Price-Whelan}, A.~M., {Lim}, P.~L., {et~al.} 2022,
  \apj, 935, 167, \dodoi{10.3847/1538-4357/ac7c74}

\bibitem[{Aydi(2018)}]{aydi2018}
Aydi, E. 2018, Phd thesis, University of Cape Town, Cape Town, South Africa.
\newblock \url{http://hdl.handle.net/11427/29576}

\bibitem[{{Aydi} {et~al.}(2020{\natexlab{a}}){Aydi}, {Sokolovsky}, {Chomiuk},
  {Steinberg}, {Li}, {Vurm}, {Metzger}, {Strader}, {Mukai}, {Pejcha}, {Shen},
  {Wade}, {Kuschnig}, {Moffat}, {Pablo}, {Pigulski}, {Popowicz}, {Weiss},
  {Zwintz}, {Izzo}, {Pollard}, {Handler}, {Ryder}, {Filipovi{\'c}}, {Alsaberi},
  {Manojlovi{\'c}}, {Lopes de Oliveira}, {Walter}, {Vallely}, {Buckley},
  {Brown}, {Harvey}, {Kawash}, {Kniazev}, {Kochanek}, {Linford},
  {Mikolajewska}, {Molaro}, {Orio}, {Page}, {Shappee}, \&
  {Sokoloski}}]{2020NatAs...4..776A}
{Aydi}, E., {Sokolovsky}, K.~V., {Chomiuk}, L., {et~al.} 2020{\natexlab{a}},
  Nature Astronomy, 4, 776, \dodoi{10.1038/s41550-020-1070-y}

\bibitem[{{Aydi} {et~al.}(2020{\natexlab{b}}){Aydi}, {Chomiuk}, {Izzo},
  {Harvey}, {Leahy-McGregor}, {Strader}, {Buckley}, {Sokolovsky}, {Kawash},
  {Kochanek}, {Linford}, {Metzger}, {Mukai}, {Orio}, {Shappee}, {Shishkovsky},
  {Steinberg}, {Swihart}, {Sokoloski}, {Walter}, \&
  {Woudt}}]{2020ApJ...905...62A}
{Aydi}, E., {Chomiuk}, L., {Izzo}, L., {et~al.} 2020{\natexlab{b}}, \apj, 905,
  62, \dodoi{10.3847/1538-4357/abc3bb}

\bibitem[{{Aydi} {et~al.}(2023){Aydi}, {Chomiuk}, {Miko{\l}ajewska}, {Brink},
  {Metzger}, {Strader}, {Buckley}, {Harvey}, {Holoien}, {Izzo}, {Kawash},
  {Linford}, {Molaro}, {Molina}, {Mr{\'o}z}, {Mukai}, {Orio}, {Panurach},
  {Senchyna}, {Shappee}, {Shen}, {Sokoloski}, {Sokolovsky}, {Urquhart}, \&
  {Williams}}]{2023MNRAS.524.1946A}
{Aydi}, E., {Chomiuk}, L., {Miko{\l}ajewska}, J., {et~al.} 2023, \mnras, 524,
  1946, \dodoi{10.1093/mnras/stad1914}

\bibitem[{{Aydi} {et~al.}(2024){Aydi}, {Chomiuk}, {Strader}, {Sokolovsky},
  {Williams}, {Buckley}, {Ederoclite}, {Izzo}, {Kyer}, {Linford}, {Kniazev},
  {Metzger}, {Miko{\l}ajewska}, {Molaro}, {Molina}, {Mukai}, {Munari}, {Orio},
  {Panurach}, {Shappee}, {Shen}, {Sokoloski}, {Urquhart}, \&
  {Walter}}]{2023arXiv230907097A}
{Aydi}, E., {Chomiuk}, L., {Strader}, J., {et~al.} 2024, \mnras, 527, 9303,
  \dodoi{10.1093/mnras/stad3342}

\bibitem[{{Aydi} {et~al.}(2026){Aydi}, {Monnier}, {M{\'e}rand}, {Schaefer},
  {Chomiuk}, {Otulakowska-Hypka}, {Fan}, {Li}, {Sokolovsky}, {Salinas},
  {Tucker}, {Shappee}, {Rudy}, {Page}, {Kuin}, {Buckley}, {Craig}, {Izzo},
  {Linford}, {Metzger}, {Mukai}, {Orio}, {Shen}, {Strader}, {Sokoloski},
  {Williams}, {Williams}, {Habtie}, {Kraus}, {Anugu}, {Bouquin}, {Chhabra},
  {Codron}, {Gardner}, {Gutierrez}, {Ibrahim}, {Lanthermann}, {Setterholm},
  {Ashall}, {Hinkle}, {Jaeger}, \& {Payne}}]{2026NatAs..10..271A}
{Aydi}, E., {Monnier}, J.~D., {M{\'e}rand}, A., {et~al.} 2026, Nature
  Astronomy, 10, 271, \dodoi{10.1038/s41550-025-02725-1}

\bibitem[{{Barnes} \& {Evans}(1970)}]{1970PASP...82..889B}
{Barnes}, T.~G., \& {Evans}, N.~R. 1970, \pasp, 82, 889, \dodoi{10.1086/128978}

\bibitem[{{Bath} \& {Shaviv}(1976)}]{1976MNRAS.175..305B}
{Bath}, G.~T., \& {Shaviv}, G. 1976, \mnras, 175, 305,
  \dodoi{10.1093/mnras/175.2.305}

\bibitem[{{Bellm} {et~al.}(2019){Bellm}, {Kulkarni}, {Graham}, {Dekany},
  {Smith}, {Riddle}, {Masci}, {Helou}, {Prince}, {Adams}, {Barbarino},
  {Barlow}, {Bauer}, {Beck}, {Belicki}, {Biswas}, {Blagorodnova}, {Bodewits},
  {Bolin}, {Brinnel}, {Brooke}, {Bue}, {Bulla}, {Burruss}, {Cenko}, {Chang},
  {Connolly}, {Coughlin}, {Cromer}, {Cunningham}, {De}, {Delacroix}, {Desai},
  {Duev}, {Eadie}, {Farnham}, {Feeney}, {Feindt}, {Flynn}, {Franckowiak},
  {Frederick}, {Fremling}, {Gal-Yam}, {Gezari}, {Giomi}, {Goldstein},
  {Golkhou}, {Goobar}, {Groom}, {Hacopians}, {Hale}, {Henning}, {Ho}, {Hover},
  {Howell}, {Hung}, {Huppenkothen}, {Imel}, {Ip}, {Ivezi{\'c}}, {Jackson},
  {Jones}, {Juric}, {Kasliwal}, {Kaspi}, {Kaye}, {Kelley}, {Kowalski},
  {Kramer}, {Kupfer}, {Landry}, {Laher}, {Lee}, {Lin}, {Lin}, {Lunnan},
  {Giomi}, {Mahabal}, {Mao}, {Miller}, {Monkewitz}, {Murphy}, {Ngeow},
  {Nordin}, {Nugent}, {Ofek}, {Patterson}, {Penprase}, {Porter}, {Rauch},
  {Rebbapragada}, {Reiley}, {Rigault}, {Rodriguez}, {van Roestel}, {Rusholme},
  {van Santen}, {Schulze}, {Shupe}, {Singer}, {Soumagnac}, {Stein}, {Surace},
  {Sollerman}, {Szkody}, {Taddia}, {Terek}, {Van Sistine}, {van Velzen},
  {Vestrand}, {Walters}, {Ward}, {Ye}, {Yu}, {Yan}, \&
  {Zolkower}}]{2019PASP..131a8002B}
{Bellm}, E.~C., {Kulkarni}, S.~R., {Graham}, M.~J., {et~al.} 2019, \pasp, 131,
  018002, \dodoi{10.1088/1538-3873/aaecbe}

\bibitem[{{Blank} {et~al.}(2011){Blank}, {Anglin}, {Beletic}, {Baia}, {Buck},
  {Bhargava}, {Chen}, {Cooper}, {Eads}, {Farris}, {Hall}, {Hodapp}, {Lavelle},
  {Loose}, {Luppino}, {Piquette}, {Ricardo}, {Sprafke}, {Starr}, {Xu}, \&
  {Zandian}}]{2011ASPC..437..383B}
{Blank}, R., {Anglin}, S., {Beletic}, J.~W., {et~al.} 2011, in Astronomical
  Society of the Pacific Conference Series, Vol. 437, Solar Polarization 6, ed.
  J.~R. {Kuhn}, D.~M. {Harrington}, H.~{Lin}, S.~V. {Berdyugina},
  J.~{Trujillo-Bueno}, S.~L. {Keil}, \& T.~{Rimmele}, 383

\bibitem[{{Bradley} {et~al.}(2016){Bradley}, {Sipocz}, {Robitaille},
  {Tollerud}, {Deil}, {Vin{\'\i}cius}, {Barbary}, {G{\"u}nther}, {Bostroem},
  {Droettboom}, {Bray}, {Bratholm}, {Pickering}, {Craig}, {Pascual}, {Greco},
  {Donath}, {Kerzendorf}, {Littlefair}, {Barentsen}, {D'Eugenio}, \&
  {Weaver}}]{2016ascl.soft09011B}
{Bradley}, L., {Sipocz}, B., {Robitaille}, T., {et~al.} 2016, {Photutils:
  Photometry tools}, Astrophysics Source Code Library, record ascl:1609.011.
\newblock \doeprint{1609.011}

\bibitem[{{Brasseur} {et~al.}(2019){Brasseur}, {Phillip}, {Fleming},
  {Mullally}, \& {White}}]{2019ascl.soft05007B}
{Brasseur}, C.~E., {Phillip}, C., {Fleming}, S.~W., {Mullally}, S.~E., \&
  {White}, R.~L. 2019, {Astrocut: Tools for creating cutouts of TESS images}.
\newblock \doeprint{1905.007}

\bibitem[{{Bruch}(2023{\natexlab{a}})}]{2023MNRAS.519..352B}
{Bruch}, A. 2023{\natexlab{a}}, \mnras, 519, 352,
  \dodoi{10.1093/mnras/stac3493}

\bibitem[{{Bruch}(2023{\natexlab{b}})}]{2023MNRAS.525.1953B}
---. 2023{\natexlab{b}}, \mnras, 525, 1953, \dodoi{10.1093/mnras/stad2089}

\bibitem[{{Chambers} {et~al.}(2016){Chambers}, {Magnier}, {Metcalfe},
  {Flewelling}, {Huber}, {Waters}, {Denneau}, {Draper}, {Farrow}, {Finkbeiner},
  {Holmberg}, {Koppenhoefer}, {Price}, {Rest}, {Saglia}, {Schlafly}, {Smartt},
  {Sweeney}, {Wainscoat}, {Burgett}, {Chastel}, {Grav}, {Heasley}, {Hodapp},
  {Jedicke}, {Kaiser}, {Kudritzki}, {Luppino}, {Lupton}, {Monet}, {Morgan},
  {Onaka}, {Shiao}, {Stubbs}, {Tonry}, {White}, {Ba{\~n}ados}, {Bell},
  {Bender}, {Bernard}, {Boegner}, {Boffi}, {Botticella}, {Calamida},
  {Casertano}, {Chen}, {Chen}, {Cole}, {Deacon}, {Frenk}, {Fitzsimmons},
  {Gezari}, {Gibbs}, {Goessl}, {Goggia}, {Gourgue}, {Goldman}, {Grant},
  {Grebel}, {Hambly}, {Hasinger}, {Heavens}, {Heckman}, {Henderson}, {Henning},
  {Holman}, {Hopp}, {Ip}, {Isani}, {Jackson}, {Keyes}, {Koekemoer}, {Kotak},
  {Le}, {Liska}, {Long}, {Lucey}, {Liu}, {Martin}, {Masci}, {McLean}, {Mindel},
  {Misra}, {Morganson}, {Murphy}, {Obaika}, {Narayan}, {Nieto-Santisteban},
  {Norberg}, {Peacock}, {Pier}, {Postman}, {Primak}, {Rae}, {Rai}, {Riess},
  {Riffeser}, {Rix}, {R{\"o}ser}, {Russel}, {Rutz}, {Schilbach}, {Schultz},
  {Scolnic}, {Strolger}, {Szalay}, {Seitz}, {Small}, {Smith}, {Soderblom},
  {Taylor}, {Thomson}, {Taylor}, {Thakar}, {Thiel}, {Thilker}, {Unger},
  {Urata}, {Valenti}, {Wagner}, {Walder}, {Walter}, {Watters}, {Werner},
  {Wood-Vasey}, \& {Wyse}}]{2016arXiv161205560C}
{Chambers}, K.~C., {Magnier}, E.~A., {Metcalfe}, N., {et~al.} 2016, arXiv
  e-prints, arXiv:1612.05560, \dodoi{10.48550/arXiv.1612.05560}

\bibitem[{{Chen} {et~al.}(2020){Chen}, {Wang}, {Deng}, {de Grijs}, {Yang}, \&
  {Tian}}]{2020ApJS..249...18C}
{Chen}, X., {Wang}, S., {Deng}, L., {et~al.} 2020, \apjs, 249, 18,
  \dodoi{10.3847/1538-4365/ab9cae}

\bibitem[{{Cheung} {et~al.}(2022){Cheung}, {Johnson}, {Jean}, {Kerr}, {Page},
  {Osborne}, {Beardmore}, {Sokolovsky}, {Teyssier}, {Ciprini},
  {Mart{\'\i}-Devesa}, {Mereu}, {Razzaque}, {Wood}, {Shore}, {Korotkiy},
  {Levina}, \& {Blumenzweig}}]{2022ApJ...935...44C}
{Cheung}, C.~C., {Johnson}, T.~J., {Jean}, P., {et~al.} 2022, \apj, 935, 44,
  \dodoi{10.3847/1538-4357/ac7eb7}

\bibitem[{{Chochol} \& {Pribulla}(1997)}]{1997CoSka..27...53C}
{Chochol}, D., \& {Pribulla}, T. 1997, Contributions of the Astronomical
  Observatory Skalnate Pleso, 27, 53

\bibitem[{{Chomiuk} {et~al.}(2021{\natexlab{a}}){Chomiuk}, {Metzger}, \&
  {Shen}}]{2021ARA&A..59..391C}
{Chomiuk}, L., {Metzger}, B.~D., \& {Shen}, K.~J. 2021{\natexlab{a}}, \araa,
  59, 391, \dodoi{10.1146/annurev-astro-112420-114502}

\bibitem[{{Chomiuk} {et~al.}(2014){Chomiuk}, {Linford}, {Yang}, {O'Brien},
  {Paragi}, {Mioduszewski}, {Beswick}, {Cheung}, {Mukai}, {Nelson}, {Ribeiro},
  {Rupen}, {Sokoloski}, {Weston}, {Zheng}, {Bode}, {Eyres}, {Roy}, \&
  {Taylor}}]{2014Natur.514..339C}
{Chomiuk}, L., {Linford}, J.~D., {Yang}, J., {et~al.} 2014, \nat, 514, 339,
  \dodoi{10.1038/nature13773}

\bibitem[{{Chomiuk} {et~al.}(2021{\natexlab{b}}){Chomiuk}, {Linford}, {Aydi},
  {Bannister}, {Krauss}, {Mioduszewski}, {Mukai}, {Nelson}, {Rupen}, {Ryder},
  {Sokoloski}, {Sokolovsky}, {Strader}, {Filipovi{\'c}}, {Finzell}, {Kawash},
  {Kool}, {Metzger}, {Nyamai}, {Ribeiro}, {Roy}, {Urquhart}, \&
  {Weston}}]{2021ApJS..257...49C}
{Chomiuk}, L., {Linford}, J.~D., {Aydi}, E., {et~al.} 2021{\natexlab{b}},
  \apjs, 257, 49, \dodoi{10.3847/1538-4365/ac24ab}

\bibitem[{{Cohen} {et~al.}(2025){Cohen}, {Guetta}, {Hillman}, {Della Valle},
  {Izzo}, {Perdelwitz}, \& {Livio}}]{2025ApJ...981..198C}
{Cohen}, A., {Guetta}, D., {Hillman}, Y., {et~al.} 2025, \apj, 981, 198,
  \dodoi{10.3847/1538-4357/adb628}

\bibitem[{{Condon}(1974)}]{1974ApJ...188..279C}
{Condon}, J.~J. 1974, \apj, 188, 279, \dodoi{10.1086/152714}

\bibitem[{{Craig} {et~al.}(2026){Craig}, {Aydi}, {Chomiuk}, {Stone}, {Strader},
  {Chong}, {Li}, {Fan}, {Bahramian}, {Buckley}, {Izzo}, {Kawash}, {Metzger},
  {Mukai}, {Linford}, {Orio}, {Sokoloski}, {Sokolovsky}, {Tremou}, {Walter},
  {Fl{\'o}}, {Boussin}, {Charbonnel}, {Garde}, {Belyakov}, {Monard}, {Hambsch},
  \& {Thomas}}]{2026MNRAS.546f2270C}
{Craig}, P., {Aydi}, E., {Chomiuk}, L., {et~al.} 2026, \mnras, 546, staf2270,
  \dodoi{10.1093/mnras/staf2270}

\bibitem[{{Cushing} {et~al.}(2004){Cushing}, {Vacca}, \&
  {Rayner}}]{2004PASP..116..362C}
{Cushing}, M.~C., {Vacca}, W.~D., \& {Rayner}, J.~T. 2004, \pasp, 116, 362,
  \dodoi{10.1086/382907}

\bibitem[{{De} {et~al.}(2020){De}, {Hankins}, {Kasliwal}, {Moore}, {Ofek},
  {Adams}, {Ashley}, {Babul}, {Bagdasaryan}, {Burdge}, {Burnham}, {Dekany},
  {Declacroix}, {Galla}, {Greffe}, {Hale}, {Jencson}, {Lau}, {Mahabal},
  {McKenna}, {Sharma}, {Shopbell}, {Smith}, {Soon}, {Sokoloski}, {Soria}, \&
  {Travouillon}}]{2020PASP..132b5001D}
{De}, K., {Hankins}, M.~J., {Kasliwal}, M.~M., {et~al.} 2020, \pasp, 132,
  025001, \dodoi{10.1088/1538-3873/ab6069}

\bibitem[{{De} {et~al.}(2021){De}, {Kasliwal}, {Hankins}, {Sokoloski}, {Adams},
  {Ashley}, {Babul}, {Bagdasaryan}, {Delacroix}, {Dekany}, {Greffe}, {Hale},
  {Jencson}, {Karambelkar}, {Lau}, {Mahabal}, {McKenna}, {Moore}, {Ofek},
  {Sharma}, {Smith}, {Soon}, {Soria}, {Srinivasaragavan}, {Tinyanont},
  {Travouillon}, {Tzanidakis}, \& {Yao}}]{2021ApJ...912...19D}
{De}, K., {Kasliwal}, M.~M., {Hankins}, M.~J., {et~al.} 2021, \apj, 912, 19,
  \dodoi{10.3847/1538-4357/abeb75}

\bibitem[{{De} {et~al.}(2022){De}, {Soria}, {Agusti}, {Kong}, {Karambelkar},
  {Hankins}, {Kasliwal}, {Sokoloski}, {Ashley}, {Babul}, {Lau}, {Moore},
  {Ofek}, {Sharma}, {Simcoe}, {Soon}, {Soria}, {Travouillon}, \&
  {Vanderburg}}]{2022ATel15587....1D}
{De}, K., {Soria}, R., {Agusti}, M.~B., {et~al.} 2022, The Astronomer's
  Telegram, 15587, 1

\bibitem[{{Deeming}(1975)}]{1975Ap&SS..36..137D}
{Deeming}, T.~J. 1975, \apss, 36, 137, \dodoi{10.1007/BF00681947}

\bibitem[{{Dubovsk{\'y}} {et~al.}(2024){Dubovsk{\'y}}, {Petr{\'\i}k}, \&
  {Breus}}]{2024CoSka..54b.128D}
{Dubovsk{\'y}}, P.~A., {Petr{\'\i}k}, K., \& {Breus}, V. 2024, Contributions of
  the Astronomical Observatory Skalnate Pleso, 54, 128,
  \dodoi{10.31577/caosp.2024.54.2.128}

\bibitem[{{Eyres} {et~al.}(2017){Eyres}, {Bewsher}, {Hillman}, {Holdsworth},
  {Rushton}, {Bresnahan}, {Evans}, \& {Mr{\'o}z}}]{2017MNRAS.467.2684E}
{Eyres}, S.~P.~S., {Bewsher}, D., {Hillman}, Y., {et~al.} 2017, \mnras, 467,
  2684, \dodoi{10.1093/mnras/stx298}

\bibitem[{{Fabian} \& {Pringle}(1977)}]{1977MNRAS.180..749F}
{Fabian}, A.~C., \& {Pringle}, J.~E. 1977, \mnras, 180, 749,
  \dodoi{10.1093/mnras/180.4.749}

\bibitem[{{Franckowiak} {et~al.}(2018){Franckowiak}, {Jean}, {Wood}, {Cheung},
  \& {Buson}}]{2018A&A...609A.120F}
{Franckowiak}, A., {Jean}, P., {Wood}, M., {Cheung}, C.~C., \& {Buson}, S.
  2018, \aap, 609, A120, \dodoi{10.1051/0004-6361/201731516}

\bibitem[{{Friedjung}(1990)}]{1990LNP...369..244F}
{Friedjung}, M. 1990, in IAU Colloq. 122: Physics of Classical Novae, ed.
  A.~{Cassatella} \& R.~{Viotti}, Vol. 369, 244,
  \dodoi{10.1007/3-540-53500-4_132}

\bibitem[{{Friedjung}(2004)}]{2004BaltA..13..116F}
---. 2004, Baltic Astronomy, 13, 116

\bibitem[{{Gonz{\'a}lez-Bol{\'\i}var}
  {et~al.}(2022){Gonz{\'a}lez-Bol{\'\i}var}, {De Marco}, {Lau}, {Hirai}, \&
  {Price}}]{2022MNRAS.517.3181G}
{Gonz{\'a}lez-Bol{\'\i}var}, M., {De Marco}, O., {Lau}, M. Y.~M., {Hirai}, R.,
  \& {Price}, D.~J. 2022, \mnras, 517, 3181, \dodoi{10.1093/mnras/stac2301}

\bibitem[{{Goranskij} {et~al.}(2007){Goranskij}, {Katysheva}, {Kusakin},
  {Metlova}, {Pogrosheva}, {Shugarov}, {Barsukova}, {Fabrika}, {Borisov},
  {Burenkov}, {Pramsky}, {Karitskaya}, \& {Retter}}]{2007AstBu..62..125G}
{Goranskij}, V.~P., {Katysheva}, N.~A., {Kusakin}, A.~V., {et~al.} 2007,
  Astrophysical Bulletin, 62, 125, \dodoi{10.1134/S1990341307020046}

\bibitem[{{Gordon} {et~al.}(2021){Gordon}, {Aydi}, {Page}, {Li}, {Chomiuk},
  {Sokolovsky}, {Mukai}, \& {Seitz}}]{2021ApJ...910..134G}
{Gordon}, A.~C., {Aydi}, E., {Page}, K.~L., {et~al.} 2021, \apj, 910, 134,
  \dodoi{10.3847/1538-4357/abe547}

\bibitem[{{Graham} {et~al.}(2019){Graham}, {Kulkarni}, {Bellm}, {Adams},
  {Barbarino}, {Blagorodnova}, {Bodewits}, {Bolin}, {Brady}, {Cenko}, {Chang},
  {Coughlin}, {De}, {Eadie}, {Farnham}, {Feindt}, {Franckowiak}, {Fremling},
  {Gezari}, {Ghosh}, {Goldstein}, {Golkhou}, {Goobar}, {Ho}, {Huppenkothen},
  {Ivezi{\'c}}, {Jones}, {Juric}, {Kaplan}, {Kasliwal}, {Kelley}, {Kupfer},
  {Lee}, {Lin}, {Lunnan}, {Mahabal}, {Miller}, {Ngeow}, {Nugent}, {Ofek},
  {Prince}, {Rauch}, {van Roestel}, {Schulze}, {Singer}, {Sollerman}, {Taddia},
  {Yan}, {Ye}, {Yu}, {Barlow}, {Bauer}, {Beck}, {Belicki}, {Biswas}, {Brinnel},
  {Brooke}, {Bue}, {Bulla}, {Burruss}, {Connolly}, {Cromer}, {Cunningham},
  {Dekany}, {Delacroix}, {Desai}, {Duev}, {Feeney}, {Flynn}, {Frederick},
  {Gal-Yam}, {Giomi}, {Groom}, {Hacopians}, {Hale}, {Helou}, {Henning},
  {Hover}, {Hillenbrand}, {Howell}, {Hung}, {Imel}, {Ip}, {Jackson}, {Kaspi},
  {Kaye}, {Kowalski}, {Kramer}, {Kuhn}, {Landry}, {Laher}, {Mao}, {Masci},
  {Monkewitz}, {Murphy}, {Nordin}, {Patterson}, {Penprase}, {Porter},
  {Rebbapragada}, {Reiley}, {Riddle}, {Rigault}, {Rodriguez}, {Rusholme}, {van
  Santen}, {Shupe}, {Smith}, {Soumagnac}, {Stein}, {Surace}, {Szkody}, {Terek},
  {Van Sistine}, {van Velzen}, {Vestrand}, {Walters}, {Ward}, {Zhang}, \&
  {Zolkower}}]{2019PASP..131g8001G}
{Graham}, M.~J., {Kulkarni}, S.~R., {Bellm}, E.~C., {et~al.} 2019, \pasp, 131,
  078001, \dodoi{10.1088/1538-3873/ab006c}

\bibitem[{{H.~E.~S.~S. Collaboration} {et~al.}(2022){H.~E.~S.~S.
  Collaboration}, {Aharonian}, {Ait Benkhali}, {Ang{\"u}ner}, {Ashkar},
  {Backes}, {Baghmanyan}, {Barbosa Martins}, {Batzofin}, {Becherini}, {Berge},
  {Bernl{\"o}hr}, {Bi}, {B{\"o}ttcher}, {Boisson}, {Bolmont}, {de Bony de
  Lavergne}, {Breuhaus}, {Brose}, {Brun}, {Caroff}, {Casanova}, {Cerruti},
  {Chand}, {Chen}, {Cotter}, {Damascene Mbarubucyeye}, {Djannati-Ata{\"\i}},
  {Dmytriiev}, {Doroshenko}, {Duffy}, {Egberts}, {Ernenwein}, {Fegan},
  {Feijen}, {Fiasson}, {Fichet de Clairfontaine}, {Fontaine},
  {F{\"u}{\ss}ling}, {Funk}, {Gabici}, {Gallant}, {Ghafourizadeh}, {Giavitto},
  {Giunti}, {Glawion}, {Glicenstein}, {Grondin}, {Hermann}, {Hinton},
  {H{\"o}rbe}, {Hofmann}, {Hoischen}, {Holch}, {Holler}, {Horns}, {Huang},
  {Jamrozy}, {Jankowsky}, {Jung-Richardt}, {Kasai}, {Katarzy{\'n}ski}, {Katz},
  {Khangulyan}, {Kh{\'e}lifi}, {Klepser}, {Klu{\'z}niak}, {Komin}, {Konno},
  {Kosack}, {Kostunin}, {Le Stum}, {Lemi{\`e}re}, {Lemoine-Goumard}, {Lenain},
  {Leuschner}, {Lohse}, {Luashvili}, {Lypova}, {Mackey}, {Malyshev},
  {Malyshev}, {Marandon}, {Marchegiani}, {Marcowith}, {Mart{\'\i}-Devesa},
  {Marx}, {Maurin}, {Meyer}, {Mitchell}, {Moderski}, {Mohrmann}, {Montanari},
  {Moulin}, {Muller}, {Murach}, {Nakashima}, {de Naurois}, {Nayerhoda},
  {Niemiec}, {Priyana Noel}, {O{\textquoteright}Brien}, {Ohm}, {Olivera-Nieto},
  {de Ona Wilhelmi}, {Ostrowski}, {Panny}, {Panter}, {Parsons}, {Peron},
  {Pita}, {Poireau}, {Prokhorov}, {Prokoph}, {P{\"u}hlhofer}, {Punch},
  {Quirrenbach}, {Reichherzer}, {Reimer}, {Reimer}, {Renaud}, {Reville},
  {Rieger}, {Rowell}, {Rudak}, {Rueda Ricarte}, {Ruiz-Velasco}, {Sahakian},
  {Sailer}, {Salzmann}, {Sanchez}, {Santangelo}, {Sasaki}, {Sch{\"a}fer},
  {Sch{\"u}ssler}, {Schutte}, {Schwanke}, {Senniappan}, {Shapopi}, {Simoni},
  {Sinha}, {Sol}, {Specovius}, {Spencer}, {Stawarz}, {Steinmassl}, {Steppa},
  {Takahashi}, {Tanaka}, {Taylor}, {Terrier}, {Thorpe-Morgan}, {Tsirou},
  {Tsuji}, {Tuffs}, {Uchiyama}, {Unbehaun}, {van Eldik}, {van Soelen}, {Veh},
  {Venter}, {Vink}, {Wagner}, {Werner}, {White}, {Wierzcholska}, {Wong},
  {Yusafzai}, {Zacharias}, {Zargaryan}, {Zdziarski}, {Zech}, {Zhu}, {Zouari},
  \& {{\.Z}ywucka}}]{2022Sci...376...77H}
{H.~E.~S.~S. Collaboration}, {Aharonian}, F., {Ait Benkhali}, F., {et~al.}
  2022, Science, 376, 77, \dodoi{10.1126/science.abn0567}

\bibitem[{{Hachisu} \& {Kato}(2004)}]{2004ApJ...612L..57H}
{Hachisu}, I., \& {Kato}, M. 2004, \apjl, 612, L57, \dodoi{10.1086/424595}

\bibitem[{{Harrison} \& {Campbell}(2016)}]{2016MNRAS.459.4161H}
{Harrison}, T.~E., \& {Campbell}, R.~K. 2016, \mnras, 459, 4161,
  \dodoi{10.1093/mnras/stw961}

\bibitem[{{Heinze} {et~al.}(2018){Heinze}, {Tonry}, {Denneau}, {Flewelling},
  {Stalder}, {Rest}, {Smith}, {Smartt}, \& {Weiland}}]{2018AJ....156..241H}
{Heinze}, A.~N., {Tonry}, J.~L., {Denneau}, L., {et~al.} 2018, \aj, 156, 241,
  \dodoi{10.3847/1538-3881/aae47f}

\bibitem[{{Herter} {et~al.}(2008){Herter}, {Henderson}, {Wilson}, {Matthews},
  {Rahmer}, {Bonati}, {Muirhead}, {Adams}, {Lloyd}, {Skrutskie}, {Moon},
  {Parshley}, {Nelson}, {Martinache}, \& {Gull}}]{2008SPIE.7014E..0XH}
{Herter}, T.~L., {Henderson}, C.~P., {Wilson}, J.~C., {et~al.} 2008, in Society
  of Photo-Optical Instrumentation Engineers (SPIE) Conference Series, Vol.
  7014, Ground-based and Airborne Instrumentation for Astronomy II, ed. I.~S.
  {McLean} \& M.~M. {Casali}, 70140X, \dodoi{10.1117/12.789660}

\bibitem[{{Hillman}(2022)}]{2022MNRAS.515.1404H}
{Hillman}, Y. 2022, \mnras, 515, 1404, \dodoi{10.1093/mnras/stac1688}

\bibitem[{{Hillman} {et~al.}(2014){Hillman}, {Prialnik}, {Kovetz}, {Shara}, \&
  {Neill}}]{2014MNRAS.437.1962H}
{Hillman}, Y., {Prialnik}, D., {Kovetz}, A., {Shara}, M.~M., \& {Neill}, J.~D.
  2014, \mnras, 437, 1962, \dodoi{10.1093/mnras/stt2027}

\bibitem[{{Hogg}(2001)}]{2001AJ....121.1207H}
{Hogg}, D.~W. 2001, \aj, 121, 1207, \dodoi{10.1086/318736}

\bibitem[{{Holdsworth} {et~al.}(2014){Holdsworth}, {Rushton}, {Bewsher},
  {Walter}, {Eyres}, {Hounsell}, \& {Darnley}}]{2014MNRAS.438.3483H}
{Holdsworth}, D.~L., {Rushton}, M.~T., {Bewsher}, D., {et~al.} 2014, \mnras,
  438, 3483, \dodoi{10.1093/mnras/stt2455}

\bibitem[{{Hounsell} {et~al.}(2016){Hounsell}, {Darnley}, {Bode}, {Harman},
  {Surina}, {Starrfield}, {Holdsworth}, {Bewsher}, {Hick}, {Jackson},
  {Buffington}, {Clover}, \& {Shafter}}]{2016ApJ...820..104H}
{Hounsell}, R., {Darnley}, M.~J., {Bode}, M.~F., {et~al.} 2016, \apj, 820, 104,
  \dodoi{10.3847/0004-637X/820/2/104}

\bibitem[{{Iben}(2003)}]{2003ASPC..303..177I}
{Iben}, Jr., I. 2003, in Astronomical Society of the Pacific Conference Series,
  Vol. 303, Symbiotic Stars Probing Stellar Evolution, ed. R.~L.~M. {Corradi},
  J.~{Mikolajewska}, \& T.~J. {Mahoney}, 177

\bibitem[{{Ivanov}(1978)}]{1978SvAL....4..141I}
{Ivanov}, L.~N. 1978, Soviet Astronomy Letters, 4, 141

\bibitem[{{Ivanova} {et~al.}(2013){Ivanova}, {Justham}, {Avendano Nandez}, \&
  {Lombardi}}]{2013Sci...339..433I}
{Ivanova}, N., {Justham}, S., {Avendano Nandez}, J.~L., \& {Lombardi}, J.~C.
  2013, Science, 339, 433, \dodoi{10.1126/science.1225540}

\bibitem[{{Kahabka} \& {van den Heuvel}(1997)}]{1997ARA&A..35...69K}
{Kahabka}, P., \& {van den Heuvel}, E.~P.~J. 1997, \araa, 35, 69,
  \dodoi{10.1146/annurev.astro.35.1.69}

\bibitem[{{Kato} \& {Hachisu}(1994)}]{1994ApJ...437..802K}
{Kato}, M., \& {Hachisu}, I. 1994, \apj, 437, 802, \dodoi{10.1086/175041}

\bibitem[{{Kato} \& {Hachisu}(2009)}]{2009ApJ...699.1293K}
---. 2009, \apj, 699, 1293, \dodoi{10.1088/0004-637X/699/2/1293}

\bibitem[{{Kato} \& {Hachisu}(2011)}]{2011ApJ...743..157K}
---. 2011, \apj, 743, 157, \dodoi{10.1088/0004-637X/743/2/157}

\bibitem[{{Kato} \& {Hachisu}(2015)}]{2015gacv.workE..52K}
{Kato}, M., \& {Hachisu}, I. 2015, in The Golden Age of Cataclysmic Variables
  and Related Objects - III (Golden2015), 52, \dodoi{10.22323/1.255.0052}

\bibitem[{{Kawash} {et~al.}(2021){Kawash}, {Chomiuk}, {Strader}, {Aydi},
  {Sokolovsky}, {Jayasinghe}, {Kochanek}, {Schmeer}, {Stanek}, {Mukai},
  {Shappee}, {Way}, {Basinger}, {Holoien}, \& {Prieto}}]{2021ApJ...910..120K}
{Kawash}, A., {Chomiuk}, L., {Strader}, J., {et~al.} 2021, \apj, 910, 120,
  \dodoi{10.3847/1538-4357/abe53d}

\bibitem[{{Kawash} {et~al.}(2022){Kawash}, {Chomiuk}, {Strader}, {Sokolovsky},
  {Aydi}, {Kochanek}, {Stanek}, {Kostrzewa-Rutkowska}, {Hodgkin}, {Mukai},
  {Shappee}, {Jayasinghe}, {Rizzo Smith}, {Holoien}, {Prieto}, \&
  {Thompson}}]{2022ApJ...937...64K}
---. 2022, \apj, 937, 64, \dodoi{10.3847/1538-4357/ac8d5e}

\bibitem[{{Kenyon} \& {Truran}(1983)}]{1983ApJ...273..280K}
{Kenyon}, S.~J., \& {Truran}, J.~W. 1983, \apj, 273, 280,
  \dodoi{10.1086/161367}

\bibitem[{{Kloppenborg}(2025)}]{AAVSODATA}
{Kloppenborg}, B.~K. 2025, {Observations from the AAVSO International Database,
  \url{https://www.aavso.org}}

\bibitem[{{K{\"o}nig} {et~al.}(2022){K{\"o}nig}, {Wilms}, {Arcodia}, {Dauser},
  {Dennerl}, {Doroshenko}, {Haberl}, {H{\"a}mmerich}, {Kirsch}, {Kreykenbohm},
  {Lorenz}, {Malyali}, {Merloni}, {Rau}, {Rauch}, {Sala}, {Schwope},
  {Suleimanov}, {Weber}, \& {Werner}}]{2022Natur.605..248K}
{K{\"o}nig}, O., {Wilms}, J., {Arcodia}, R., {et~al.} 2022, \nat, 605, 248,
  \dodoi{10.1038/s41586-022-04635-y}

\bibitem[{{Kornilov} {et~al.}(2012){Kornilov}, {Lipunov}, {Gorbovskoy},
  {Belinski}, {Kuvshinov}, {Tyurina}, {Shatsky}, {Sankovich}, {Krylov},
  {Balanutsa}, {Chazov}, {Kuznetsov}, {Zimnuhov}, {Senik}, {Tlatov},
  {Parkhomenko}, {Dormidontov}, {Krushinsky}, {Zalozhnyh}, {Popov}, {Yazev},
  {Budnev}, {Ivanov}, {Konstantinov}, {Gress}, {Chvalaev}, {Yurkov},
  {Sergienko}, \& {Kudelina}}]{2012ExA....33..173K}
{Kornilov}, V.~G., {Lipunov}, V.~M., {Gorbovskoy}, E.~S., {et~al.} 2012,
  Experimental Astronomy, 33, 173, \dodoi{10.1007/s10686-011-9280-z}

\bibitem[{{Krishnamurthy} {et~al.}(2019){Krishnamurthy}, {Villasenor},
  {Seager}, {Ricker}, \& {Vanderspek}}]{2019AcAau.160...46K}
{Krishnamurthy}, A., {Villasenor}, J., {Seager}, S., {Ricker}, G., \&
  {Vanderspek}, R. 2019, Acta Astronautica, 160, 46,
  \dodoi{10.1016/j.actaastro.2019.04.016}

\bibitem[{{Kuiper}(1941)}]{1941ApJ....93..133K}
{Kuiper}, G.~P. 1941, \apj, 93, 133, \dodoi{10.1086/144252}

\bibitem[{{Lamb} {et~al.}(1983){Lamb}, {Aly}, {Cook}, \&
  {Lamb}}]{1983ApJ...274L..71L}
{Lamb}, F.~K., {Aly}, J.-J., {Cook}, M.~C., \& {Lamb}, D.~Q. 1983, \apjl, 274,
  L71, \dodoi{10.1086/184153}

\bibitem[{{Li} {et~al.}(2017){Li}, {Metzger}, {Chomiuk}, {Vurm}, {Strader},
  {Finzell}, {Beloborodov}, {Nelson}, {Shappee}, {Kochanek}, {Prieto}, {Kafka},
  {Holoien}, {Thompson}, {Luckas}, \& {Itoh}}]{2017NatAs...1..697L}
{Li}, K.-L., {Metzger}, B.~D., {Chomiuk}, L., {et~al.} 2017, Nature Astronomy,
  1, 697, \dodoi{10.1038/s41550-017-0222-1}

\bibitem[{{Lightkurve Collaboration} {et~al.}(2018){Lightkurve Collaboration},
  {Cardoso}, {Hedges}, {Gully-Santiago}, {Saunders}, {Cody}, {Barclay}, {Hall},
  {Sagear}, {Turtelboom}, {Zhang}, {Tzanidakis}, {Mighell}, {Coughlin}, {Bell},
  {Berta-Thompson}, {Williams}, {Dotson}, \& {Barentsen}}]{2018ascl.soft12013L}
{Lightkurve Collaboration}, {Cardoso}, J.~V.~d.~M., {Hedges}, C., {et~al.}
  2018, {Lightkurve: Kepler and TESS time series analysis in Python},
  Astrophysics Source Code Library.
\newblock \doeprint{1812.013}

\bibitem[{{Lindegren} \& {Lindgren}(1975)}]{1975Natur.258..501L}
{Lindegren}, L., \& {Lindgren}, H. 1975, \nat, 258, 501,
  \dodoi{10.1038/258501a0}

\bibitem[{{Lipunov} {et~al.}(2010){Lipunov}, {Kornilov}, {Gorbovskoy},
  {Shatskij}, {Kuvshinov}, {Tyurina}, {Belinski}, {Krylov}, {Balanutsa},
  {Chazov}, {Kuznetsov}, {Kortunov}, {Sankovich}, {Tlatov}, {Parkhomenko},
  {Krushinsky}, {Zalozhnyh}, {Popov}, {Kopytova}, {Ivanov}, {Yazev}, \&
  {Yurkov}}]{2010AdAst2010E..30L}
{Lipunov}, V., {Kornilov}, V., {Gorbovskoy}, E., {et~al.} 2010, Advances in
  Astronomy, 2010, 349171, \dodoi{10.1155/2010/349171}

\bibitem[{{Livio} {et~al.}(1990){Livio}, {Shankar}, {Burkert}, \&
  {Truran}}]{1990ApJ...356..250L}
{Livio}, M., {Shankar}, A., {Burkert}, A., \& {Truran}, J.~W. 1990, \apj, 356,
  250, \dodoi{10.1086/168836}

\bibitem[{{Lomb}(1976)}]{1976Ap&SS..39..447L}
{Lomb}, N.~R. 1976, \apss, 39, 447, \dodoi{10.1007/BF00648343}

\bibitem[{{Luna} {et~al.}(2026{\natexlab{a}}){Luna}, {Dobrotka}, \&
  {Orio}}]{2026A&A...708A.352L}
{Luna}, G.~J.~M., {Dobrotka}, A., \& {Orio}, M. 2026{\natexlab{a}}, \aap, 708,
  A352, \dodoi{10.1051/0004-6361/202557972}

\bibitem[{{Luna} {et~al.}(2024){Luna}, {Lima}, \& {Orio}}]{2023arXiv231002220L}
{Luna}, G.~J.~M., {Lima}, I.~J., \& {Orio}, M. 2024, Boletin de la Asociacion
  Argentina de Astronomia La Plata Argentina, 65, 60,
  \dodoi{10.48550/arXiv.2310.02220}

\bibitem[{{Luna} {et~al.}(2026{\natexlab{b}}){Luna}, {Rawat}, {Angeloni},
  {Orio}, {Scaringi}, {Dobrotka}, \& {Magdolen}}]{2026arXiv260522802L}
{Luna}, G.~J.~M., {Rawat}, N., {Angeloni}, R., {et~al.} 2026{\natexlab{b}},
  arXiv e-prints, arXiv:2605.22802, \dodoi{10.48550/arXiv.2605.22802}

\bibitem[{{MacLeod} {et~al.}(2018){MacLeod}, {Ostriker}, \&
  {Stone}}]{2018ApJ...863....5M}
{MacLeod}, M., {Ostriker}, E.~C., \& {Stone}, J.~M. 2018, \apj, 863, 5,
  \dodoi{10.3847/1538-4357/aacf08}

\bibitem[{{Malanchev} {et~al.}(2023){Malanchev}, {Kornilov}, {Pruzhinskaya},
  {Ishida}, {Aleo}, {Korolev}, {Lavrukhina}, {Russeil}, {Sreejith}, {Volnova},
  {Voloshina}, \& {Krone-Martins}}]{2023PASP..135b4503M}
{Malanchev}, K., {Kornilov}, M.~V., {Pruzhinskaya}, M.~V., {et~al.} 2023,
  \pasp, 135, 024503, \dodoi{10.1088/1538-3873/acb292}

\bibitem[{{Masci} {et~al.}(2019){Masci}, {Laher}, {Rusholme}, {Shupe}, {Groom},
  {Surace}, {Jackson}, {Monkewitz}, {Beck}, {Flynn}, {Terek}, {Landry},
  {Hacopians}, {Desai}, {Howell}, {Brooke}, {Imel}, {Wachter}, {Ye}, {Lin},
  {Cenko}, {Cunningham}, {Rebbapragada}, {Bue}, {Miller}, {Mahabal}, {Bellm},
  {Patterson}, {Juri{\'c}}, {Golkhou}, {Ofek}, {Walters}, {Graham}, {Kasliwal},
  {Dekany}, {Kupfer}, {Burdge}, {Cannella}, {Barlow}, {Van Sistine}, {Giomi},
  {Fremling}, {Blagorodnova}, {Levitan}, {Riddle}, {Smith}, {Helou}, {Prince},
  \& {Kulkarni}}]{2019PASP..131a8003M}
{Masci}, F.~J., {Laher}, R.~R., {Rusholme}, B., {et~al.} 2019, \pasp, 131,
  018003, \dodoi{10.1088/1538-3873/aae8ac}

\bibitem[{{Mikolajewska}(2008)}]{2008ASPC..401...42M}
{Mikolajewska}, J. 2008, in Astronomical Society of the Pacific Conference
  Series, Vol. 401, RS Ophiuchi (2006) and the Recurrent Nova Phenomenon, ed.
  A.~{Evans}, M.~F. {Bode}, T.~J. {O'Brien}, \& M.~J. {Darnley}, 42,
  \dodoi{10.48550/arXiv.0803.3685}

\bibitem[{{Miko{\l}ajewska}(2012)}]{2012BaltA..21....5M}
{Miko{\l}ajewska}, J. 2012, Baltic Astronomy, 21, 5,
  \dodoi{10.1515/astro-2017-0352}

\bibitem[{{Miko{\l}ajewska} \& {Shara}(2017)}]{2017ApJ...847...99M}
{Miko{\l}ajewska}, J., \& {Shara}, M.~M. 2017, \apj, 847, 99,
  \dodoi{10.3847/1538-4357/aa87b6}

\bibitem[{{M{\"o}ller} {et~al.}(2021){M{\"o}ller}, {Peloton}, {Ishida},
  {Arnault}, {Bachelet}, {Blaineau}, {Boutigny}, {Chauhan}, {Gangler},
  {Hernandez}, {Hrivnac}, {Leoni}, {Leroy}, {Moniez}, {Pateyron}, {Ramparison},
  {Turpin}, {Ansari}, {Allam}, {Bajat}, {Biswas}, {Boucaud}, {Bregeon},
  {Campagne}, {Cohen-Tanugi}, {Coleiro}, {Dornic}, {Fouchez}, {Godet}, {Gris},
  {Karpov}, {Nebot Gomez-Moran}, {Neveu}, {Plaszczynski}, {Savchenko}, \&
  {Webb}}]{2021MNRAS.501.3272M}
{M{\"o}ller}, A., {Peloton}, J., {Ishida}, E. E.~O., {et~al.} 2021, \mnras,
  501, 3272, \dodoi{10.1093/mnras/staa3602}

\bibitem[{{Munari}(2025)}]{2025CoSka..55c..47M}
{Munari}, U. 2025, Contributions of the Astronomical Observatory Skalnate
  Pleso, 55, 47, \dodoi{10.31577/caosp.2025.55.3.47}

\bibitem[{{Munari} {et~al.}(2017){Munari}, {Hambsch}, \&
  {Frigo}}]{2017MNRAS.469.4341M}
{Munari}, U., {Hambsch}, F.~J., \& {Frigo}, A. 2017, \mnras, 469, 4341,
  \dodoi{10.1093/mnras/stx1116}

\bibitem[{{Ness} {et~al.}(2013){Ness}, {Osborne}, {Henze}, {Dobrotka}, {Drake},
  {Ribeiro}, {Starrfield}, {Kuulkers}, {Behar}, {Hernanz}, {Schwarz}, {Page},
  {Beardmore}, \& {Bode}}]{2013A&A...559A..50N}
{Ness}, J.~U., {Osborne}, J.~P., {Henze}, M., {et~al.} 2013, \aap, 559, A50,
  \dodoi{10.1051/0004-6361/201322415}

\bibitem[{{Norton} {et~al.}(2004){Norton}, {Wynn}, \&
  {Somerscales}}]{2004ApJ...614..349N}
{Norton}, A.~J., {Wynn}, G.~A., \& {Somerscales}, R.~V. 2004, \apj, 614, 349,
  \dodoi{10.1086/423333}

\bibitem[{{Ofek}(2019)}]{2019PASP..131e4504O}
{Ofek}, E.~O. 2019, \pasp, 131, 054504, \dodoi{10.1088/1538-3873/ab04df}

\bibitem[{{Oke} \& {Gunn}(1982)}]{1982PASP...94..586O}
{Oke}, J.~B., \& {Gunn}, J.~E. 1982, \pasp, 94, 586, \dodoi{10.1086/131027}

\bibitem[{{Olbemo} {et~al.}(2026){Olbemo}, {Errando}, \&
  {Gokus}}]{2026arXiv260506917O}
{Olbemo}, T., {Errando}, M., \& {Gokus}, A. 2026, \apj, 1004, 38,
  \dodoi{10.3847/1538-4357/ae6a8c}

\bibitem[{{Paegert} {et~al.}(2021){Paegert}, {Stassun}, {Collins}, {Pepper},
  {Torres}, {Jenkins}, {Twicken}, \& {Latham}}]{2021arXiv210804778P}
{Paegert}, M., {Stassun}, K.~G., {Collins}, K.~A., {et~al.} 2021, arXiv
  e-prints, arXiv:2108.04778, \dodoi{10.48550/arXiv.2108.04778}

\bibitem[{{Patterson}(1979)}]{1979ApJ...231..789P}
{Patterson}, J. 1979, \apj, 231, 789, \dodoi{10.1086/157244}

\bibitem[{{Patterson} {et~al.}(2022){Patterson}, {Enenstein}, {de Miguel},
  {Epstein-Martin}, {Kemp}, {Sabo}, {Cooney}, {Vanmunster}, {Dubovsky},
  {Hambsch}, {Myers}, {Lemay}, {Sokolovsky}, {Collins}, {Campbell}, {Roberts},
  {Richmond}, {Brincat}, {Ulowetz}, {Dvorak}, {Tordai}, {Dufoer}, {Cahaly},
  {Galdies}, {Goff}, {Wilkin}, \& {Wood}}]{2022ApJ...940L..56P}
{Patterson}, J., {Enenstein}, J., {de Miguel}, E., {et~al.} 2022, \apjl, 940,
  L56, \dodoi{10.3847/2041-8213/ac9ebe}

\bibitem[{{Pavana}(2020)}]{2020PhDT........48P}
{Pavana}, M. 2020, PhD thesis, Indian Institute of Astrophysics, Bangalore

\bibitem[{{Pavlenko} {et~al.}(2018){Pavlenko}, {Mason}, {Sosnovskij},
  {Shugarov}, {Babina}, {Antonyuk}, {Andreev}, {Pit}, {Antonyuk}, \&
  {Baklanov}}]{2018MNRAS.479..341P}
{Pavlenko}, E.~P., {Mason}, P.~A., {Sosnovskij}, A.~A., {et~al.} 2018, \mnras,
  479, 341, \dodoi{10.1093/mnras/sty1494}

\bibitem[{{Pejcha}(2014)}]{2014ApJ...788...22P}
{Pejcha}, O. 2014, \apj, 788, 22, \dodoi{10.1088/0004-637X/788/1/22}

\bibitem[{{Pejcha} {et~al.}(2016{\natexlab{a}}){Pejcha}, {Metzger}, \&
  {Tomida}}]{2016MNRAS.461.2527P}
{Pejcha}, O., {Metzger}, B.~D., \& {Tomida}, K. 2016{\natexlab{a}}, \mnras,
  461, 2527, \dodoi{10.1093/mnras/stw1481}

\bibitem[{{Pejcha} {et~al.}(2016{\natexlab{b}}){Pejcha}, {Metzger}, \&
  {Tomida}}]{2016MNRAS.455.4351P}
---. 2016{\natexlab{b}}, \mnras, 455, 4351, \dodoi{10.1093/mnras/stv2592}

\bibitem[{{Pejcha} {et~al.}(2017){Pejcha}, {Metzger}, {Tyles}, \&
  {Tomida}}]{2017ApJ...850...59P}
{Pejcha}, O., {Metzger}, B.~D., {Tyles}, J.~G., \& {Tomida}, K. 2017, \apj,
  850, 59, \dodoi{10.3847/1538-4357/aa95b9}

\bibitem[{{Poggiani}(2018)}]{2018arXiv180707947P}
{Poggiani}, R. 2018, arXiv e-prints, arXiv:1807.07947,
  \dodoi{10.48550/arXiv.1807.07947}

\bibitem[{{Prialnik} \& {Kovetz}(1995)}]{1995ApJ...445..789P}
{Prialnik}, D., \& {Kovetz}, A. 1995, \apj, 445, 789, \dodoi{10.1086/175741}

\bibitem[{{Prialnik} \& {Livio}(1995)}]{1995PASP..107.1201P}
{Prialnik}, D., \& {Livio}, M. 1995, \pasp, 107, 1201, \dodoi{10.1086/133678}

\bibitem[{{Qian} \& {Zhao}(2026)}]{2026NewA..12602540Q}
{Qian}, M.-Y., \& {Zhao}, E.-G. 2026, \na, 126, 102540,
  \dodoi{10.1016/j.newast.2026.102540}

\bibitem[{{Quimby} {et~al.}(2024){Quimby}, {Metzger}, {Shen}, {Shafter},
  {Corbett}, \& {Overton}}]{2024ApJ...977...17Q}
{Quimby}, R.~M., {Metzger}, B.~D., {Shen}, K.~J., {et~al.} 2024, \apj, 977, 17,
  \dodoi{10.3847/1538-4357/ad887f}

\bibitem[{{Rayner} {et~al.}(1998){Rayner}, {Toomey}, {Onaka}, {Denault},
  {Stahlberger}, {Watanabe}, \& {Wang}}]{1998SPIE.3354..468R}
{Rayner}, J.~T., {Toomey}, D.~W., {Onaka}, P.~M., {et~al.} 1998, in Society of
  Photo-Optical Instrumentation Engineers (SPIE) Conference Series, Vol. 3354,
  Infrared Astronomical Instrumentation, ed. A.~M. {Fowler}, 468--479,
  \dodoi{10.1117/12.317273}

\bibitem[{{Rector} {et~al.}(2022){Rector}, {Shafter}, {Burris}, {Walentosky},
  {Viafore}, {Strom}, {Cool}, {Sola}, {Crayton}, {Pilachowski}, {Jacoby},
  {Corbett}, {Rene}, \& {Hernandez}}]{2022ApJ...936..117R}
{Rector}, T.~A., {Shafter}, A.~W., {Burris}, W.~A., {et~al.} 2022, \apj, 936,
  117, \dodoi{10.3847/1538-4357/ac87ad}

\bibitem[{{Ren} {et~al.}(2021){Ren}, {de Grijs}, {Zhang}, {Deng}, {Chen},
  {Matsunaga}, {Liu}, {Sun}, {Maehara}, {Ukita}, \&
  {Kobayashi}}]{2021AJ....161..176R}
{Ren}, F., {de Grijs}, R., {Zhang}, H., {et~al.} 2021, \aj, 161, 176,
  \dodoi{10.3847/1538-3881/abe30e}

\bibitem[{{Ricker} {et~al.}(2015){Ricker}, {Winn}, {Vanderspek}, {Latham},
  {Bakos}, {Bean}, {Berta-Thompson}, {Brown}, {Buchhave}, {Butler}, {Butler},
  {Chaplin}, {Charbonneau}, {Christensen-Dalsgaard}, {Clampin}, {Deming},
  {Doty}, {De Lee}, {Dressing}, {Dunham}, {Endl}, {Fressin}, {Ge}, {Henning},
  {Holman}, {Howard}, {Ida}, {Jenkins}, {Jernigan}, {Johnson}, {Kaltenegger},
  {Kawai}, {Kjeldsen}, {Laughlin}, {Levine}, {Lin}, {Lissauer}, {MacQueen},
  {Marcy}, {McCullough}, {Morton}, {Narita}, {Paegert}, {Palle}, {Pepe},
  {Pepper}, {Quirrenbach}, {Rinehart}, {Sasselov}, {Sato}, {Seager},
  {Sozzetti}, {Stassun}, {Sullivan}, {Szentgyorgyi}, {Torres}, {Udry}, \&
  {Villasenor}}]{2015JATIS...1a4003R}
{Ricker}, G.~R., {Winn}, J.~N., {Vanderspek}, R., {et~al.} 2015, Journal of
  Astronomical Telescopes, Instruments, and Systems, 1, 014003,
  \dodoi{10.1117/1.JATIS.1.1.014003}

\bibitem[{{Ricker} \& {Taam}(2012)}]{2012ApJ...746...74R}
{Ricker}, P.~M., \& {Taam}, R.~E. 2012, \apj, 746, 74,
  \dodoi{10.1088/0004-637X/746/1/74}

\bibitem[{{Savitzky} \& {Golay}(1964)}]{1964AnaCh..36.1627S}
{Savitzky}, A., \& {Golay}, M.~J.~E. 1964, Analytical Chemistry, 36, 1627,
  \dodoi{10.1021/ac60214a047}

\bibitem[{{Scargle}(1982)}]{1982ApJ...263..835S}
{Scargle}, J.~D. 1982, \apj, 263, 835, \dodoi{10.1086/160554}

\bibitem[{{Schaefer}(2020)}]{2020MNRAS.492.3323S}
{Schaefer}, B.~E. 2020, \mnras, 492, 3323, \dodoi{10.1093/mnras/stz3325}

\bibitem[{{Schaefer}(2022{\natexlab{a}})}]{2022MNRAS.517.3640S}
---. 2022{\natexlab{a}}, \mnras, 517, 3640, \dodoi{10.1093/mnras/stac2089}

\bibitem[{{Schaefer}(2022{\natexlab{b}})}]{2022MNRAS.517.6150S}
---. 2022{\natexlab{b}}, \mnras, 517, 6150, \dodoi{10.1093/mnras/stac2900}

\bibitem[{{Schaefer}(2023)}]{2023MNRAS.525..785S}
---. 2023, \mnras, 525, 785, \dodoi{10.1093/mnras/stad2223}

\bibitem[{{Schaefer}(2025)}]{2025ApJ...993..232S}
---. 2025, \apj, 993, 232, \dodoi{10.3847/1538-4357/ae0616}

\bibitem[{{Schenker}(1998)}]{1998ASPC..137..483S}
{Schenker}, K. 1998, in Astronomical Society of the Pacific Conference Series,
  Vol. 137, Wild Stars in the Old West, ed. S.~{Howell}, E.~{Kuulkers}, \&
  C.~{Woodward}, 483

\bibitem[{{Schenker}(2002)}]{2002ASPC..259..580S}
{Schenker}, K. 2002, in Astronomical Society of the Pacific Conference Series,
  Vol. 259, IAU Colloq. 185: Radial and Nonradial Pulsationsn as Probes of
  Stellar Physics, ed. C.~{Aerts}, T.~R. {Bedding}, \&
  J.~{Christensen-Dalsgaard}, 580, \dodoi{10.48550/arXiv.astro-ph/0109206}

\bibitem[{{Schenker} \& {Gautschy}(1998)}]{1998ASPC..135..116S}
{Schenker}, K., \& {Gautschy}, A. 1998, in Astronomical Society of the Pacific
  Conference Series, Vol. 135, A Half Century of Stellar Pulsation
  Interpretation, ed. P.~A. {Bradley} \& J.~A. {Guzik}, 116

\bibitem[{{Schlegel} {et~al.}(1998){Schlegel}, {Finkbeiner}, \&
  {Davis}}]{1998ApJ...500..525S}
{Schlegel}, D.~J., {Finkbeiner}, D.~P., \& {Davis}, M. 1998, \apj, 500, 525,
  \dodoi{10.1086/305772}

\bibitem[{{Schmidt}(2016)}]{2016MPBu...43..129S}
{Schmidt}, R.~E. 2016, Minor Planet Bulletin, 43, 129

\bibitem[{{Schmidt}(2020)}]{2020JAVSO..48...13S}
---. 2020, \jaavso, 48, 13

\bibitem[{{Schmidt}(2022)}]{2022JAVSO..50..260S}
---. 2022, \jaavso, 50, 260

\bibitem[{{Schwarz} {et~al.}(2011){Schwarz}, {Ness}, {Osborne}, {Page},
  {Evans}, {Beardmore}, {Walter}, {Helton}, {Woodward}, {Bode}, {Starrfield},
  \& {Drake}}]{2011ApJS..197...31S}
{Schwarz}, G.~J., {Ness}, J.-U., {Osborne}, J.~P., {et~al.} 2011, \apjs, 197,
  31, \dodoi{10.1088/0067-0049/197/2/31}

\bibitem[{{Semeniuk} {et~al.}(1976){Semeniuk}, {Kruszewski}, \&
  {Schwarzenberg-Czerny}}]{1976IBVS.1157....1S}
{Semeniuk}, I., {Kruszewski}, A., \& {Schwarzenberg-Czerny}, A. 1976,
  Information Bulletin on Variable Stars, 1157, 1

\bibitem[{{Shafter}(2017)}]{2017ApJ...834..196S}
{Shafter}, A.~W. 2017, \apj, 834, 196, \dodoi{10.3847/1538-4357/834/2/196}

\bibitem[{{Shafter} \& {Hornoch}(2026)}]{2026ApJS..283...24S}
{Shafter}, A.~W., \& {Hornoch}, K. 2026, \apjs, 283, 24,
  \dodoi{10.3847/1538-4365/ae3a86}

\bibitem[{{Shafter} {et~al.}(2009){Shafter}, {Rau}, {Quimby}, {Kasliwal},
  {Bode}, {Darnley}, \& {Misselt}}]{2009ApJ...690.1148S}
{Shafter}, A.~W., {Rau}, A., {Quimby}, R.~M., {et~al.} 2009, \apj, 690, 1148,
  \dodoi{10.1088/0004-637X/690/2/1148}

\bibitem[{{Shara} {et~al.}(2018){Shara}, {Prialnik}, {Hillman}, \&
  {Kovetz}}]{2018ApJ...860..110S}
{Shara}, M.~M., {Prialnik}, D., {Hillman}, Y., \& {Kovetz}, A. 2018, \apj, 860,
  110, \dodoi{10.3847/1538-4357/aabfbd}

\bibitem[{{Shaviv}(2001)}]{2001MNRAS.326..126S}
{Shaviv}, N.~J. 2001, \mnras, 326, 126,
  \dodoi{10.1046/j.1365-8711.2001.04574.x}

\bibitem[{{Shaviv}(2002)}]{2002ASPC..261..585S}
{Shaviv}, N.~J. 2002, in Astronomical Society of the Pacific Conference Series,
  Vol. 261, The Physics of Cataclysmic Variables and Related Objects, ed. B.~T.
  {G{\"a}nsicke}, K.~{Beuermann}, \& K.~{Reinsch}, 585

\bibitem[{{Shen} \& {Quataert}(2022)}]{2022ApJ...938...31S}
{Shen}, K.~J., \& {Quataert}, E. 2022, \apj, 938, 31,
  \dodoi{10.3847/1538-4357/ac9136}

\bibitem[{{Shore}(2014)}]{Shore_etal_2014}
{Shore}, S.~N. 2014, in Astronomical Society of the Pacific Conference Series,
  Vol. 490, Stellar Novae: Past and Future Decades, ed. P.~A. {Woudt} \&
  V.~A.~R.~M. {Ribeiro}, 145

\bibitem[{{Shu} {et~al.}(1979){Shu}, {Lubow}, \&
  {Anderson}}]{1979ApJ...229..223S}
{Shu}, F.~H., {Lubow}, S.~H., \& {Anderson}, L. 1979, \apj, 229, 223,
  \dodoi{10.1086/156948}

\bibitem[{{Shugarov}(1967)}]{1967ATsir.447....7S}
{Shugarov}, S.~Y. 1967, Astronomicheskij Tsirkulyar, 447, 7

\bibitem[{{Shugarov} {et~al.}(2005){Shugarov}, {Goranskij}, {Katysheva},
  {Kusakin}, {Metlova}, {Volkov}, {Chochol}, {Pribulla}, {Karitskaja},
  {Retter}, {Shemmer}, \& {Lipkin}}]{2005Ap&SS.296..431S}
{Shugarov}, S.~Y., {Goranskij}, V.~P., {Katysheva}, N.~A., {et~al.} 2005,
  \apss, 296, 431, \dodoi{10.1007/s10509-005-4864-6}

\bibitem[{{Skrutskie} {et~al.}(2006){Skrutskie}, {Cutri}, {Stiening},
  {Weinberg}, {Schneider}, {Carpenter}, {Beichman}, {Capps}, {Chester},
  {Elias}, {Huchra}, {Liebert}, {Lonsdale}, {Monet}, {Price}, {Seitzer},
  {Jarrett}, {Kirkpatrick}, {Gizis}, {Howard}, {Evans}, {Fowler}, {Fullmer},
  {Hurt}, {Light}, {Kopan}, {Marsh}, {McCallon}, {Tam}, {Van Dyk}, \&
  {Wheelock}}]{2006AJ....131.1163S}
{Skrutskie}, M.~F., {Cutri}, R.~M., {Stiening}, R., {et~al.} 2006, \aj, 131,
  1163, \dodoi{10.1086/498708}

\bibitem[{{Smith} {et~al.}(2020){Smith}, {Smartt}, {Young}, {Tonry}, {Denneau},
  {Flewelling}, {Heinze}, {Weiland}, {Stalder}, {Rest}, {Stubbs}, {Anderson},
  {Chen}, {Clark}, {Do}, {F{\"o}rster}, {Fulton}, {Gillanders}, {McBrien},
  {O'Neill}, {Srivastav}, \& {Wright}}]{2020PASP..132h5002S}
{Smith}, K.~W., {Smartt}, S.~J., {Young}, D.~R., {et~al.} 2020, \pasp, 132,
  085002, \dodoi{10.1088/1538-3873/ab936e}

\bibitem[{{Sokolovsky} \& {Lebedev}(2018)}]{2018A&C....22...28S}
{Sokolovsky}, K.~V., \& {Lebedev}, A.~A. 2018, Astronomy and Computing, 22, 28,
  \dodoi{10.1016/j.ascom.2017.12.001}

\bibitem[{{Sokolovsky} {et~al.}(2022){Sokolovsky}, {Strader}, {Swihart},
  {Aydi}, {Bahramian}, {Chomiuk}, {Heinke}, {Hughes}, {Li}, {de Oliveira},
  {Miller-Jones}, {Mukai}, {Sand}, {Shishkovsky}, {Tremou}, \&
  {Voggel}}]{2022ApJ...934..142S}
{Sokolovsky}, K.~V., {Strader}, J., {Swihart}, S.~J., {et~al.} 2022, \apj, 934,
  142, \dodoi{10.3847/1538-4357/ac7b25}

\bibitem[{{Sokolovsky} {et~al.}(2023){Sokolovsky}, {Aydi}, {Malanchev},
  {Burke}, {Mukai}, {Sokoloski}, {Metzger}, {Atapin}, {Belinski}, {Chen},
  {Chomiuk}, {Dubovsky}, {Faucher-Giguere}, {Hounsell}, {Ikonnikova}, {Lander},
  {Li}, {Linford}, {Mioduszewski}, {Molina}, {Munari}, {Potanin}, {Quimby},
  {Rupen}, {Scaringi}, {Shatsky}, {Shen}, {Steinberg}, {Stone}, {Tatarnikov},
  {Vurm}, {Williams}, {Agudo Azcona}, {Boyd}, {Bean}, {Braunwarth},
  {Blackwell}, {Bolzoni}, {Casas}, {Cejudo Fernandez}, {Dubois}, {Foster},
  {Farfan}, {Galdies}, {Hodge}, {Prieto Gallego}, {Lane}, {Larsson}, {Lindner},
  {Logie}, {Mantero}, {Morales Aimar}, {Menzies}, {Nakonechny}, {Philpot},
  {Padilla Filho}, {Ramey}, {Rau}, {Reina}, {Romanov}, {Ruocco}, {Shears},
  {Serreau}, {Schmidt}, {Solomonov}, {Tracy}, {Tulloch}, {Tomlin}, {Tordai},
  {Vanaverbeke}, {Wenzel}, {Maitan}, \& {Moretti}}]{2023arXiv231104903S}
{Sokolovsky}, K.~V., {Aydi}, E., {Malanchev}, K., {et~al.} 2023, arXiv
  e-prints, arXiv:2311.04903, \dodoi{10.48550/arXiv.2311.04903}

\bibitem[{{Sparks} \& {Sion}(2021)}]{2021ApJ...914....5S}
{Sparks}, W.~M., \& {Sion}, E.~M. 2021, \apj, 914, 5,
  \dodoi{10.3847/1538-4357/abf2bc}

\bibitem[{{Starrfield} {et~al.}(2016){Starrfield}, {Iliadis}, \&
  {Hix}}]{2016PASP..128e1001S}
{Starrfield}, S., {Iliadis}, C., \& {Hix}, W.~R. 2016, \pasp, 128, 051001,
  \dodoi{10.1088/1538-3873/128/963/051001}

\bibitem[{{Stockman} {et~al.}(1988){Stockman}, {Schmidt}, \&
  {Lamb}}]{1988ApJ...332..282S}
{Stockman}, H.~S., {Schmidt}, G.~D., \& {Lamb}, D.~Q. 1988, \apj, 332, 282,
  \dodoi{10.1086/166652}

\bibitem[{{Strope} {et~al.}(2010){Strope}, {Schaefer}, \&
  {Henden}}]{2010AJ....140...34S}
{Strope}, R.~J., {Schaefer}, B.~E., \& {Henden}, A.~A. 2010, \aj, 140, 34,
  \dodoi{10.1088/0004-6256/140/1/34}

\bibitem[{{Taguchi} {et~al.}(2023){Taguchi}, {Maeda}, {Maehara}, {Tajitsu},
  {Yamanaka}, {Arai}, {Isogai}, {Shibata}, {Tampo}, {Kojiguchi}, {Nogami}, \&
  {Kato}}]{2023ApJ...958..156T}
{Taguchi}, K., {Maeda}, K., {Maehara}, H., {et~al.} 2023, \apj, 958, 156,
  \dodoi{10.3847/1538-4357/ad0133}

\bibitem[{{Tavleev} {et~al.}(2024){Tavleev}, {Ducci}, {Suleimanov}, {Maitra},
  {Werner}, {Santangelo}, \& {Doroshenko}}]{2024A&A...689A.335T}
{Tavleev}, A., {Ducci}, L., {Suleimanov}, V.~F., {et~al.} 2024, \aap, 689,
  A335, \dodoi{10.1051/0004-6361/202451195}

\bibitem[{{Tempesti}(1975)}]{1975IBVS.1052....1T}
{Tempesti}, P. 1975, Information Bulletin on Variable Stars, 1052, 1

\bibitem[{{Terzan} {et~al.}(1974){Terzan}, {Bally}, \&
  {Durand}}]{1974A&AS...15..107T}
{Terzan}, A., {Bally}, M., \& {Durand}, A. 1974, \aaps, 15, 107

\bibitem[{{Thompson}(2017)}]{2017MNRAS.470.4061T}
{Thompson}, W.~T. 2017, \mnras, 470, 4061, \dodoi{10.1093/mnras/stx1552}

\bibitem[{{Thoroughgood} {et~al.}(2001){Thoroughgood}, {Dhillon}, {Littlefair},
  {Marsh}, \& {Smith}}]{2001MNRAS.327.1323T}
{Thoroughgood}, T.~D., {Dhillon}, V.~S., {Littlefair}, S.~P., {Marsh}, T.~R.,
  \& {Smith}, D.~A. 2001, \mnras, 327, 1323,
  \dodoi{10.1046/j.1365-8711.2001.04828.x}

\bibitem[{{Tonry} {et~al.}(2018){Tonry}, {Denneau}, {Heinze}, {Stalder},
  {Smith}, {Smartt}, {Stubbs}, {Weiland}, \& {Rest}}]{2018PASP..130f4505T}
{Tonry}, J.~L., {Denneau}, L., {Heinze}, A.~N., {et~al.} 2018, \pasp, 130,
  064505, \dodoi{10.1088/1538-3873/aabadf}

\bibitem[{{Tylenda} {et~al.}(2011){Tylenda}, {Hajduk}, {Kami{\'n}ski},
  {Udalski}, {Soszy{\'n}ski}, {Szyma{\'n}ski}, {Kubiak}, {Pietrzy{\'n}ski},
  {Poleski}, {Wyrzykowski}, \& {Ulaczyk}}]{2011A&A...528A.114T}
{Tylenda}, R., {Hajduk}, M., {Kami{\'n}ski}, T., {et~al.} 2011, \aap, 528,
  A114, \dodoi{10.1051/0004-6361/201016221}

\bibitem[{{Vacca} {et~al.}(2003){Vacca}, {Cushing}, \&
  {Rayner}}]{2003PASP..115..389V}
{Vacca}, W.~D., {Cushing}, M.~C., \& {Rayner}, J.~T. 2003, \pasp, 115, 389,
  \dodoi{10.1086/346193}

\bibitem[{{Valisa} {et~al.}(2023){Valisa}, {Munari}, {Dallaporta}, {Maitan}, \&
  {Vagnozzi}}]{2023arXiv230204656V}
{Valisa}, P., {Munari}, U., {Dallaporta}, S., {Maitan}, A., \& {Vagnozzi}, A.
  2023, arXiv e-prints, arXiv:2302.04656, \dodoi{10.48550/arXiv.2302.04656}

\bibitem[{{VanderPlas}(2018)}]{2018ApJS..236...16V}
{VanderPlas}, J.~T. 2018, \apjs, 236, 16, \dodoi{10.3847/1538-4365/aab766}

\bibitem[{{Vanderspek} {et~al.}(2018){Vanderspek}, {Doty}, {Fausnaugh},
  {Villase{\~n}or}, {Jenkins}, {Berta-Thompson}, {Burke}, \&
  {Ricker}}]{tesshandbook}
{Vanderspek}, R., {Doty}, J.~P., {Fausnaugh}, M., {et~al.} 2018, {TESS
  Instrument Handbook} (TESS Science Office).
\newblock
  \url{https://archive.stsci.edu/files/live/sites/mast/files/home/missions-and-data/active-missions/tess/_documents/TESS_Instrument_Handbook_v0.1.pdf}

\bibitem[{{Vogt}(1990)}]{1990ApJ...356..609V}
{Vogt}, N. 1990, \apj, 356, 609, \dodoi{10.1086/168866}

\bibitem[{{Warner}(1996)}]{1996Ap&SS.241..263W}
{Warner}, B. 1996, \apss, 241, 263, \dodoi{10.1007/BF00645229}

\bibitem[{{Williams} {et~al.}(2022){Williams}, {Walter}, {Rudy}, {Munari},
  {Luckas}, {Subasavage}, \& {Mauerhan}}]{2022ApJ...941..138W}
{Williams}, R., {Walter}, F.~M., {Rudy}, R.~J., {et~al.} 2022, \apj, 941, 138,
  \dodoi{10.3847/1538-4357/aca2a9}

\bibitem[{{Williams} {et~al.}(2016){Williams}, {Darnley}, {Bode}, \&
  {Shafter}}]{2016ApJ...817..143W}
{Williams}, S.~C., {Darnley}, M.~J., {Bode}, M.~F., \& {Shafter}, A.~W. 2016,
  \apj, 817, 143, \dodoi{10.3847/0004-637X/817/2/143}

\bibitem[{{Yaron} {et~al.}(2005){Yaron}, {Prialnik}, {Shara}, \&
  {Kovetz}}]{2005ApJ...623..398Y}
{Yaron}, O., {Prialnik}, D., {Shara}, M.~M., \& {Kovetz}, A. 2005, \apj, 623,
  398, \dodoi{10.1086/428435}

\bibitem[{{Zuckerman} {et~al.}(2023){Zuckerman}, {De}, {Eilers}, {Meisner}, \&
  {Panagiotou}}]{2023MNRAS.523.3555Z}
{Zuckerman}, L., {De}, K., {Eilers}, A.-C., {Meisner}, A.~M., \& {Panagiotou},
  C. 2023, \mnras, 523, 3555, \dodoi{10.1093/mnras/stad1625}

\end{thebibliography}
\bibliographystyle{aasjournal}



\end{document}